\documentclass[spanish,12pt]{book}

 \usepackage{babel}
 \title{Monograf\'ia de Estad\'istica Bayesiana}
 \author{Dr. Arturo Erdely Ruiz $^1$ { } \\ Dr. Eduardo Guti\'errez Pe\~na $^2$ }
 \date{\small $^1$ FES -- Acatl\'an, Universidad Nacional Aut\'onoma de M\'exico \hspace{1cm} \\ $^2$ IIMAS, Universidad Nacional Aut\'onoma de M\'exico }

 \usepackage{amsthm}
 \usepackage{amssymb}
 \usepackage{amsmath}
 \usepackage{enumerate}
 \usepackage{graphics,graphpap,epsfig,pb-diagram}
 
 \newcommand{\prob}{\mathbb{P}}
 \newcommand{\esper}{\mathbb{E}}
 \newcommand{\vari}{\mathbb{V}}
 \newcommand{\indic}{\textbf{\textsf{\large{1}}}}
 
 \newcommand{\autor}[1]{\noindent\hspace{-5mm}{#1} }
 \newcommand{\anio}[1]{({#1}). }
 \newcommand{\articulo}[1]{{#1}. }
 \newcommand{\revista}[1]{\textit{#1} }
 \newcommand{\volumen}[1]{\textbf{#1}, }
 \newcommand{\pags}[1]{{#1}.}
 \newcommand{\libro}[1]{\textit{#1}, }
 \newcommand{\editorial}[1]{{#1}.}
 \newcommand{\salta}{\par\smallskip}

\begin{document}

 \maketitle
 \tableofcontents
 
 \theoremstyle{plain}
 \newtheorem{enun}{}[chapter]
 \newenvironment{defn}{\emph{\textbf{Definici\'on. }}\rm}{}
 \newenvironment{lema}{\emph{\textbf{Lema. }}}{}
 \newenvironment{prop}{\emph{\textbf{Proposici\'on. }}}{}
 \newenvironment{teor}{\emph{\textbf{Teorema. }}}{}
 \newenvironment{cor}{\emph{\textbf{Corolario. }}}{}
   
 \theoremstyle{remark}
 \newtheorem{ejem}{\textbf{Ejemplo}}
 
 \newenvironment{obs}{\textsl{Observaci\'on:}\sl}{}
 
 \newenvironment{cita}{\begin{quotation}\begin{sffamily}}{\end{sffamily}\end{quotation}}
 
 \newenvironment{grande}{\large}{}

 %\frontmatter
 
 %\chapter{\prefacename}
 
 %[ Se escribir\'a posteriormente ]

 \mainmatter

 \chapter{Introducci\'{o}n}

 \section{Breve rese\~na hist\'orica}
 
 Considerando el conjunto de libros de la bibliograf\'ia que se incluye al final de esta monograf\'ia, sin duda el libro de Press (2003) es el que m\'as espacio destina a hacer una rese\~na hist\'orica sobre los or\'igenes y desarrollo de la \textit{Estad\'istica Bayesiana}, particularmente sobre Thomas Bayes, a quien dedica parte importante del primer cap\'itulo, adem\'as de cuatro ap\'endices. Aprovecharemos ese invaluable trabajo para lograr una breve rese\~na, en el entendido de que el lector interesado en m\'as detalles debiera, en principio, consultar la mencionada referencia.\par
 
 \smallskip
 
 El trabajo de Thomas Bayes, publicado de manera p\'ostuma en 1763, ha tenido una importante consecuencia en la forma de hacer inferencia estad\'istica: provee una manera formal de combinar el conocimiento \textit{a priori} (o inicial) que se tiene sobre un fen\'omeno, con el nuevo conocimiento que se adquiere a partir de nuevos datos y mediciones sobre el mismo, obteniendo as\'i un conocimiento \textit{a posteriori} (o final), es decir, el conocimiento a priori se \textit{actualiza} con la nueva informaci\'on, y dicho conocimiento a posteriori se convertir\'a en el nuevo conocimiento a priori, a la espera, otra vez, de nueva informaci\'on que lo actualice.
 
 \smallskip
 
 Sin embargo, existieron otras figuras en la comunidad cient\'ifica que contribuyeron de manera importante al desarrollo del enfoque bayesiano, e incluso algunos de ellos precedieron a Thomas Bayes. Por ejemplo, la idea esencial subyacente en el teorema atribuido a Bayes, el problema de la \textit{probabilidad inversa}, se encuentra en el libro de James Bernoulli (1713), y aunque su autor no le dio una estructura matem\'atica, abord\'o problemas particulares al respecto. De hecho, el trabajo de Bayes (1763) aborda directamente el problema de hacer inferencias sobre el par\'ametro de una distribuci\'on binomial, condicional en algunas observaciones de dicha distribuci\'on, y es este caso particular lo que originalmente debiera conocerse como el \textit{Teorema de Bayes.} Seg\'un Press (2003), parece ser que Bayes se dio cuenta de que su resultado pod\'ia generalizarse m\'as all\'a de la distribuci\'on binomial, pero en lo que de \'el se public\'o se limita a la distribuci\'on binomial. Fue Laplace (1774) quien enunci\'o el teorema sobre probabilidad inversa en su forma general (caso discreto), y seg\'un Stigler (1986 a,b), es muy probable que Laplace jam\'as haya conocido el famoso ensayo de Bayes, y por tanto que haya descubierto este resultado de forma totalmente independiente, ya que Bayes llev\'o a cabo su trabajo en Inglaterra, y sus resultados fueron ignorados por m\'as de 20 a\~nos, y Laplace desarroll\'o su trabajo en Francia. De hecho fue Jeffreys (1939) quien redescubri\'o el trabajo de Laplace en este campo. Quiz\'as ser\'ia igualmente justo hacer referencia a la \textit{Estad\'istica Laplaciana.}
 
 \smallskip
 
 Respecto al desarrollo formal de lo que hoy se conoce como \textit{Estad\'istica Bayesiana,} en uno de los ap\'endices del libro de Press (2003) se presenta la metodolog\'ia y resultados de una consulta a un panel de expertos para seleccionar a las personalidades que debieran, a juicio de dicho panel, integrar el ``Sal\'on Bayesiano de la Fama'', con los siguientes resultados: 
 \begin{itemize}
   \item Thomas Bayes (1701? - 1761),
   \item Bruno De Finetti (1906 - 1985),
   \item Morris De Groot (1931 - 1989),
   \item Harold Jeffreys (1891 - 1989),
   \item Dennis V. Lindley (1923 - \hspace{0.5cm}),
   \item Leonard J. Savage (1917 - 1971).
 \end{itemize}

  De acuerdo a la metodolog\'ia de Press (2003), las siguientes personalidades lograron una ``Menci\'on Honor\'ifica'': James O. Berger, George E.P. Box, Pierre Simone de Laplace, Adrian F.M. Smith, Arnold Zellner. \par
  
  \smallskip
  
  Aunque la omisi\'on de algunos nombres podr\'ia generar una controversia que no es objeto de la presente monograf\'ia, lo importante a destacar son justamente algunos nombres importantes y de obligada referencia para conocer sobre el origen y cimientos de la Estad\'istica Bayesiana. 
  
  \smallskip
  
  Para completar la lista de referencias obligadas sobre el enfoque bayesiano, se recomienda consultar la secci\'on 1.5 del libro de Bernardo y Smith (1994), lista a la que sin duda habr\'ia que a\~nadir esta \'ultima.\par
  
 \section{Enfoques bayesiano versus frecuentista}
 
 \hyphenation{con-fia-bles}
 \hyphenation{ope-rar}
 
 El enfoque de la llamada \textit{estad\'istica frecuentista} no permite incorporar de manera coherente en el an\'alisis estad\'istico la \textit{informaci\'on extra-muestral} disponible, se apoya \'unicamente en datos muestrales observados. Si no hay datos, la estad\'istica frecuentista est\'a imposibilitada para operar. Si hay muy pocos datos, la estad\'istica frecuentista presenta fuertes problemas tambi\'en, pues muchos de sus m\'etodos se apoyan en resultados asint\'oticos, tales como la ley de los grandes n\'umeros, el teorema central del l\'imite, y sus consecuencias, y por ello por lo general requiere de muestras ``grandes'' para que sus resultados sean ``confiables''. En cambio, la \textit{Estad\'istica Bayesiana} aprovecha tanto la informaci\'on que nos proporcionan los datos muestrales as\'i como la \textit{informaci\'on extra-muestral} disponible, entendiendo por esto \'ultimo, de manera informal, toda aquella informaci\'on relevante, adem\'as de los datos, que nos ayude a disminuir nuestra incertidumbre o ignorancia en torno a un fen\'omeno aleatorio de inter\'es. Esquem\'aticamente:\par
 $$\textit{Estad\'istica frecuentista}\qquad \longrightarrow \qquad \textit{s\'olo datos}$$
 $$\textit{          Estad\'istica Bayesiana}\qquad \longrightarrow \qquad \textit{datos }+\textit{ info extra-muestral}$$
 
 En estad\'istica bayesiana, el t\'ermino com\'un para referirse a la informaci\'on extra-muestral es el de \textit{informaci\'on subjetiva,} y es importante aclarar, al menos brevemente, qu\'e se entiende en este contexto por el adjetivo \textit{subjetiva,} ya que en el lenguaje corriente puede tener una connotaci\'on distinta a la que se requiere bajo el enfoque bayesiano. Al hablar de \textit{informaci\'on subjetiva} nos referimos a toda aquella informaci\'on a priori que se tiene en relaci\'on  al fen\'omento aleatorio de inter\'es, antes de recolectar o realizar nuevas mediciones sobre el mismo, y esto incluye: datos hist\'oricos, teor\'ias, opiniones y conjeturas de expertos, conclusiones basadas en estudios previos, etc. El primer paso en la inferencia estad\'istica bayesiana es traducir todo lo anterior en una \textit{distribuci\'on de probabilidad a priori} (o inicial). El segundo paso consiste en recolectar o realizar nuevas mediciones, y \textit{actualizar} la distribuci\'on de probabilidad a priori, para obtener, mediante la \textit{Regla de Bayes,} una \textit{distribuci\'on de probabilidad a posteriori} (o final) y ser\'a esta \'ultima la mejor descripci\'on posible de \textit{nuestra incertidumbre}, de acuerdo a toda \textit{nuestra informaci\'on} disponible, y por tanto ser\'a la herramienta fundamental a partir de la cual se realiza inferencia estad\'istica.\par\smallskip
 
 De una u otra manera, la subjetividad siempre ha estado presente en la actividad cient\'ifica, comenzando por los supuestos sobre los cuales se decide abordar un determinado problema, t\'ipicamente se les denomina \textit{supuestos razonables}, pero son ``razonables'' de acuerdo a la experiencia e informaci\'on (subjetiva) particular de quien o quienes estudian un fen\'omeno en un momento dado. De acuerdo a Wolpert (1992):
    \begin{cita}
      \noindent [\ldots] la idea de una ``objetividad cient\'ifica'' tiene tan solo un valor limitado, ya que el proceso mediante el cual se generan las ideas [o hip\'otesis] cient\'ificas puede ser bastante subjetivo [\ldots] Es una ilusi\'on creer que los cient\'ificos no tienen un cierto v\'inculo emocional con sus convicciones cient\'ificas [\ldots] las teor\'ias cient\'ificas implican una continua interacci\'on con otros cient\'ificos y el conocimiento previamente adquirido [\ldots] as\'i como una explicaci\'on que tenga posibilidades de ser aceptada [o al menos considerada seriamente] por el resto de la comunidad cient\'ifica.
    \end{cita}
    
  De acuerdo a Press (2003) la subjetividad es una parte inherente y requerida para la inferencia estad\'istica, y para el \textit{m\'etodo cient\'ifico.} Sin embargo, ha sido la manera informal y desorganizada en la que ha se ha permitido la presencia de la subjetividad, la responsable de diversos errores y malas interpretaciones en la historia de la ciencia. La estad\'istica bayesiana incorpora de manera formal y fundamentada la informaci\'on subjetiva, y es por ello que Press y Tanur (2001) opinan que la ciencia en general avanzar\'a m\'as r\'apido en la medida en que los m\'etodos modernos del an\'alisis estad\'istico bayesiano reemplacen varios de los m\'etodos cl\'asicos del siglo XX. De hecho, en el libro de Press y Tanur (2001) se hace un recuento hist\'orico sobre la presencia de la subjetividad en el trabajo de relevantes cient\'ificos como Kepler, Mendel, Arist\'oteles, Galileo Galilei, Newton, Darwin, Pasteur, Freud, Einstein, entre otros.\par\smallskip
  
  Un ejemplo t\'ipico es la investigaci\'on cient\'ifica en medicina, en donde comunmente se cuenta con pocos datos, pero al mismo tiempo se cuenta con la valiosa experiencia de m\'edicos muy familiarizados con una enfermedad en estudio. Otro ejemplo es el caso de los mercados financieros, cuya din\'amica es impulsada mucho m\'as por las expectativas futuras de los inversionistas participantes (apreciaciones subjetivas de lo que creen que va a suceder) que por lo que describen las series de tiempo hist\'oricas de los indicadores financieros. En ambos ejemplos, la estad\'istica bayesiana permite el aprovechamiento de la valiosa informaci\'on subjetiva, en contraste con lo poco que puede hacer el enfoque frecuentista de la estad\'istica, con pocos datos (en el ejemplo de medicina) o con datos hist\'oricos con poco poder predictivo (en el ejemplo de mercados financieros).

 \section{Interpretaciones de la Probabilidad}
 
 \hyphenation{te-ne-mos}
 \hyphenation{ob-ser-va-bles}
 
 El objeto de estudio de la \textit{Teor\'ia de la Probabilidad} son los fen\'omenos (o experimentos) aleatorios, mismos que son representados mediante un \textit{espacio de probabilidad,} digamos $(\Omega,\mathcal{F},\prob),$ en donde $\Omega$ es el \textit{espacio muestral} (conjunto cuyos elementos representan resultados posibles del fen\'omeno aleatorio), $\mathcal{F}$ el \textit{espacio de eventos} (de hecho un $\sigma$-\'algebra de subconjuntos de $\Omega$), y una \textit{medida de probabilidad} $\prob:\mathcal{F}\rightarrow\mathbb{R}^+\cup\{0\},$ esto es, una \textit{medida} que como tal debe satisfacer $\prob(\varnothing)=0,$ y para cualesquiera $E_1,E_2,\ldots\in\mathcal{F}$ disjuntos satisface ${\prob(\,\bigcup_{\,n} E_n)=\sum_{\,n}\prob(E_n)},$ y adem\'as $\prob(\Omega)=1.$\par\smallskip
 
  Recordemos que, en estricto sentido, un \textit{evento} es una proposici\'on l\'ogica que puede evaluarse como verdadera o falsa, despu\'es de conocer el resultado de una realizaci\'on (u observaci\'on) del fen\'omeno aleatorio en cuesti\'on, y como consecuencia del \textit{Axioma de Comprensi\'on} de la teor\'ia de conjuntos, a todo evento puede asoci\'arsele un subconjunto de $\Omega.$ Normalmente no se hace esta distinci\'on, y al hablar de un evento, realmente se trabaja directamente con el conjunto que lo representa. En general, no es cierto que todo subconjunto de $\Omega$ representa alg\'un evento, s\'olo es cierto para el caso en que $\Omega$ sea, cuando m\'as, numerable.\par\smallskip
  
  Si el espacio muestral es cuando m\'as numerable, digamos $\Omega:=\{\omega_1,\omega_2,\ldots\},$ basta poder definir una medida de probabilidad $\prob$ sobre la clase  de los \textit{eventos simples} $\mathcal{C}:=\{\{\omega\}:\omega\in\Omega\},$ ya que en este caso particular cualquier elemento del $\sigma$-\'algebra $\mathcal{F}$ se puede expresar como uni\'on disjunta cuando m\'as numerable de elementos de la clase $\mathcal{C},$ y por la propiedad $\sigma$-aditiva de medidas se tiene que
        $$\prob(E)=\prob\bigg(\bigcup_{\,\omega\,\in\,E}\{\omega\}\bigg)=\sum_{\omega\,\in\,E}\prob(\{\omega\})\,,\quad \text{para todo evento }\,E\in\mathcal{F}.$$
        
  Si $\Omega$ tiene al menos dos elementos, existe una infinidad no numerable de distintas medidas de probabilidad que podr\'ian definirse sobe el espacio de probabilidad correspondiente. ?`Cu\'al de todas ellas debe utilizarse? El problema de determinar una medida de probabilidad adecuada para un fen\'omeno aleatorio dado, y en dado caso verificar su unicidad, depende de \textit{nuestra informaci\'on} acerca del mismo.\par\smallskip  
  
  Considere el experimento aleatorio de lanzar un dado una vez (no sabemos si es un dado equilibrado o no). Como un dado tiene 6 caras podemos etiquetar cada cara con un n\'umero y expresar su espacio muestral como $\Omega:=\{1,2,3,4,5,6\}.$ Definiremos tres funciones distintas sobre el espacio de eventos $\mathcal{F}$, esto es, $\prob_1,\prob_2,\prob_3:\mathcal{F}\rightarrow[\,0,1\,]$ tales que para todo evento $E\in\mathcal{F}:$
     $$\prob_1(E):=\frac{|E|}{6}\,,$$
     $$\prob_2(E):=\frac{|E|}{10}+\frac{2}{5}\,\indic_E(4)\,,$$
     $$\prob_3(E):=\indic_E(4)\,.$$
en donde $|E|$ representa el n\'umero de elementos del conjunto $E.$ Es f\'acil verificar que las tres funciones anteriores son medidas de probabilidad para el mismo espacio medible $(\Omega,\mathcal{F})$ arriba descrito. Y de hecho existe una infinidad no numerable de medidas de probabilidad que se podr\'ian definir para este experimento aleatorio.\par\smallskip 
 
     En lo anterior surge la inquietud sobre cu\'al medida de probabilidad resulta ser la m\'as adecuada, de entre esa infinidad no numerable de medidas de probabilidad posibles. Esto depende de la informaci\'on que se tenga en un momento dado sobre el experimento aleatorio. Es un sencillo ejercicio verificar lo siguiente: si se tratara de un dado equilibrado (o si al menos creemos razonable suponer que as\'i es) entonces $\prob_1$ es la medida de probabilidad adecuada; si se tratara de un dado cargado de modo que exista una probabilidad de $\frac{1}{2}$ de que salga un 4 y una probabilidad de $\frac{1}{2}$ de que salga cualquier otro n\'umero que no sea 4 (o si al menos creemos razonable suponer que as\'i es), entonces $\prob_2$ resulta ser la medida de probabilidad adecuada. ?`Qu\'e estado de informaci\'on representa $\prob_3$? Existe otro estado de informaci\'on bajo el cual $\prob_1$ podr\'ia considerarse una medida de probabilidad adecuada: si no se tiene informaci\'on alguna (?`por qu\'e?)
   
 \bigskip
 
 \textit{\textbf{Enfoque cl\'asico.}} Si un experimento o fen\'omeno aleatorio puede ocurrir de $n$ maneras diferentes mutuamente excluyentes e igualmente probables, nos encontramos ante el caso de un espacio muestral finito $\Omega:=\{\omega_1,\omega_2,\ldots,\omega_n\}$ en donde cualquier medida de probabilidad que se defina sobre el espacio de eventos $\mathcal{F}$ debe satisfacer una \textit{condici\'on de equiprobabilidad} sobre la clase de los eventos simples $\mathcal{C}:=\{\{\omega\}:\omega\in\Omega\},$ esto es
    $$\prob(\{\omega_1\})=\prob(\{\omega_2\})=\cdots=\prob(\{\omega_n\})$$
 
 Es un sencillo ejercicio de probabilidad demostrar que la \'unica soluci\'on posible bajo la condici\'on anterior es
     $$\prob(E)=\frac{|E|}{n}\,,\quad\text{para todo }\,E\in\mathcal{F}.$$

   Lo anterior implica que para todos aquellos fen\'omenos aleatorios con espacio muestral finito en los que, de acuerdo a \textit{nuestra informaci\'on,} resulte razonable suponer o considerar una condici\'on de equiprobabilidad, el universo de posibles medidas de probabilidad se reduce a una sola.\par\smallskip
     
  El caso anterior es lo que se conoce como el enfoque cl\'asico de la probabilidad, y tiene la limitante de que relativamente pocos problemas reales de inter\'es pueden reducirse a lo anterior.
 
 \bigskip
 
 \textit{\textbf{Enfoque frecuentista.}} Bajo este enfoque, normalmente se dice que la probabilidad de un evento $A$ est\'a dada por:
 $$\mathbb{P}(A)=\lim_{n\,\rightarrow\,\infty}\frac{f_A(n)}{n}$$
 donde $f_A(n)$ es el n\'umero de veces que ocurre el evento $A$ en $n$ repeticiones id\'enticas e independientes del experimento o fen\'omeno aleatorio. Este enfoque presume de ser objetivo porque se basa s\'olo en datos observables pero:
 \begin{itemize}
 \item Tenemos que $\lim_{n\rightarrow\infty}\frac{f_A(n)}{n}=\mathbb{P}(A)$ si y s\'olo si para todo valor $\varepsilon>0$ existe un n\'umero natural $k$ tal que si $n>k$ entonces $|\frac{f_A(n)}{n}-\mathbb{P}(A)|<\varepsilon,$ lo cual \textbf{NO} se puede garantizar ya que bien puede existir $n_0>k$ tal que $|\frac{f_A(n_0)}{n_0}-\mathbb{P}(A)\Big|>\varepsilon\,$ (por ejemplo, alguna racha de ocurrencias sucesivas del evento $A$ suficientemente grande).
 \item $f_A$ no es una funci\'on determinista como las que se utilizan en la teor\'ia del c\'alculo para definir l\'imites, as\'i que primero se tendr\'ia que aclarar qu\'e definici\'on de l\'imite se est\'a ocupando.
 \item El decir que se tienen repeticiones id\'enticas e independientes, fuera de los juegos de azar, es una apreciaci\'on subjetiva.  
 \item En la pr\'actica $n$ nunca se va a $\infty$, as\'i que no hay manera de comprobar emp\'iricamente dicho l\'imite.
 \end{itemize}
 
 \bigskip
 
 \textit{\textbf{Enfoque subjetivo.}} La probabilidad de un evento $A$ es una medida del grado de creencia que tiene un individuo en la ocurrencia de $A$ con base en la informaci\'on $K$ que dicho individuo posee. Bajo este enfoque toda probabilidad es condicional en la informaci\'on de la cual se dispone.\par
 Por ejemplo, sea $A$ el evento de que est\'e lloviendo en el centro de la Ciudad de M\'exico. Para un individuo que vive en el Polo Sur, totalmente aislado del resto del mundo, tendr\'iamos que si $K_1$ denota la informaci\'on (total ignorancia en este caso) que tiene el individuo respecto a lo que sucede en la Ciudad de M\'exico, y al no haber raz\'on alguna para asignar mayor probabilidad al evento $A$ o a su complemento, s\'olo queda establecer $\mathbb{P}(A\,|\,K_1)=\mathbb{P}(A^c\,|\,K_1)$ y como se debe cumplir $\mathbb{P}(A\,|\,K_1)+\mathbb{P}(A^c\,|\,K_1)=1$ esto inmediatamente nos lleva a que $\mathbb{P}(A\,|\,K_1)=\frac{1}{2}$. \par
 \smallskip
 Si pensamos ahora en un individuo que vive en los suburbios de la Ciudad de M\'exico es claro que posee una informaci\'on $K_2$ distinta a la del individuo en el Polo Sur y quiz\'as podr\'iamos hablar de algo como:
 $$\mathbb{P}(A\,|\,K_2) = \left\{ \begin{array}{cc}
    \frac{3}{4} & \textrm{si est\'a lloviendo en los suburbios} \\
    { }         & { } \\
    \frac{1}{4} & \textrm{si no est\'a lloviendo en los suburbios}
    \end{array} \right. $$
    
 Si bien es cierto que el hecho de que est\'e lloviendo en los suburbios de la Ciudad de M\'exico no es garant\'ia de que est\'e lloviendo en el centro de la ciudad, dada la cercan\'ia es m\'as probable que as\'i sea. Podemos decir, informalmente, que $K_2$ representa un mayor nivel de informaci\'on que $K_1$. Y si ahora pensamos en un individuo que vive justamente en el centro de la Ciudad de M\'exico tenemos entonces que este inidividuo posee un nivel de informaci\'on $K_3$ que de hecho es el m\'aximo nivel de informaci\'on que se puede tener respecto al evento $A$ y por lo tanto dicho individuo esta en posici\'on de reportar uno de dos resultados: $\mathbb{P}(A\,|\,K_3)=1$ o bien $\mathbb{P}(A\,|\,K_3)=0$.\par\smallskip
 Lo importante a destacar en este ejemplo es el hecho de la existencia de distinitas medidas de probabilidad para un mismo evento, dependiendo de la cantidad de informaci\'on con la que se cuente.\par\smallskip
 Son muy diversos los factores que incrementan nuestro nivel de informaci\'on en relaci\'on a un fen\'omeno o experimento aleatorio. Van desde la informaci\'on que proveen datos hist\'oricos observados hasta la apreciaci\'on y experiencia de especialistas en dicho fen\'omeno.\par\smallskip
 
 Este enfoque de la probabilidad es ampliamente aprovechado por la \linebreak metodolog\'ia bayesiana y es por ello que podemos decir que la estad\'istica bayesiana va m\'as all\'a que la estad\'istica frecuentista al buscar aprovechar \textbf{toda la informaci\'on disponible}, as\'i se trate de datos observados o de informaci\'on de otro tipo que nos ayude a disminuir de manera coherente nuestra incertidumbre en torno a un fen\'omeno aleatorio de inter\'es.
 Un buen ejemplo para ilustrar que efectivamente la pura experiencia de las personas puede contener informaci\'on muy valiosa consiste en el siguiente ejercicio. En un sal\'on de clase se solicita a cada estudiante que anote en un papel tres cosas: su estatura y las estaturas m\'axima y m\'inima que \textbf{\'el (o ella) creen} que hay en el sal\'on. A\'un cuando no se hayan practicado mediciones de estatura en el sal\'on es sorprendente corroborar que en general los alumnos tendr\'an una idea de las estaturas m\'axima y m\'inima bastante cercana a la realidad, lo que nos da idea de la cantidad de informaci\'on valiosa que puede llegar a tener una apreciaci\'on subjetiva.\par

 \section{La Regla de Bayes}
 
 \hyphenation{pro-ba-bi-li-dad}
 
 La estad\'istica \textit{bayesiana} toma su nombre del resultado de probabilidad conocido como la \textit{Regla de Bayes} as\'i que brevemente enunciaremos los principales resultados al respecto.\par\smallskip
 Dado un espacio de probabilidad $(\Omega,\mathcal{F},\mathbb{P}),$ si $A,B\in\mathcal{F}$ y $\mathbb{P}(B)>0$ entonces la \textit{probabilidad condicional} del evento $A$ dado el evento $B$ se define como:
 $$\mathbb{P}(A\,|\,B):=\frac{\mathbb{P}(A\cap B)}{\mathbb{P}(B)}\,.$$
 Cabe recordar que a\'un cuando se tuviera que $\mathbb{P}(B)=0$ existen resultados de probabilidad que nos permiten calcular probabilidades condicionales en eventos de probabilidad cero. Quiz\'as una notaci\'on m\'as adecuada, en vez de $\prob(A\,|\,B)$ ser\'ia $\prob_B(A)$ para hacer \'enfasis en que la medida de probabilidad condicional $\prob_B$ sustituye (o actualiza) a la medida original $\prob,$ dado el hecho de que ha ocurrido (o se desea suponer que ha ocurrido) el evento $B.$\par
 Si $\{B_{\,n}\}$ es una partici\'on del espacio muestral $\Omega$ y para toda $n$ tenemos que $B_{\,n}\in \mathcal{F}$ y $\mathbb{P}(B_{\,n})>0$, entonces:
 $$\mathbb{P}(A)=\sum\limits_n \mathbb{P}(A\,|\,B_{\,n})\,\mathbb{P}(B_{\,n})\,,\quad\,\mbox{para toda}\; A\in \mathcal{F}\,.$$
 Un corolario importante del resultado anterior es:
 $$\textrm{Si } B\in \mathcal{F}\,\Rightarrow\, \mathbb{P}(A)=\mathbb{P}(A\,|\,B)\mathbb{P}(B)+\mathbb{P}(A\,|\,B^{\,c})\,\mathbb{P}(B^{\,c})\,.$$
 Bajo los supuestos anteriores tenemos la \textit{Regla de Bayes}:
 $$\mathbb{P}(B_{\,k}\,|\,A)=\frac{\mathbb{P}(A\,|\,B_{\,k})\,\mathbb{P}(B_{\,k})}{\sum\limits_n \mathbb{P}(A\,|\,B_{\,n})\,\mathbb{P}(B_{\,n})} \quad\,,\, A\in \mathcal{F}\,,\, \mathbb{P}(A)>0\,.$$
 Y como corolario importante:
$$\mathbb{P}(B\,|\,A)=\frac{\mathbb{P}(A\,|\,B)\,\mathbb{P}(B)}{\mathbb{P}(A\,|\,B)\,\mathbb{P}(B)+\mathbb{P}(A\,|\,B^{\,c})\,\mathbb{P}(B^{\,c})}\,.$$

 Si bien las demostraciones y ejemplos de lo anterior son propios de un curso de probabilidad bien vale la pena abordar un ejemplo que nos servir\'a m\'as adelante para ilustrar el por qu\'e algunos c\'alculos de la estad\'istica frecuentista resultan cuestionables.\par\smallskip
 Supongamos que un grupo de ingenieros biom\'edicos mexicanos ha dise\~nado un nuevo aparato para diagn\'ostico del SIDA (en realidad es para diagn\'ostico de presencia de VIH pero coloquialmente nos referimos a esta inmunodeficiencia simplemente como SIDA, aunque los m\'edicos nos rega\~nen). Haremos el experimento de escoger a un individuo para probar dicho aparato y consideremos los eventos de inter\'es $A$ y $B$ en donde $A$ sea el evento de que el aparato diagnostique SIDA y $B$ el evento de tener efectivamente SIDA. Los ingenieros que dise\~naron este aparato presumen de que \'este es muy bueno pues nos reportan que lo probaron con un grupo de 100 portadores del virus del SIDA y en 99\% de los casos el aparato dio positivo y que tambi\'en lo probaron con 100 individuos sanos y que tambi\'en en el 99\% de los casos el aparato dio negativo. Probabil\'isticamente esto se expresa:
 $$\mathbb{P}(A\,|\,B) = \frac{99}{100} = \mathbb{P}(A^c\,|\,B^{\,c})$$
 
 Sea $p$ la proporci\'on de mexicanos con virus del SIDA (en este contexo, a $p$ se le conoce como \textit{prevalencia}). Tenemos entonces que $\mathbb{P}(B)=p$. Con la informaci\'on anterior y utilizando la Regla de Bayes estamos en posici\'on de calcular $\mathbb{P}(B\,|\,A)$, que quedar\'a expresada en funci\'on de $p$ :
$$\varphi(p):=\mathbb{P}(B\,|\,A)=\frac{\frac{99}{100}}{\frac{98}{100}+\frac{1}{100p}}\,.$$
 Ahora tabulamos algunos valores $(p,\varphi(p))$:
 $$ \begin{array}{c|c}
 p & \prob(B\,|\,A) \\ \hline
 0{.}002 & 0{.}1656 \\
 0{.}010 & 0{.}5000 \\
 0{.}100 & 0{.}9167 \\
 0{.}500 & 0{.}9900
 \end{array} $$
 
 De lo anterior se observa que \'unicamente en el caso de que $p=\frac{1}{2}$ se cumple $\mathbb{P}(B\,|\,A)=\mathbb{P}(A\,|\,B),$ y que conforme $p\rightarrow 0$ estas dos probabilidades se alejan (siendo m\'as precisos, $\mathbb{P}(B\,|\,A)$ se aleja de $\mathbb{P}(A\,|\,B)$ que permanece constante en 0.99). A primera vista pareciera que estamos discutiendo una trivialidad, ya que es bien sabido que por lo general $\mathbb{P}(B\,|\,A)$ suele ser distinta de $\mathbb{P}(A\,|\,B),$ pero si nos detenemos un poco a analizar el ejemplo anterior esto tiene consecuencias terribles para los ingenieros que inventaron un aparato que creen es muy efectivo y en realidad no lo es. En M\'exico (2008) se estima una tasa de prevalencia de SIDA de 0.2\% de la poblaci\'on total, esto es, $p=0{.}002$, lo cual significa que la probabilidad (en M\'exico) de que un individuo tenga SIDA dado que el aparato dice que lo tiene !`es de tan solo $0{.}1656\,!$\par\smallskip
 
 \hyphenation{ante-rior}
 
 ?`Qu\'e sucedi\'o? Sucede que el aparato fue probado con personas de las cuales conoc\'iamos previamente su estado de salud, pero ya en la pr\'actica cotidiana esto no sucede, las personas que llegan a practicarse un an\'alisis lo hacen porque justamente desconocen cu\'al es su estado de salud (es decir si tienen o no el virus) y es por ello que el que $\mathbb{P}(A\,|\,B)$ sea de 99\% no implica necesariamente que $\mathbb{P}(B\,|\,A)$ sea igualmente alta.\par\smallskip
 
 El ejemplo anterior nos ser\'a de mucha utilidad para comprender por qu\'e algunos m\'etodos de inferencia de la estad\'istica frecuentista son cuestionables, en particular cuando se busca hacer inferencias sobre un par\'ametro $\theta$ y en lugar de calcular la probabilidad de que $\theta$ tome ciertos valores dada una muestra, es decir $\mathbb{P}[\,\theta\in \Theta_0\,|\,(x_1,\ldots,x_n)\,]$, la  estad\'istica frecuentista utiliza la probabilidad de observar una determinada muestra bajo el supuesto de un valor espec\'ifico del par\'ametro, es decir $\mathbb{P}[\,(x_1,\ldots,x_n)\,|\,\,\theta=\theta_0\,],$ mejor conocida como la ``verosimilitud'', y esto es tanto como pretender que $\mathbb{P}(B\,|\,A)=\mathbb{P}(A\,|\,B)$ siempre se cumple, o que $\mathbb{P}(A\,|\,B)$ es siempre una buena aproximaci\'on de $\mathbb{P}(B\,|\,A).$ Para una discusi\'on a detalle sobre inconsistencias en la inferencia estad\'istica al utilizar ``verosimilitud'' se recomienda Le Cam (1990).

 \section{La filosof\'ia bayesiana}
 
 La teor\'ia de la probabilidad se ocupa del estudio de la incertidumbre, y del comportamiento de los fen\'omenos o experimentos aleatorios. La probabilidad depende de dos elementos: el evento incierto y las condiciones bajo las cuales es considerado, por lo que desde este punto de vista la probabilidad es siempre condicional. La estad\'istica es una herramienta para la toma de decisiones bajo condiciones de incertidumbre.\par\smallskip
 
 Un enfoque cient\'ifico sobre la incertidumbre es la medici\'on de la misma. El c\'elebre f\'isico ingl\'es Sir William Thomson, mejor conocido como Lord Kelvin, dijo que s\'olo asociando n\'umeros con el concepto cient\'ifico es como se puede comprender adecuadamente dicho concepto. La raz\'on de querer medir no es s\'olo para ser m\'as precisos respecto a la intensidad de la incertidumbre sino tambi\'en para combinar incertidumbres: En un problema t\'ipico de estad\'istica encontramos combinadas la incertidumbre de los datos y la del par\'ametro.\par\smallskip
 
 La medici\'on de la incertidumbre puede realizarse por medio del c\'alculo de probabilidades. En sentido inverso, las reglas de la probabilidad se reducen de manera simple a las reglas sobre proporciones. Esto explica por qu\'e los argumentos frecuentistas son en muchos casos \'utiles: La combinaci\'on de incertidumbres puede ser estudiada por medio de proporciones o frecuencias. El objetivo de recolectar datos es precisamente reducir el nivel de incertidumbre, pero bajo la perspectiva bayesiana se aprovechan tanto los datos muestrales como otro tipo de informaciones que de manera coherente nos ayuden tambi\'en a reducir nuestro nivel de incertidumbre en torno a los par\'ametros de inter\'es.\par\smallskip
 En resumen:
 \begin{itemize}
 \item La estad\'istica es una herramienta para la toma de decisiones bajo condiciones de incertidumbre.
 \item La incertidumbre debe ser medida por medio de la probabilidad.
 \item La incertidumbre sobre los datos debe ser medida condicionalmente en los par\'ametros.
 \item La incertidumbre sobre los par\'ametros es similarmente medida por medio de la probabilidad.
 \item La inferencia se lleva a cabo mediante c\'alculo de probabilidades, haciendo uso particular de la Regla de Bayes.
 \end{itemize}
 
 Las discusiones en contra del enfoque bayesiano se centran en el punto de medir la incertidumbre sobre el par\'ametro probabil\'isticamente, esto es, darle tratamiento de variable aleatoria. Para la estad\'istica frecuentista existe algo que llaman ``el verdadero valor del par\'ametro'' que consideran fijo y que ``s\'olo Dios conoce'' pero que resulta desconocido para nosotros los mortales. Lo anterior, adem\'as de que los estad\'isticos frecuentistas rechazan la utilizaci\'on de cualquier otro tipo de informaci\'on que no provenga de los datos muestrales para hacer inferencias. Para profundizar m\'as a detalle en las ideas anteriores se recomienda la lectura del art\'iculo de Lindley (2000).

 \chapter{El paradigma bayesiano}

 \section{El modelo general}
 
 \hyphenation{va-lor}
 \hyphenation{bien}
 
 Para referirnos a un \textit{modelo probabil\'istico param\'etrico} general lo denotamos $p(x\,|\,\theta)$ en donde la funci\'on $p(\,\cdot\,|\,\theta)$ puede ser una funci\'on de masa de probabilidades de una variable (o vector) aleatoria (v.a.) discreta o bien una funci\'on de densidad de una v.a. continua. El escribir dicha funci\'on condicional en el par\'ametro (o vector de par\'ametros) $\theta$ se debe al hecho de que una vez dado un valor espec\'ifico de $\theta$ la funci\'on de probabilidad queda totalmente determinada. Para referirnos a una muestra aleatoria (m.a.) utilizamos la notaci\'on $\mathbf{X}:=(X_1,\ldots,X_n)$ y para referirnos a una observaci\'on muestral utilizamos $\mathbf{x}:=(x_1,\ldots,x_n)$. Por \textit{espacio param\'etrico} entendemos el conjunto $\Theta$ de todos los valores que puede tomar $\theta$ y por \textit{familia param\'etrica} entendemos un conjunto $\mathcal{P}=\{p(x\,|\,\theta):\;\theta\in \Theta\}$.\par

 \smallskip
 
 Al empezar a estudiar un fen\'omeno o experimento aleatorio recurrimos a la teor\'ia de la probabilidad para escoger o definir alguna familia param\'etrica que modele razonablemente el fen\'omeno. Una vez hecho esto queda la \linebreak incertidumbre sobre el par\'ametro del modelo (no olvidemos que el par\'ametro puede ser un vector) pues de entre todos los elementos de la familia \linebreak param\'etrica $\mathcal{P}=\{p(x\,|\,\theta):\;\theta\in \Theta\}$ ?`Cu\'al utilizamos para hacer inferencias?\par
 
 \hyphenation{pro-ba-bi-lis-ti-ca-mente}
 
 La Estad\'istica Bayesiana modela la incertidumbre que tenemos sobre $\theta$ probabil\'isticamente, esto es, consideramos al valor de $\theta$ como una variable (o vector) aleatoria (v.a.) con una \textit{\textbf{distribuci\'on de probabilidad a priori (o inicial)}} $p(\theta)$. Se trata de una distribuci\'on basada en experiencia previa (experiencia de especialistas, datos hist\'oricos, etc.) antes de obtener datos muestrales nuevos.\par
 Luego procedemos a observar los nuevos datos (obtenci\'on de la muestra) $\mathbf{x}:=(x_1,\ldots,x_n)$ y combinamos esta informaci\'on con la distribuci\'on a priori mediante la Regla de Bayes y obtenemos una \textit{\textbf{distribuci\'on de probabilidad a posteriori (o final)}} :  

\begin{grande}
\begin{equation}\label{eq:posteriori}
p(\theta\,|\,\mathbf{x})=\frac{p(\mathbf{x},\theta)}{p(\mathbf{x})}=\frac{p(\mathbf{x}\,|\,\theta)p(\theta)}{\int_{\Theta}p(\mathbf{x}\,|\,\tilde{\theta})p(\tilde{\theta})\,d\tilde{\theta}}
\end{equation}
\end{grande}

 Tenemos que $p(\theta\,|\,\mathbf{x})$ es tambi\'en una distribuci\'on de probabilidad de $\theta$ pero que a diferencia de la distribuci\'on a priori $p(\theta)$ toma en cuenta tanto la informaci\'on contemplada en $p(\theta)$ as\'i como la informaci\'on contenida en los datos observados $\mathbf{x}=(x_1,\ldots,x_n)$. La distribuci\'on a posteriori de $\theta$ es la base para hacer inferencias sobre $\theta$. \par
 Es importante tener presente que, por un lado, $p(\mathbf{x}\,|\,\theta)$ y $p(\theta)$ son distribuciones de probabilidad, y por otro:
 $$p(\mathbf{x})=\int_{\Theta}p(\mathbf{x}\,|\,\tilde{\theta})p(\tilde{\theta})\,d\tilde{\theta}$$
es la probabilidad (o densidad) conjunta de la muestra \linebreak $\mathbf{x}=(x_1,\ldots,x_n)$ observada a partir del vector aleatorio $\mathbf{X}=(X_1,\ldots,X_n)$. Pero lo m\'as importante es estar consciente de que $p(\mathbf{x})$ es constante respecto a $\theta$, por lo que podemos escribir:
 
 \begin{grande}
 \begin{equation}\label{eq:posteriori propor}
 p(\theta\,|\,\mathbf{x})\propto p(\mathbf{x}\,|\,\theta)p(\theta)
 \end{equation}
 \end{grande}
 
 \hyphenation{pro-ba-bi-li-dad}

 Respecto a $p(\mathbf{x}\,|\,\theta)=p((x_1,\ldots,x_n)\,|\,\theta)$ tenemos que se trata de la probabilidad conjunta de la muestra condicional en $\theta$ (usualmente llamada \textit{verosimilitud}). En el caso particular de que los componentes del vector aleatorio $\mathbf{X}=(X_1,\ldots,X_n)$ resulten ser independientes (esto es, observaciones independientes) tenemos que:
 
 $$p(\mathbf{x}\,|\,\theta)=\prod_{j=1}^{n} p(x_j\,|\,\theta)$$
   
 \smallskip

 Aunque ser\'a hasta el cap\'itulo 5 en donde veamos a detalle la metodolog\'ia para la inferencia bayesiana, conviene adelantar un poco al respecto para \linebreak tener una idea general. Podemos proponer como \textit{estimador puntual} de $\theta$ a alguna medida de tendencia central, por ejemplo la mediana o la esperanza:
 
 $$\hat{\theta}:=\mathbb{E}(\theta)=\int_{\Theta}\theta p(\theta\,|\,\mathbf{x})\,d\theta$$
 
 Y a\'un en el caso de que no se cuente con infomaci\'on muestral se puede calcular $\hat{\theta}$ utilizando $p(\theta)$ en lugar de $p(\theta\,|\,\mathbf{x})$.\par
 
 \smallskip
 
  Para hacer \textit{estimaci\'on por regiones}, por ejemplo, si deseamos calcular la probabilidad de que el vector de par\'ametros $\theta$ pertenezca a una regi\'on $A\subset \Theta\,:$
 
 $$\mathbb{P}(\theta\in A)=\int_A p(\theta\,|\,\mathbf{x})\,d\theta$$

\noindent o bien, dado un valor $\alpha\in\,]\,0,1\,[$ se busca un $A\subset\Theta$ tal que $\prob(\theta\in A)=\alpha\,.$ Con frecuencia la soluci\'on para $A$ no es \'unica, y se recurre a ciertos criterios de decisi\'on como los que se ver\'an en el cap\'itulo 4. Cabe aclarar que si $\dim\Theta=1$ las \textit{regiones} son subconjuntos de $\mathbb{R}$ y que un caso particular de estas regiones son los intervalos. En este sentido la estimaci\'on por regiones en estad\'istica bayesiana es m\'as general que la estimaci\'on por intervalos de la estad\'istica frecuentista.\par

 Y ya que estamos dando ideas preliminares de lo que es la inferencia bayesiana podemos introducir a un nivel muy simple c\'omo se hace el \textit{contraste de $k$ hip\'otesis}. Supongamos que se desea contrastar las hip\'otesis:
 
 $$\begin{array}{cc}
 H_1: & \theta\in \Theta_1 \\
 H_2: & \theta\in \Theta_2 \\
 \vdots & \vdots \\
 H_k: & \theta\in \Theta_k 
 \end{array}$$
 
 Una manera de hacerlo es calcular directamente la probabilidad de cada hip\'otesis y escoger aquella que tenga la m\'as alta probabilidad, y calcular la probabilidad de una hip\'otesis $H_j$ puede ser tan simple como:
 
 $$\mathbb{P}(H_j)=\int_{\Theta_j}p(\theta\,|\,\mathbf{x})\,d\theta$$
 
  La anterior es una manera muy simple de hacer contraste de hip\'otesis, en el cap\'itulo 5 se ver\'a que este esquema se puede enriquecer mucho mediante el uso de funciones de utilidad. \par
  
  \smallskip
  
  Muchos son los casos en los que, m\'as que estar interesados en el vector de par\'ametros $\theta\,,$ lo que queremos es describir el comportamiento de observaciones futuras del fen\'omeno aleatorio en cuesti\'on, esto es, hacer \textit{\textbf{predicci\'on}}. \par
 Dado un valor de $\theta$, la distribuci\'on que describe el comportamiento de la observaci\'on futura $X$ es $p(x\,|\,\theta)$. El problema es que por lo general el valor de $\theta$ es desconocido. Por lo regular la estad\'istica frecuentista aborda este problema estimando puntualmente a $\theta$ con base en la muestra observada y dicho estimador $\hat{\theta}$ es sustituido en $p(x\,|\,\theta)$, es decir, utilizan $p(x\,|\,\hat{\theta})$.
 Desde la perspectiva bayesiana el modelo $p(x\,|\,\theta)$ junto con la distribuci\'on a priori $p(\theta)$ inducen una distribuci\'on conjunta para el vector aleatorio $(X,\theta)$ mediante el concepto de probabilidad condicional:
\hyphenation{ob-te-ne-mos}
 
 $$p(x,\theta)=p(x\,|\,\theta)p(\theta)$$
y marginalizando la distribuci\'on de probabilidad conjunta anterior obtenemos:
 
 $$p(x)=\int_{\Theta}p(x,\theta)\,d\theta$$
 
 Combinando los dos resultados anteriores:
 
 \begin{grande}
 \begin{equation} \label{eq:predictiva}
 p(x)=\int_{\Theta}p(x|\theta)p(\theta)\,d\theta
 \end{equation}
 \end{grande}
 
 A $p(x)$ la denominamos \textit{\textbf{distribuci\'on predictiva a priori (o inicial)}}, y describe nuestro conocimiento acerca de una observaci\'on futura $X$ basado \'unicamente en la informaci\'on contenida en $p(\theta)$. N\'otese que $p(x)$ no depende ya de $\theta$. \par
 \smallskip
 Una vez obtenida la muestra, el modelo $p(x\,|\,\theta)$ y la distribuci\'on a posteriori $p(\theta\,|\,\mathbf{x})$ inducen una distribuci\'on conjunta para $(X,\theta)$ condicional en los valores observados $\mathbf{x}=(x_1,\ldots,x_n)$ :
 
 \begin{eqnarray*}
 p(x,\theta\,|\,\mathbf{x}) & = & \frac{p(x,\theta,\mathbf{x})}{p(\mathbf{x})} \\
   & = & \frac{p(x\,|\,\theta,\mathbf{x})p(\theta,\mathbf{x})}{p(\mathbf{x})} \\
   & = & p(x\,|\,\theta,\mathbf{x})p(\theta\,|\,\mathbf{x}) \\
   & = & p(x\,|\,\theta)p(\theta\,|\,\mathbf{x})
 \end{eqnarray*}
 
 En lo inmediato anterior $p(x\,|\,\theta,\mathbf{x})=p(x\,|\,\theta)$ se justifica por la independencia condicional de $X$ y $\mathbf{X}=(X_1,\ldots,X_n)$ dado $\theta$. Marginalizando la distribuci\'on conjunta condicional anterior:
 
 $$p(x\,|\,\mathbf{x})=\int_{\Theta}p(x,\theta\,|\,\mathbf{x})\,d\theta$$
 
 Combinando los dos resultados anteriores:
 
 \begin{grande}
 \begin{equation} \label{eq:predictiva post}
 p(x\,|\,\mathbf{x})=\int_{\Theta}p(x\,|\,\theta)p(\theta\,|\,\mathbf{x})\,d\theta
 \end{equation}
 \end{grande}
 
 A $p(x\,|\,\mathbf{x})$ la denominamos \textit{\textbf{distribuci\'on predictiva a posteriori (o final)}}, y describe nuestro conocimiento acerca de una observaci\'on futura $X$ basado tanto en la informaci\'on contenida en $p(\theta)$ como en la informaci\'on muestral $\mathbf{x}=(x_1,\ldots,x_n)$. N\'otese nuevamente que $p(x\,|\,\mathbf{x})$ no depende de $\theta$. \par
 As\'i que para hacer predicci\'on sobre observaciones futuras del fen\'omeno aleatorio que estemos modelando usamos $p(x)$ o bien $p(x\,|\,\mathbf{x})$, seg\'un sea el caso. Y de manera an\'aloga a lo brevemente mencionado sobre inferencia bayesiana, una manera simple de hacer predicci\'on puntual, por ejemplo, de una observaci\'on futura $X$ podr\'ia ser mediante alguna medida de tendencia central, como la mediana o la esperanza:
 
 $$\hat{x}:=\mathbb{E}(X)=\int_{Ran\,X}xp(x\,|\,\mathbf{x})\,dx$$
 
\noindent donde $Ran\,X$ es el rango de la v.a. $X$. Tambi\'en, una manera de calcular la probabilidad de que una observaci\'on futura caiga en un conjunto $A\subset Ran\,X$ ser\'ia:
 
 $$\mathbb{P}(\{X\in A\})=\int_{A}p(x\,|\,\mathbf{x})\,dx$$
 
 Y algo an\'alogo para contraste de hip\'otesis.
 
 \bigskip
 
 Las ecuaciones (\ref{eq:posteriori}), (\ref{eq:predictiva}) y (\ref{eq:predictiva post}) constituyen \textit{el modelo general de la estad\'istica bayesiana}. Cualquier problema estad\'istico tratado bajo el enfoque bayesiano implica la obtenci\'on y utilizaci\'on de las f\'ormulas mencionadas.
 
 \bigskip
 
 \hyphenation{mo-de-lar}

 \section{Un primer ejemplo}

 El siguiente ejemplo, a pesar de ser muy simple, resulta muy ilustrativo para dos cosas: primero, comenzar a entender por qu\'e es v\'alido modelar nuestra incertidumbre sobre $\theta$ probabil\'isticamente; segundo, para irnos familiarizando con el enfoque bayesiano.
 
 \bigskip
 
 \begin{ejem} \label{e1}
 Consideremos una urna que contiene dos monedas: una cargada y la otra equilibrada. Supongamos que la moneda cargada est\'a cient\'ificamente construida para tener una probabilidad de $\frac{3}{4}$ de que salga \'aguila. Una persona tomar\'a una de las dos monedas de la urna (no necesariamente al azar) y echar\'a un volado con apuesta de por medio. Haremos lo siguiente:
 \begin{enumerate}
 \item Proponer una familia param\'etrica para el experimento anterior,
 \item proponer una distribuci\'on a priori para el par\'ametro del modelo, tomando especialmente en cuenta que no estamos seguros de que la moneda fue tomada al azar,
 \item obtener la distribuci\'on predictiva a priori,
 \item obtener la distribuci\'on a posteriori,
 \item obtener la distribuci\'on predictiva a posteriori.
 \end{enumerate}
 
 En este sencillo ejemplo es f\'acil identificar que la familia param\'etrica\linebreak Bernoulli es la adecuada:
 $$\mathcal{P}=\{Ber(x\,|\,\theta):\theta\in\Theta\}$$
donde
 
 $$Ber(x\,|\,\theta)=\theta^x (1-\theta)^{1-x}\indic_{\{0,1\}}(x)$$
 
 S\'olo que en este caso el espacio param\'etrico se reduce a $\Theta=\{\frac{3}{4},\frac{1}{2}\}$.\par
 \medskip
 
 Desde el enfoque bayesiano, nuestra incertidumbre sobre el par\'ametro $\theta$ la modelamos probabil\'isticamente, es decir, trataremos a $\theta$ como variable aleatoria y en tal caso $Ran\,\theta=\Theta,$ y como en este caso $\Theta$ es finito entonces tenemos que $\theta$ es una variable aleatoria discreta que s\'olo toma dos valores: $\frac{3}{4}$ o $\frac{1}{2}$.\par
 Sean $\prob(\theta=\frac{3}{4})=\alpha$ y $\prob(\theta=\frac{1}{2})=1-\alpha$, para alg\'un $\alpha$ entre $0$ y $1$. Entonces la distribuci\'on a priori queda como sigue:
 $$p(\theta)=\alpha\indic_{\{\frac{3}{4}\}}(\theta)+(1-\alpha)\indic_{\{\frac{1}{2}\}}(\theta) \qquad \alpha\in ]0,1[$$
 
 La distribuci\'on inicial propuesta permite modelar la parte de los supuestos del problema en donde se dijo que se toma una moneda de la urna no necesariamente al azar. En particular es por medio de $\alpha$ que reflejaremos en la distribuci\'on a priori nuestro grado de informaci\'on acerca de c\'omo fue escogida la moneda de la urna. As\'i por ejemplo si estamos o nos sentimos seguros de que fue tomada al azar entonces $\alpha=\frac{1}{2}$. Har\'iamos la misma asignaci\'on si carecemos totalmente de informaci\'on al respecto ya que no habr\'ia raz\'on alguna para suponer que alguna de las dos monedas tiene mayor probabilidad que la otra de ser escogida. Y si por alguna raz\'on contamos con cierta informaci\'on que nos haga pensar que alguna de las monedas tiene mayor probabilidad de ser escogida tambi\'en podemos reflejarlo por medio de $\alpha.$ Por ejemplo, suponiendo que el procedimiento para elegir la moneda de la urna es lanzando un dado y que si sale un seis entonces escogemos la moneda equilibrada. En este caso claramente $\alpha=\frac{5}{6}$. Es importante destacar el hecho de que restringimos a $\alpha$ al intervalo abierto $]\,0,1\,[$ ya que si ocurriese que $\alpha=0$ o bien $\alpha=1$ entonces querr\'ia decir que estamos seguros del valor de $\theta$ y en tal caso no tendr\'ia sentido hacer inferencias sobre $\theta$.\par
 \medskip
 Por medio de la f\'ormula (\ref{eq:predictiva}) obtenemos la distribuci\'on predictiva a priori:
 \begin{eqnarray*}
 p(x) & = & \sum_{\theta\in\Theta}p(x\,|\,\theta)p(\theta) \\
      & = & \alpha\Big(\frac{3}{4}\Big)^x \Big(\frac{1}{4}\Big)^{(1-x)}\indic_{\{0,1\}}(x) + (1-\alpha)\Big(\frac{1}{2}\Big)^x \Big(\frac{1}{2}\Big)^{(1-x)}\indic_{\{0,1\}}(x)
 \end{eqnarray*}
 La expresi\'on anterior se simplifica a:
 $$p(x)=\frac{\alpha}{4}\Big(\indic_{\{1\}}(x)-\indic_{\{0\}}(x)\Big) + \frac{1}{2}\indic_{\{0,1\}}(x)$$
es decir:
 \begin{equation*}
 p(1)=\frac{1}{2}+\frac{\alpha}{4}=\textrm{probabilidad de que salga \'aguila}
 \end{equation*}
 \begin{equation*}
 p(0)=\frac{1}{2}-\frac{\alpha}{4}=\textrm{probabilidad de que salga sol}
 \end{equation*}
 
 De lo anterior cabe destacar que si $\alpha\rightarrow 1$ (lo cual se interpreta como que nos sentimos muy seguros de que se escogi\'o la moneda cargada) entonces \linebreak $p(1)\rightarrow \frac{3}{4}$, tal cual se esperar\'ia.\par
 \smallskip
 Pensando en que se va a hacer una apuesta sobre el resultado del primer volado, para tener un juego justo, definimos la variable aleatoria $U$ como la ganancia/p\'erdida resultante de apostar en favor de que salga sol:
 $$\prob[U=u]=p(0)\indic_{\{a\}}(u)+p(1)\indic_{\{-b\}}(u)\qquad a,b>0$$
es decir, $U$ es una v.a. que toma el valor $+a$ (ganancia) con probabilidad $p(0)$ o bien el valor $-b$ (p\'erdida) con probabilidad $p(1)$. Para tener un juego justo se requiere que $\esper(U)=0$ :
 \begin{eqnarray*}
 \esper(U)=0 & \Leftrightarrow & ap(0)-bp(1)=0 \\
             & \Leftrightarrow & a=\frac{p(1)}{p(0)}\,b=\frac{2+\alpha}{2-\alpha}\,b 
 \end{eqnarray*}
 Es decir, que la cantidad justa a apostar en favor de que salga sol debe ser igual a $\displaystyle{\frac{p(1)}{p(0)}}$ veces la cantidad que se apueste en favor de que salga \'aguila. Si bien lo inmediato anterior es m\'as un problema t\'ipico de un curso elemental de probabilidad, resultar\'a interesante analizar c\'omo se modifica el esquema de apuestas conforme se van lanzando volados, esto es, ante la presencia de informaci\'on muestral.\par
 \medskip
 Supongamos ahora que ya se escogi\'o una de las monedas de la urna y que se efectuaron $n$ lanzamientos con ella y los resultados de cada lanzamiento se registran como un vector $n-$dimensional de unos y ceros $\mathbf{x}:=(x_1,\ldots,x_n)$. La informaci\'on contenida en la muestra observada $\mathbf{x}$ modifica nuestra incertidumbre sobre $\theta$ pasando de la distribuci\'on a priori $p(\theta)$ a la distribuci\'on a posteriori:
 $$p(\theta\,|\,\mathbf{x})=\frac{p(\mathbf{x}\,|\,\theta)p(\theta)}{p(\mathbf{x}\,|\,\frac{3}{4})p(\frac{3}{4})+p(\mathbf{x}\,|\,\frac{1}{2})p(\frac{1}{2})}$$
 
 Si resulta razonable suponer que se hacen lanzamientos independientes entonces:
 
 \begin{eqnarray*}
 p(\mathbf{x}\,|\,\theta) &=& \prod_{j=1}^n p(x_j\,|\,\theta) \\
                      &=& \prod_{j=1}^n \theta^{x_j}(1-\theta)^{1-x_j}\indic_{\{0,1\}}(x_j) \\
                      &=& \theta^{\sum x_j}(1-\theta)^{n-\sum x_j}\prod_{j=1}^n\indic_{\{0,1\}}(x_j)\\
                      &=& \theta^{\sum x_j}(1-\theta)^{n-\sum x_j}g(\mathbf{x})
 \end{eqnarray*}
 
 Por otro lado:
 
 $$p(\mathbf{x}\,|\,\mbox{\small{$\frac{3}{4}$}})p(\mbox{\small{$\frac{3}{4}$}})= \frac{3^{\sum x_j}}{4^n}\,\alpha\,g(\mathbf{x}) \qquad p(\mathbf{x}\,|\,\mbox{\small{$\frac{1}{2}$}})p(\mbox{\small{$\frac{1}{2}$}})= \frac{1-\alpha}{2^n}\,g(\mathbf{x})$$
 
 De lo anterior:
 
 $$p(\theta\,|\,\mathbf{x})=\frac{[2(1-\theta)]^n \big(\frac{\theta}{1-\theta}\big)^{\sum x_j} [\alpha\indic_{\{\frac{3}{4}\}}(\theta)+(1-\alpha)\indic_{\{\frac{1}{2}\}}(\theta)]} {\alpha\big(\frac{3^{\sum x_j}}{2^n}-1 \big)+1}$$
 
 Si definimos:
 $$\nu=\nu(\alpha,\mathbf{x}):=\frac{3^{\sum x_j}}{2^n}\Big(\frac{\alpha}{1-\alpha}\Big)$$
reescribimos $p(\theta\,|\,\mathbf{x})$ como:
 $$p(\mbox{\small{$\frac{3}{4}$}}\,|\,\mathbf{x})=\frac{1}{1+\nu^{-1}} \qquad p(\mbox{\small{$\frac{1}{2}$}}|\mathbf{x})=\frac{1}{1+\nu}$$
 
 Supongamos que la moneda con que se lanzar\'an los volados es tomada de la urna al azar. En este caso tendr\'iamos que $\alpha=\frac{1}{2}$. La probabilidad a priori de que la moneda escogida sea la cargada es:
 $$p(\mbox{\small{$\frac{3}{4}$}})= 0{.}5$$
 Se lanza un primer volado y observamos que sale \'aguila. En este caso $n=1$, $\mathbf{x}=(x_1)=(1)$ y por lo tanto $p(\frac{3}{4}\,|\,(1))=0{.}6$. Es decir, a la luz de la informaci\'on muestral con la que se cuenta hasta el momento nos vemos obligados a revisar o \textit{actualizar} la probabilidad de que sea la moneda cargada la que se est\'a utilizando. Cabe destacar que con la informaci\'on muestral obtenida ahora es m\'as probable que sea la moneda cargada la que se est\'a utilizando. Se podr\'ia pensar que no es dif\'icil que salga un \'aguila con una moneda equilibrada pero el que haya salido \'aguila es evidencia m\'as a favor de que se est\'e usando la moneda cargada que la equilibrada.\par
 Ahora efectuamos un segundo lanzamiento con la moneda en cuesti\'on y resulta que obtenemos un sol. Ahora $n=2$ , $\mathbf{x}=(x_1,x_2)=(1,0)$ y obtenemos $p(\frac{3}{4}\,|\,(1,0))=0{.}4286$. Es decir, a la luz de la informaci\'on muestral con la que se cuenta hasta el momento nos vemos obligados a \textit{actualizar} nuevamente la probabilidad de que sea la moneda cargada la que se est\'a utilizando. Cabe destacar que con la informaci\'on muestral obtenida hasta ahora es m\'as probable que sea la moneda equilibrada la que se est\'a utilizando. Se podr\'ia pensar que no es dif\'icil que salga un \'aguila y luego un sol con una moneda cargada pero el que haya salido \'aguila y luego sol es evidencia m\'as a favor de que se est\'e usando la moneda equilibrada que la cargada.\par
 Podr\'ia pensarse que nos la podemos pasar as\'i oscilando de un valor a otro a capricho de los resultados muestrales, pero no es as\'i pues conforme el tama\~no de la muestra $n$ crece el valor de $p(\frac{3}{4}\,|\,\mathbf{x})$ se va estabilizando:
 \begin{eqnarray*}
 \textrm{Si}\;n\rightarrow\infty & \Rightarrow & \sum x_j\rightarrow \mbox{\small{$\frac{3}{4}$}}\,n \quad \mbox{o} \quad \sum x_j\rightarrow \mbox{\small{$\frac{1}{2}$}}\,n \\
      & \Rightarrow &  \nu\rightarrow\infty \quad \mbox{o} \quad \nu\rightarrow 0 \\
      & \Rightarrow &  p(\mbox{\small{$\frac{3}{4}$}}\,|\,\mathbf{x})\rightarrow 1 \quad\mbox{o} \quad p(\mbox{\small{$\frac{3}{4}$}}\,|\,\mathbf{x})\rightarrow 0
 \end{eqnarray*}
 Lo anterior quiere decir que conforme el tama\~no de la muestra se vuelve m\'as y m\'as grande iremos acumulando informaci\'on que ir\'a reduciendo nuestra incertidumbre respecto a $\theta$ (es decir, nuestra incertidumbre respecto a qu\'e moneda se est\'a usando) hasta llegar a un nivel muy cercano a la certeza.\par
 El siguiente es el resultado de una simulaci\'on del presente ejemplo con $n=20$ y $\alpha=\frac{1}{2}$ :
 $$\mathbf{x}=(0,1,1,1,0,1,0,1,0,1,1,1,1,1,1,1,1,1,1,1)$$
 $$\Rightarrow p(\mbox{\small{$\frac{3}{4}$}}\,|\,\mathbf{x})=0{.}9762$$
 Y de hecho, efectivamente result\'o ser la moneda cargada la que se estaba utilizando.\par
 \medskip
 La informaci\'on contenida en la muestra observada $\mathbf{x}$ modifica nuestra\linebreak incertidumbre sobre un siguiente lanzamiento $X_{n+1}$ pasando de la distribuci\'on predictiva a priori $p(x)$ a la distribuci\'on predictiva a posteriori:
 
 \begin{eqnarray*} p(x\,|\,\mathbf{x}) &=& p(x\,|\,\mbox{\small{$\frac{3}{4}$}})p(\mbox{\small{$\frac{3}{4}$}}\,|\,\mathbf{x})+p(x\,|\,\mbox{\small{$\frac{1}{2}$}})p(\mbox{\small{$\frac{1}{2}$}}\,|\,\mathbf{x}) \\
 &=& \frac{3^x \nu +2}{4(\nu + 1)}\indic_{\{0,1\}}(x)
 \end{eqnarray*}
 
 Regresando al esquema de la apuesta justa, ante la presencia de informaci\'on muestral $\mathbf{x}=(x_1,\ldots,x_n)$ es necesario ir revisando o \textit{actualizando} los c\'alculos para determinar cu\'al ser\'ia la apuesta justa, esto es, sin tener informaci\'on muestral y contando tan solo con la informaci\'on inicial de que la moneda fue escogida al azar ten\'iamos que la apuesta justa en favor de que salga sol era:
 $$a=\frac{p(1)}{p(0)}\,b$$
 Esto es, que la apuesta en favor de que salga sol debe ser $\displaystyle{\frac{p(1)}{p(0)}}$ veces la apuesta a favor de que salga \'aguila. Despu\'es de observar el resultado $x_1$ de un primer volado la apuesta justa para el segundo volado se debe actualizar a:
 $$a=\frac{p(1|x_1)}{p(0|x_1)}\,b$$
 Y despu\'es de $n$ volados la apuesta justa para el $(n+1)$-\'esimo volado se debe actualizar a:
 $$a=\frac{p(1|(x_1,\ldots,x_n))}{p(0|(x_1,\ldots,x_n))}\,b \qquad \qquad \diamond$$
 \end{ejem}
  
 \newpage
  
 $\S$ \textbf{EJERCICIOS}
 \bigskip
 \begin{enumerate}
  \item Sea $p(x\,|\,\theta)$ el modelo param\'etrico Bernoulli con par\'ametro desconocido $\theta$ y supongamos que la informaci\'on a priori sobre $\theta$ est\'a dada por $p(\theta)=\mbox{Beta}(\theta\,|\,\alpha,\beta)$, con $\alpha$ y $\beta$ conocidos. Suponiendo observaciones muestrales independientes, obtenga la distribuci\'on a posteriori de $\theta$, as\'i como las predictivas a priori y posteriori.
  \item Sea $p(x\,|\,\theta)=\mbox{Unif}(x\,|\,0,\theta)$ con $\theta$ desconocido y con distribuci\'on a priori sobre $\theta$ dada por:
        $$p_{\,1}(\theta)=2(\theta-1)\indic_{[\,1,2\,]}(\theta)$$
Suponiendo observaciones muestrales independientes:
    \begin{enumerate}
       \item Obtener la distribuci\'on predictiva a priori y graficarla.
       \item Obtener la distribuci\'on a posteriori de $\theta$ as\'i como la predictiva a posteriori, en ambos casos para tama\~no de muestra $n\geq 3$.
       \item Obtener nuevamente la distribuci\'on predictiva a priori y graficarla, pero ahora utilizando:
           $$p_{\,2}(\theta)=2(2-\theta)\indic_{[\,1,2\,]}(\theta)$$
y compara con lo obtenido en el inciso a). Compare e interprete las gr\'aficas anteriores.
       \item Utilizando lo obtenido en los incisos a), b) y c) calcula $\hat{\theta}=\esper(\theta)$ sin y con informaci\'on muestral.
       \item Calcula $\prob(\theta<\frac{3}{2})$ sin y con informaci\'on muestral.
    \end{enumerate}
  \item Calcula la distribuci\'on a posteriori del par\'ametro en cuesti\'on as\'i como las predictivas a priori y a posteriori para los siguientes casos:
   \begin{enumerate}
     \item $p(x\,|\,\lambda)=\mbox{Poisson}(x\,|\lambda)$ con $\lambda$ desconocida, $p(\lambda)=\mbox{Gamma}(\lambda\,|\alpha,\beta)$, con $\alpha$ y $\beta$ conocidos.
     \item $p(x\,|\,\theta)=\mbox{Unif}(x\,|\,0,\theta)$ con $\theta$ desconocida, $p(\theta)=\mbox{Pareto}(\theta\,|\,\alpha,\beta)$, con $\alpha$ y $\beta$ conocidos.
   \end{enumerate}
  \item Considera una urna con bolas del mismo tama\~no y numeradas de la $1$ a la $N$, donde $N$ es desconocido. Sea la variable aleatoria $X\sim$ Poisson$(\lambda)$, con $\lambda$ desconocida y sea $N:=X+1$. Se cuenta con la informaci\'on a priori (inicial) de que el valor m\'as probable para $N$ es un valor $k$, conocido. Obtener:
  \begin{enumerate}
    \item Una distribuci\'on a priori para $N$.
    \item La distribuci\'on predictiva a priori.
    \item La distribuci\'on a posteriori de $N$ para una muestra aleatoria de tama\~no $1$.
    \item La distribuci\'on predictiva a posteriori para una muestra aleatoria de tama\~no $1$, suponiendo que las bolas de la muestra son regresadas a la urna antes de hacer predicci\'on.
    \item Suponiendo que $k=3$ y que se tiene la muestra $x_1=2$ calcula las probabilidades a priori y a posteriori de que la urna contenga m\'as de $2$ bolas y explique por qu\'e una es menor que la otra.
    \item Continuando con el inciso anterior, suponiendo que la bola de la muestra se regresa a la urna, calcula las probabilidades a priori y a posteriori de que una bola tomada al azar tenga el n\'umero $3$.
   \end{enumerate}
  \item Si los datos muestrales provienen de observaciones independientes utilizamos $p(\mathbf{x}\,|\,\theta)=\prod p(x_j\,|\,\theta)$, pero si las observaciones no son independientes el modelo general sigue siendo v\'alido, pero en este caso $p(\mathbf{x}\,|\,\theta)\neq\prod p(x_j\,|\,\theta)$. Supongamos que tenemos una urna con bolas numeradas de la $1$ a la $N$ y que lo \'unico que sabemos sobre $N$ es que es $4$ o $5$. 
   \begin{enumerate}
    \item Prop\'on y justifica una distribuci\'on a priori razonable para $N$.
    \item Deduce la distribuci\'on predictiva a priori y calcula la probabilidad de que una bola tomada al azar tenga el n\'umero $5$.
    \item Si se va a tomar una muestra de tama\~no $2$ \textit{sin reemplazo} deduce la distribuci\'on a posteriori de $N$. Luego, suponiendo que la muestra obtenida fueron las bolas $1$ y $3$ calcula la probabilidad de que haya $5$ bolas en la urna sin y con informaci\'on muestral, explicando el por qu\'e de la diferencia.
    \item Supongamos ahora que las dos bolas de la muestra se regresan a la urna. Deduce la distribuci\'on predictiva a posteriori y con base en toda la informaci\'on disponible calcula la probabilidad de que una bola tomada al azar tenga el n\'umero $5$ y compara este resultado con el obtenido en el inciso b), explicando el por qu\'e de la diferencia.
   \end{enumerate}
  \item Sea $\{p_i(\theta)\}_{i=1}^k$ una sucesi\'on de distribuciones de probabilidad sobre $\theta$. Definimos la siguiente distribuci\'on de probabilidad sobre $\theta$ :
      $$p(\theta):=\sum_{i=1}^k \alpha_i\,p_i(\theta)\quad,\quad\sum_{i=1}^k\alpha_i = 1\,\,,\,\alpha_i>0$$
Sea la familia param\'etrica $\mathcal{P}:=\{p(x\,|\,\theta)\,:\,\theta\in\Theta\}$. Si utilizamos como distribuci\'on a priori a la $p(\theta)$ definida anteriormente, demuestre que la distribuci\'on a posteriori de $\theta$ se puede expresar tambi\'en como la combinaci\'on lineal convexa:
   $$p(\theta\,|\,\mathbf{x})=\sum_{i=1}^k \beta_i\,p_i(\theta\,|\,\mathbf{x})$$
exhibiendo la f\'ormula general de $\beta_i$ y $p_i(\theta\,|\,\mathbf{x})$.
 \end{enumerate}

 \chapter{Informaci\'on a priori}

 \section{Determinaci\'on de la distribuci\'on a priori}
 
 Utilizamos el enfoque subjetivo de la probabilidad mencionado en el cap\'itulo 1 para especificar la distribuci\'on a priori $p(\theta)$ con base en la informaci\'on que se tiene en un momento dado, como puede ser: informaci\'on hist\'orica, la experiencia de especialistas, etc. La elecci\'on de una distribuci\'on para modelar nuestro nivel de incertidumbre (o informaci\'on) sobre $\theta$ no resulta crucial, en tanto cualquiera de ellas (o ambas) tengan la capacidad de reflejar la informaci\'on que se tiene sobre $\theta$. \par
 
 \bigskip
 
 \begin{ejem} \label{e2}
  Una compa\~n\'ia de seguros va a ser objeto de una auditor\'ia por parte de la Comisi\'on Nacional de Seguros y Fianzas (CNSyF). La auditor\'ia consistir\'a en revisar los expedientes de los asegurados y determinar qu\'e porcentaje de ellos est\'an incompletos. En caso de que dicho porcentaje exceda el $10\%$ la CNSyF proceder\'a a multar a la compa\~n\'ia de seguros. Antes de que esto suceda, la mencionada compa\~n\'ia decide apoyarse en su \'area de auditor\'ia interna para darse idea del porcentaje de expedientes incompletos. Supongamos que la cantidad de expedientes es tal que s\'olo dar\'ia tiempo de revisar el $0{.}75\%$ de ellos antes de que los audite la CNSyF. Aqu\'i podemos intentar aprovechar la experiencia de los auditores internos de la compa\~n\'ia, formulando algunas preguntas como:
   \begin{enumerate}
    \item De acuerdo a su experiencia y conocimiento de la compa\~n\'ia ?`Alrededor de qu\'e cantidad estiman se ubica el porcentaje de expedientes incompletos? \textit{Respuesta: El a\~no pasado estimamos dicho porcentaje en $8\%$; sin embargo, este a\~no el volumen de ventas ha superado nuestras expectativas y esto generalmente juega un poco en contra en lo que a eficiencia administrativa se refiere por lo que para este a\~no estimamos que dicho porcentaje estar\'a alrededor del $9\%$}.
    \item ?`Cu\'ales ser\'ian sus escenarios optimista y pesimista para dicho porcentaje? \textit{Respuesta: En el mejor de los casos ubicar\'iamos dicho porcentaje en $8\%$ y en el peor de los casos vemos dif\'icil que exceda el $11\%$}.
    \item ?`Qu\'e cantidad de expedientes da tiempo de revisar antes de la auditor\'ia de la CNSyF? \textit{Respuesta: $150$}.
   \end{enumerate}
   Sea $\theta$ el porcentaje de expedientes incompletos. Podemos modelar lo anterior mediante la familia param\'etrica Bernoulli, ya utilizada en el Ejemplo \ref{e1}, pero aqu\'i el espacio param\'etrico $\Theta=\,]\,0,1\,[\,$. Las respuestas a las preguntas 1 y 2 nos dan idea de la centralidad y dispersi\'on de $\theta$. Modelaremos dicha informaci\'on mediante:
   $$p(\theta)=\mbox{Beta}(\theta\,|\,\alpha,\beta)$$
    
   Pudimos haber elegido alguna otra distribuci\'on, la elecci\'on anterior se debe a dos razones: primero, que es una distribuci\'on de probabilidad en el intervalo $]\,0,1\,[$ tal cual la necesitamos; segundo, cuenta con dos \linebreak par\'ametros (que llamaremos \textit{hiperpar\'ametros,} para distinguirlos del par\'ametro de inter\'es en el modelo) que nos permiten controlar de manera amplia la centralidad y dispersi\'on de la distribuci\'on. Habremos de traducir la informaci\'on a priori que se tiene en $p(\theta)$, esto es, a trav\'es de los hiperpar\'ametros $\alpha$ y $\beta$ asign\'andoles valores que reflejen la informaci\'on que se tiene.\par
   La respuesta a la pregunta 1 nos permite establecer la siguiente ecuaci\'on:
      $$\esper(\theta)=0{.}09\,,$$
y la respuesta a la pregunta 2 la podemos expresar como:
      $$\prob[\,0{.}08<\theta<0{.}11\,]=0{.}95\,,$$
as\'i que para asignar valores a $\alpha$ y $\beta$ que reflejen lo anterior basta resolver el siguiente sistema de ecuaciones: 

 \pagebreak

 \hyphenation{ex-pe-dien-tes}

      $$\frac{\alpha}{\alpha+\beta}=0{.}09\,,$$
      $$\int_{0{.}08}^{\,0{.}11}\mbox{Beta}(\theta\,|\,\alpha,\beta)\,d\theta = 0{.}95\,,$$
y obtenemos $\alpha=193{.}090$ y $\beta=1952{.}354$. Bas\'andonos \'unicamente en la informaci\'on a priori disponible podemos calcular la probabilidad de que el porcentaje de expedientes incompletos rebase el $10\%$ :
  $$\prob[\,\theta>0{.}10\,]=\int_{0{.}10}^1 \mbox{Beta}(\theta\,|\,193{.}09  ,1952{.}354)\,d\theta =0{.}0561\,.$$
  
  De la pregunta 3 tenemos que s\'olo queda tiempo para revisar $150$ expedientes que representan tan solo el $0{.}75\%$ de un total de veinte mil expedientes que tiene la compa\\~n\'ia por lo que aprovecharemos esta otra fuente de informaci\'on (informaci\'on muestral) escogiendo al azar $150$ expedientes y obtendremos la informaci\'on muestral $\mathbf{x}=(x_1,\ldots,x_n)$, en donde $x_j\in\{0,1\}$ y $x_j=1$ representa un expediente incompleto. Utilizando el resultado del Ejercicio 1 del Cap\'itulo 2 obtenemos la distribuci\'on a posteriori de $\theta$ :
  $$p(\theta\,|\,\mathbf{x})=\mbox{Beta}(\theta\,|\,\alpha+r,\beta+n-r)\,,$$
en donde 
  $$r:=\sum_{j=1}^n x_j\,,$$
esto es, $r$ es el n\'umero de expedientes incompletos de una muestra aleatoria de tama\~no $n=150$. Ya con toda la informaci\'on disponible (a priori y muestral) \textit{actualizamos} la probabilidad de que el porcentaje de expedientes incompletos rebase el $10\%$ :
  $$\prob[\,\theta>0{.}10\,|\,\mathbf{x}\,]=\int_{0{.}10}^1 \mbox{Beta}(\theta\,|\,193{.}09+r  ,2102{.}354-r)\,d\theta\qquad\diamond$$
 \end{ejem}
 
 \medskip
 
 Como comentario general, al proceso de traducir informaci\'on a priori en una distribuci\'on a priori se le conoce en ingl\'es como \textit{to elicit a prior distribution}. Aunque una traducci\'on de la palabra \textit{elicit} con la misma ra\'iz etimol\'ogica no existe (a\'un) en espa\~nol, en el resto del texto definimos con este mismo sentido \textit{elicitar}.\footnote{En lat\'in \textit{elicitum} quiere decir ``hacer salir'' o ``sacar de''.}
 
 \smallskip
 
 En el ejemplo anterior, result\'o relativamente sencillo elicitar una distribuci\'on a priori para $\theta$, especialmente por el hecho de que la familia param\'etrica es univariada, pero trat\'andose de modelos multivariados elicitar una distribuci\'on a priori puede resultar bastante complicado. De hecho, Lindley (2000) pronostica que uno de los temas m\'as importantes en la investigaci\'on estad\'istica del nuevo milenio ser\'a el desarrollo de metodolog\'ias adecuadas para la asignaci\'on de probabilidades [subjetivas], y caso particular de esto es el c\'omo elicitar una distribuci\'on a priori.\par
 
 \smallskip
 
 Es importante destacar en el ejemplo anterior que al haber elegido una distribuci\'on beta para modelar la informaci\'on a priori sobre $\theta$ bajo la familia param\'etrica Bernoulli nos arroj\'o como resultado que la distribuci\'on a posteriori sobre $\theta$ es tambi\'en una distribuci\'on beta, aunque con hiperpar\'ametros distintos. En ocasiones y bajo ciertas familias param\'etricas la elecci\'on de ciertas distribuciones a priori trae como consecuencia que la distribuci\'on a posteriori del par\'ametro sea de la misma familia que la distribuci\'on a priori (por ejemplo, Ejercicio 3, Cap\'itulo 2), pero esto no siempre es as\'i (Ejercicios 2, 4 y 5 del Cap\'itulo 2). De esto nos ocupamos en la siguiente secci\'on.

 \section{Familias conjugadas}
 
 Tanto $p(\theta)$ como $p(\theta\,|\,\mathbf{x})$ son distribuciones de probabilidad sobre $\theta$ : la primera s\'olo incorpora informaci\'on a priori y la segunda \textit{actualiza} dicha informaci\'on con la informaci\'on muestral que se pueda obtener. Si bien dijimos que la elecci\'on de una distribuci\'on de probabilidad para modelar nuestra incertidumbre sobre $\theta$ no resulta crucial en tanto sea factible elicitar con cualquiera de ellas una distribuci\'on a priori, resulta conveniente tanto para el an\'alisis como desde un punto de vista computacional el que $p(\theta)$ y $p(\theta\,|\,\mathbf{x})$ pertenezcan a la misma familia.\par
 
 \bigskip
 
 \begin{enun} 
  \begin{defn} \label{defn familia conjugada}
   Sea $\mathcal{P}:=\{p(x\,|\,\theta)\,:\,\theta\in\Theta\}$ una familia param\'etrica. Una clase (o colecci\'on) de distribuciones de probabilidad $\mathcal{F}$ es una \textit{familia conjugada} para $\mathcal{P}$ si para todo $p(x\,|\,\theta)\in\mathcal{P}$ y $p(\theta)\in\mathcal{F}$ se cumple que $p(\theta\,|\,\mathbf{x})\in\mathcal{F}$.
  \end{defn}
 \end{enun}
 
 \smallskip
 
 Como ejemplos de lo anterior est\'an los resultados de los Ejercicios 1 y 3 del Cap\'itulo 2. Es inmediato notar que si $p(\theta)$ es conjugada para una familia param\'etrica $\mathcal{P}$ entonces las distribuciones predictivas a priori y a posteriori pertenecen a una misma familia de distribuciones $\mathcal{F}\,'$.
 
 \bigskip
 
 A continuaci\'on se presentan algunos modelos param\'etricos univariados con sus respectivas familias conjugadas:
 
 \begin{table}[h]\label{tabla familias conjugadas}
  \caption{Algunas familias conjugadas}
  \begin{center}
   \begin{tabular}{cc} \hline
    { } & { } \\
    Fam. param\'etrica & Fam. conjugada \\ \hline
    { } & { } \\
    Bernoulli($\theta$)  & Beta $(\theta\,|\,\alpha,\beta)$ \\
    Poisson $(\lambda)$   &  Gamma $(\lambda\,|\,\alpha,\beta)$ \\
    Geom\'etrica $(\theta)$ & Beta $(\theta\,|\,\alpha,\beta)$ \\ 
    Exponencial $(\lambda)$   &  Gamma $(\lambda\,|\,\alpha,\beta)$ \\
    Uniforme $(0,\theta)$ & Pareto $(\theta\,|\,\alpha,\beta)$\\
    Normal $(\mu)$ & Normal $(\mu\,|\,\mu_0,\lambda_0)$ \\
    Normal $(\lambda)$ & Gamma $(\lambda\,|\,\alpha,\beta)$ \\
    Normal $(\mu,\lambda)$ & Normal-Gamma $(\mu,\lambda\,|\,\mu_0,n_0,\alpha,\beta)$ \\
    { } & { } \\ \hline
   \end{tabular}
  \end{center}
 \end{table}
 
 En lo anterior, para el caso de la Normal usamos como $\lambda$ el inverso de la varianza y por ello la llamamos \textit{precisi\'on}. Se hace este cambio para utilizar la distribuci\'on gamma en vez de la gamma invertida.\par
 
 \bigskip
 
 \begin{ejem} \label{ejem fam conjugada}
   Sea la familia param\'etrica $\mathcal{P}:=\{\mbox{Poisson}(x\,|\,\lambda)\,:\,\lambda\in\mathbb{R}^{+}\}$. Si utilizamos como distribuci\'on a priori $p(\lambda)\in\mathcal{F}:=\{Gamma(\lambda\,|\,\alpha,\beta)\,:\,\alpha,\beta \in\mathbb{R}^{+}\}$ entonces para una muestra aleatoria $\mathbf{x}=(x_1,\ldots,x_n)$ :
   $$p(\lambda\,|\,\mathbf{x})=\mbox{Gamma}(\lambda\,|\,\alpha+r,\beta+n)$$
en donde
   $$r:=\sum_{j=1}^n x_j$$
y adem\'as
   $$p(x)=\mbox{Pg}(x\,|\,\alpha,\beta,1)$$
   $$p(x\,|\,\mathbf{x})=\mbox{Pg}(x\,|\,\alpha+r,\beta+n,1)$$
en donde Pg se refiere a la distribuci\'on Poisson-gamma:
   $$\mbox{Pg}(x\,|\,\alpha,\beta,n)=\frac{\beta^{\alpha}}{\Gamma(\alpha)} \frac{\Gamma(\alpha+x)}{x!} \frac{n^x}{(\beta+n)^{\alpha+x}} \indic_{\{0,1,\ldots\}}(x)$$
cuya esperanza y varianza est\'an dadas por $\esper(X)=n\frac{\alpha}{\beta}$ y  $\vari(X)=\frac{n\alpha}{\beta} \big[1+\frac{n}{\beta}\big]$, respectivamente. $\quad\diamond$
 \end{ejem}
 
 En el ejemplo anterior, $\alpha$ y $\beta$ son los hiperpar\'ametros de la distribuci\'on a priori de $\lambda$, y por tanto se les debe asignar valores que reflejen la informaci\'on a priori que se tenga, como se hizo en el Ejemplo \ref{e2}. La distribuci\'on a posteriori $p(\lambda\,|\,\mathbf{x})$ es una gamma con par\'ametros $\alpha+r$ y $\beta+n$, lo cual ilustra c\'omo se combinan informaci\'on a priori e informaci\'on muestral. Aunque se ver\'a a mayor detalle m\'as adelante c\'omo hacer estimaci\'on puntual, en el Cap\'itulo 2 se mencion\'o que una manera de hacer estimaci\'on puntual sobre el par\'ametro es calculando su esperanza, aprovechando el hecho de que se tiene una distribuci\'on de probabilidad sobre $\lambda$ y en este caso:
  $$\hat{\lambda}=\esper(\lambda\,|\,\mathbf{x})=\frac{\alpha+r}{\beta+n}$$
Respecto a lo anterior es importante la siguiente observaci\'on. Por simplicidad supongamos por un momento que $\beta=\alpha$. Si la magnitud de $\alpha$ es ``muy grande'' en comparaci\'on con $r$ y $n$ tendremos el caso en que la informaci\'on a priori tendr\'a m\'as peso que la informaci\'on muestral en las inferencias que se realicen con la distribuci\'on a posteriori. En el ejemplo anterior se tendr\'ia que $\esper(\lambda\,|\,\mathbf{x})$ ser\'ia aproximadamente igual a $1$ y $\vari(\lambda\,|\,\mathbf{x})$ aproximadamente igual a $\alpha^{-1}$, varianza que para valores ``grandes'' de $\alpha$ se acercar\'ia a cero, lo cual nos hablar\'ia de una distribuci\'on cuya densidad (o masa) est\'a muy concentrada alrededor de un punto, En este caso diremos que la distribuci\'on a priori es \textit{muy informativa}. Si, por el contrario, $\alpha$ es cercana a cero tendremos que la distribuci\'on a priori tiene una varianza muy grande y en tal caso diremos que se trata de una distribuci\'on a priori \textit{poco informativa}. En el ejemplo anterior se tendr\'ia que $\esper(\lambda\,|\,\mathbf{x})$ es aproximadamente igual a la media de los datos muestrales, que quiere decir que ante tan poca informaci\'on a priori las inferencias que se hagan se apoyar\'an pr\'acticamente s\'olo en la informaci\'on que provean los datos muestrales. Esto \'ultimo es materia de la siguiente secci\'on.\par

 \section{Distribuciones a priori no informativas}
  
 La estad\'istica bayesiana proporciona una metodolog\'ia que permite combinar de manera consistente informaci\'on a priori con informaci\'on experimental (i.e. muestral). Ante esto surge la pregunta de c\'omo realizar inferencias cuando no se dispone de informaci\'on a priori, o bien cuando dicha informaci\'on no se quiere o no se puede utilizar. \par
 
 \smallskip
 
 El problema quedar\'ia resuelto si pudi\'esemos determinar una distribuci\'on a priori que describa la situaci\'on en que los datos experimentales contienen toda la informaci\'on relevante, en lugar de proporcionar tan s\'olo parte de ella como sucede cuando se dispone de informaci\'on a priori. Una forma pragm\'atica (pero incompleta) de atacar este problema ser\'ia asignar arbitrariamente una distribuci\'on a priori con la \'unica condici\'on de que tenga una varianza ``muy grande'', con todo lo relativo que esto \'ultimo puede resultar. \par
 
 \bigskip
 
 Otra forma ser\'ia la siguiente. Supongamos que tenemos un n\'umero finito de sucesos inciertos (o hip\'otesis) $E_1,\ldots,E_k$. Una distribuci\'on a priori que describe un estado de ignorancia o carencia de informaci\'on es la siguiente:
    $$\prob(E_j)=\frac{1}{k}\indic_{\{1,\ldots,k\}}(j)$$
esto es, una distribuci\'on uniforme discreta.\par

 \smallskip
 
 Thomas Bayes propuso esta distribuci\'on con base en lo que \'el llam\'o el \textit{Principio de la Raz\'on Insuficiente}: Si no sabemos cosa alguna sobre \linebreak $\{E_1,\ldots,E_k\}$ no hay raz\'on para asignarle a alguno de los sucesos inciertos una probabilidad diferente que a los otros.\par
 
 Sin embargo, este principio no es aplicable en situaciones donde el n\'umero de sucesos inciertos no es finito. Volviendo al Ejemplo \ref{e2}, supongamos ahora que no se tiene informaci\'on alguna acerca de la proporci\'on $\theta$ de expedientes incompletos. Bajo el principio de la raz\'on insuficiente propondr\'iamos como distribuci\'on a priori no informativa para $\theta$ :
 $$p(\theta)=\indic_{]\,0,1\,[}(\theta)$$
es decir, una distribuci\'on uniforme continua en $]\,0,1\,[$ . Supongamos ahora que, m\'as que estar interesados en $\theta$ directamente, estamos interesados en una funci\'on uno-a-uno de $\theta$, digamos $\varphi:=-\log\theta$. Si no tenemos informaci\'on alguna de $\theta$ entonces tampoco tenemos informaci\'on sobre $\varphi\,.$ Bajo el principio de la raz\'on insuficiente asignar\'iamos tambi\'en una distribuci\'on uniforme continua para $\varphi$, pero aqu\'i aparece el primer problema porque $\varphi$ toma valores en $]\,0,\infty\,[$. De hecho, resultado de probabilidad elemental es que si $\theta$ de distribuye como uniforme continua en $]\,0,1\,[$ entonces $\varphi$ se distribuye exponencial con par\'ametro $1$, la cual, por ejemplo, asigna mayor probabilidad a valores de $\varphi$ en un intervalo $]\,0,1\,[$ que en el intervalo $]\,1,2\,[$ , violando as\'i el principio de la raz\'on insuficiente, por lo que dicho principio no produce distribuciones a priori consistentes en el sentido de que no resultan ser invariantes ante reparametrizaciones uno-a-uno.\par

 \bigskip

 \section{Regla de Jeffreys}
 
 S\'olo en el caso en que el espacio par\'ametrico $\Theta$ sea finito el principio de la raz\'on insuficiente provee distribuciones a priori no informativas que son invariantes ante transformaciones uno-a-uno del par\'ametro (o vector de par\'ametros). Sin embargo, esto es poco \'util ya que por lo general nos enfrentamos a espacios param\'etricos infinitos.\par
 
 \medskip

 Jeffreys (1961) propuso una clase de distribuciones a priori no informativas para el caso de espacios param\'etricos infinitos. En t\'erminos generales, la construcci\'on de esta clase consiste en buscar simult\'aneamente invariancia ante transformaciones y proveer la menor informaci\'on a priori en relaci\'on a la informaci\'on muestral, v\'ia la informaci\'on de Fisher.\par
 
 \medskip
 
 Citaremos algunos resultados de probabilidad para recordar el concepto de la informaci\'on de Fisher. Las demostraciones se pueden consultar en libros como el de Casella y Berger (2002) y el de Lehmann y Casella (1998), entre muchos otros.\par
 \smallskip
 
 \hyphenation{va-ria-bles}
 
 \begin{enun}\label{teor cramer-rao}
  \begin{teor}
   \textbf{\textrm{(cota inferior de Cr\'amer-Rao)}} Sean $X_1,\ldots,X_n$ variables aleatorias con funci\'on de densidad conjunta $p(\mathbf{x}\,|\,\theta)$, $\theta\in\Theta\subset\mathbb{R}$. Sea $W(\mathbf{X})=W(X_1,\ldots,X_n)$ cualquier funci\'on tal que $\esper_{\theta}(W(\mathbf{X}))$ sea una funci\'on diferenciable de $\theta$. Suponiendo que $p(\mathbf{x}\,|\,\theta)=p(x_1,\ldots,x_n\,|\,\theta)$ satisface:
   $$\frac{d}{d\theta}\,\esper_{\theta}(W(\mathbf{X}))=\int_{\quad\mathbb{R}^{\,n}}\!\!\!\!\!\!\!\cdots\int W(\mathbf{x})\frac{\partial}{\partial\theta}\,p(\mathbf{x}\,|\,\theta)\,dx_1\cdots dx_n$$
y adem\'as $\vari_{\theta}(W(\mathbf{X}))<\infty\,,$ entonces se cumple que:
  $$\vari_{\theta}\big(W(\mathbf{X})\big)\geq\frac{\Big[\frac{d}{d\theta} \esper_{\theta}\big(W(\mathbf{X})\big)\Big]^2}{\esper_{\theta}\Big[\big( \frac{\partial}{\partial\theta}\log p(\mathbf{X}\,|\,\theta)\big)^2\Big]}$$
  \end{teor}
 \end{enun}
 
 \smallskip
 
 \begin{enun}\label{corolario cramer-rao}
  \begin{cor}
   Si adem\'as $X_1,\ldots,X_n$ son independientes e id\'enticamente distribuidas con funci\'on de densidad com\'un $p(x\,|\,\theta)$ entonces:
   $$\vari_{\theta}\big(W(\mathbf{X})\big)\geq\frac{\Big[\frac{d}{d\theta} \esper_{\theta}\big(W(\mathbf{X})\big)\Big]^2}{n\esper_{\theta}\Big[\big( \frac{\partial}{\partial\theta}\log p(X\,|\,\theta)\big)^2\Big]}$$
  \end{cor}
 \end{enun}
 
 \smallskip
 
 El teorema de Cr\'amer-Rao es v\'alido tambi\'en para variables aleatorias discretas siempre y cuando sea v\'alido intercambiar diferenciaci\'on y sumas, as\'i como que la funci\'on de masa de probabilidades $p(x\,|\,\theta)$ sea diferenciable respecto a $\theta$.\par
 
 Cuando se tiene una muestra aleatoria $\mathbf{X}=(X_1,\ldots,X_n)$, a la cantidad $\displaystyle{n\esper_{\theta}\Big[\Big( \frac{\partial}{\partial\theta}\log p(X\,|\,\theta)\Big)^2\Big]}$ se le conoce como la \textit{informaci\'on de Fisher de la muestra} y a la cantidad $\displaystyle{\esper_{\theta}\Big[\Big( \frac{\partial}{\partial\theta}\log p(X\,|\,\theta)\Big)^2\Big]}$ se le conoce como \textit{informaci\'on de Fisher por unidad muestral} y la denotamos $I(\theta)$. Para facilitar el c\'alculo de $I(\theta)$ se tiene el siguiente resultado:\par
 
 \smallskip
 
 \begin{enun}\label{lema info de fisher}
  \begin{lema}
   Si adem\'as de lo anterior $p(x\,|\,\theta)$ satisface:
   \begin{eqnarray*}
     \frac{d}{d\theta}\,\esper_{\theta}\Big[\frac{\partial}{\partial\theta}\log p(X\,|\,\theta)\Big] &=& \int_{\mathcal{X}}\!\frac{\partial}{\partial\theta} \Big[\Big(\frac{\partial}{\partial\theta}\log p(x\,|\,\theta)\Big)p(x\,|\,\theta)\Big]\,dx \\ &=& \int_{\mathcal{X}}\! \frac{\partial^2}{\partial\theta^2}p(x\,|\,\theta)\,dx
   \end{eqnarray*}
en donde $\mathcal{X}:=\mbox{Ran}\,X$, entonces se cumple:
   $$\esper_{\theta}\Big[\Big( \frac{\partial}{\partial\theta}\log p(X\,|\,\theta)\Big)^2\Big]=-\esper_{\theta}\Big[\frac{\partial^2} {\partial\theta^2}\log p(X\,|\,\theta)\Big]$$
  \end{lema}
 \end{enun}
 
 Por lo anterior es com\'un simplemente definir la \textit{informaci\'on de Fisher} del modelo param\'etrico $p(x\,|\,\theta)$ como:
   $$I(\theta):=-\esper_{\theta}\Big[\frac{\partial^2} {\partial\theta^2}\log p(X\,|\,\theta)\Big]$$\par
   
  Con lo anterior, el Corolario \ref{corolario cramer-rao} puede expresarse como:
  $$\vari_{\theta}\big(W(\mathbf{X})\big)\geq\frac{\Big[\frac{d}{d\theta} \esper_{\theta}\big(W(\mathbf{X})\big)\Big]^2}{nI(\theta)}$$\par
  
  Para el caso de muestras aleatorias, $W(\mathbf{X})$ es lo que se utiliza en estad\'istica frecuentista para estimar $\theta$ o alguna funci\'on de $\theta$. Recu\'erdese que bajo el enfoque frecuentista la \'unica fuente de informaci\'on sobre $\theta$ es la muestra aleatoria. En caso de que $\esper_{\theta}(W(\mathbf{X}))=\theta$ (que en estad\'istica frecuentista se dice en tal caso que $W(\mathbf{X})$ es un estimador \textit{insesgado} de $\theta$), la varianza de dicho estimador satisface:
  $$\vari_{\theta}\big(W(\mathbf{X})\big)\geq\frac{1}{nI(\theta)}$$
Si se tienen varios estimadores de $\theta$ cuya esperanza es justamente $\theta$ se prefieren aquellos que tengan menor varianza (i.e. los que sean m\'as informativos), y el mejor ser\'a aqu\'el o aqu\'ellos cuya varianza coincida con la cota inferior de Cr\'amer-Rao ya que por dicho teorema es la m\'inima. Notemos pues que si $I(\theta)$ es ``grande'' entonces hay posibilidades de obtener un estimador con menor varianza y en tal caso decimos que el modelo param\'etrico $p(x\,|\,\theta)$ es muy informativo. Si $I(\theta)$ es peque\~no entonces la m\'inima varianza posible de un estimador de $\theta$ ser\'a grande y en tal caso diremos que el modelo param\'etrico es poco informativo. Como $I(\theta)$ depende justamente de $\theta$ entonces la varianza de los estimadores (frecuentistas) de $\theta$ depender\'a del supuesto ``verdadero'' valor de $\theta$.\par

  La pregunta natural que surge es cu\'ales familias param\'etricas satisfacen las condiciones del teorema de Cr\'amer-Rao (conocidas usualmente como \textit{condiciones de regularidad}) as\'i como la condici\'on del Lema \ref{lema info de fisher}. Afortunadamente varias de las familias conocidas satisfacen dichas condiciones, entre las que se encuentran las pertenecientes a la \textit{Familia Exponencial,} como son la normal, gamma, beta, lognormal, binomial, Poisson, binomial negativa, entre otras. Las distribuciones cuyo rango de la variable aleatoria depende del par\'ametro no satisfacen dichas condiciones, como es el caso de la distribuci\'on uniforme o Pareto.\par
  
 \smallskip
  
 \begin{ejem}\label{ejem info fisher binomial}
  Sea $p(x\,|\,\theta)=\mbox{Binomial}(x\,|\,m,\theta)$ con $m$ fija y $\theta\in\Theta=\,]\,0,1\,[$ . Es inmediato verificar que:
     $$I(\theta)=\frac{m}{\theta(1-\theta)}$$
Observemos que $I(\theta)\rightarrow\infty$ conforme $\theta\rightarrow 0$ o bien $\theta\rightarrow 1$ y que alcanza un m\'inimo en $\theta=\frac{1}{2}$. Esto quiere decir que para valores de $\theta$ cercanos a $0$ o a $1$ un estimador (muestral) $W(\mathbf{X})$ de m\'inima varianza se vuelve m\'as informativo, y menos informativo conforme $\theta$ se aproxime a $\frac{1}{2}. \quad\diamond$
 \end{ejem}
 
 \smallskip
  
  Todo lo anterior fue para el caso en que el espacio param\'etrico sea unidimensional, pero para el caso multidimensional se tiene un resultado an\'alogo (ver Lehmann (1998)), s\'olo que en este caso $\mathbf{\theta}$ es un vector de par\'ametros y obtenemos una \textit{matriz de informaci\'on de Fisher} $\mathbf{I}(\mathbf{\theta})=\|I_{ij}(\mathbf{\theta})\|$ cuyas entradas est\'an dadas por:
  $$I_{ij}(\mathbf{\theta})=-\esper\Big[\frac{\partial^2}{\partial\theta_i \partial\theta_j}\log p(X\,|\,\mathbf{\theta})\Big]$$\par
  
 \medskip
 
 \begin{enun}\label{regla de jeffreys}
  \begin{defn}
   Para un modelo param\'etrico $p(x\,|\,\theta)$ la \textit{distribuci\'on a priori no informativa de Jeffreys} para $\theta$ est\'a dada por:
 $$p(\theta)\propto\sqrt{I(\theta)}\,\,,\quad\theta\in\Theta\subset\mathbb{R}$$
En el caso multidimensional se tiene:
  $$p(\mathbf{\theta})\propto\sqrt{\det \mathbf{I}(\mathbf{\theta})}$$
En cualquier caso la denotaremos por $\pi(\theta)$.
  \end{defn}
 \end{enun}
 
 La idea es favorecer los valores de $\theta$ para los cuales $I(\theta)$, o en su caso $\det\mathbf{I}(\theta)$, es grande, lo que resta influencia a la distribuci\'on a priori dando mayor peso a la informaci\'on muestral. La ra\'iz cuadrada aparece para que resulte invariante bajo transformaciones uno-a-uno. La anterior \textit{Regla de Jeffreys} tiene la desventaja de que en ocasiones produce \textit{distribuciones impropias} (i.e. no integran a 1) lo cual no es del todo grave si realmente lo que se quiere es trabajar con una distribuci\'on a posteriori $\pi(\theta\,|\,\mathbf{x})$ que describa la incertidumbre sobre $\theta$ bas\'andose \'unicamente en la informaci\'on muestral $\mathbf{x}=(x_1,\ldots,x_n)\,.$ Bastar\'a que se cumpla la condici\'on
$$\int_{\Theta}\!p(\mathbf{x}\,|\,\tilde{\theta})\,\pi(\tilde{\theta})\, d\tilde{\theta} <\infty$$
para que la distribuci\'on a posteriori sea en efecto una distribuci\'on de probabilidad sobre $\theta$ :
  $$\pi(\theta\,|\,\mathbf{x})=\frac{p(\mathbf{x}\,|\,\theta)\,\pi(\theta)} {\int_{\Theta}p(\mathbf{x}\,|\,\tilde{\theta})\,\pi(\tilde{\theta})\, d\tilde{\theta}}\,\,.$$
  
  \medskip
  
  \begin{enun}
   \begin{lema} \label{invariancia de jeffreys}
    La distribuci\'on a priori no informativa de Jeffreys \linebreak $\pi(\theta)\propto \sqrt{I(\theta)}$ es invariante ante transformaciones uno-a-uno, esto es, si $\varphi=\varphi(\theta)$ es una transformaci\'on uno-a-uno de $\theta$ entonces la distribuci\'on a priori no informativa de Jeffreys para $\varphi$ es $p(\varphi)\propto\sqrt{I(\varphi)}$.
   \end{lema}
  \end{enun}
  \begin{proof}
   Sea $\varphi=\varphi(\theta)$ una transformaci\'on uno-a-uno de $\theta$. Entonces:
   $$\frac{\partial\log p(X\,|\,\theta)}{\partial\varphi} = \frac{\partial\log p(X\,|\,\varphi(\theta))}{\partial\theta} \frac{\partial\theta}{\partial\varphi}$$
en donde $\theta=\theta(\varphi)$ es la inversa de la transformaci\'on $\varphi$. Para obtener la informaci\'on de Fisher de $\varphi$ calculamos:
 $$\frac{\partial^2 \log p(X\,|\,\varphi)}{\partial\varphi^2} = \frac{\partial\log p(X\,|\,\varphi(\theta))}{\partial\theta} \frac{\partial^2\theta}{\partial\varphi^2} + \frac{\partial^2\log p(X\,|\,\varphi(\theta))}{\partial\theta^2} \Big(\frac{\partial\theta}{\partial\varphi}\Big)^2$$
Multiplicando ambos miembros por $-1$ y calculando esperanza respecto a $p(x\,|\,\theta)$ :
   \begin{eqnarray*}
     I(\varphi) & = & -\esper\bigg[\frac{\partial\log p(X\,|\,\theta)} {\partial\theta}\bigg]\frac{\partial^2\theta}{\partial\varphi^2} + I(\theta)\Big(\frac{\partial\theta}{\partial\varphi}\Big)^2
   \end{eqnarray*}
pero tenemos que:
   \begin{eqnarray*}
     \esper\bigg[\frac{\partial\log p(X\,|\,\theta)}{\partial\theta}\bigg] &=&
         \esper\bigg[\frac{\frac{\partial}{\partial\theta}p(X\,|\,\theta)} {p(X\,|\,\theta)}\bigg] \\
         &=& \int_{-\infty}^{\infty}\frac{\frac{\partial}{\partial\theta}p(x\,|\,\theta)} {p(x\,|\,\theta)} p(x\,|\,\theta)\,dx \\
         &=& \int_{-\infty}^{\infty}\frac{\partial}{\partial\theta}p(x\,|\,\theta)\,dx \\
         &=& \frac{d}{d\theta}\int_{-\infty}^{\infty}p(x\,|\,\theta)\,dx\quad \mbox{(por las condiciones de regularidad)} \\
         &=& \frac{d}{d\theta}(1)\,=\,0
   \end{eqnarray*}
por lo que:
   $$I(\varphi)=I(\theta)\Big(\frac{\partial\theta}{\partial\varphi}\Big)^2$$
esto es:
   $$\sqrt{I(\varphi)} = \sqrt{I(\theta)}\, |\partial\theta/\partial\varphi|$$
pero $|\partial\theta/\partial\varphi|$ es el valor absoluto del jacobiano de la transformaci\'on inversa por lo que si $\pi(\theta)\propto\sqrt{I(\theta)}$ entonces:
   $$p(\varphi)\propto\sqrt{I(\theta(\varphi))}\, |\partial\theta/\partial\varphi|=\sqrt{I(\varphi)}$$
y por lo tanto la distribuci\'on a priori de Jeffreys es invariante ante transformaciones uno-a-uno.
  \end{proof}
 
  \begin{enun}\label{jeffreys multidimensional}
   \begin{teor}
    El mismo resultado es v\'alido para el caso en que $\theta$ tiene un espacio param\'etrico multidimensional. Ver Lehmann y Casella (1998).
   \end{teor}
  \end{enun}
  
  \medskip
  
  \begin{ejem}\label{ejem a priori de jeffreys para la bernoulli}
    Utilizando el resultado del Ejemplo \ref{ejem info fisher binomial} con $m=1$ obtenemos la distribuci\'on a priori de Jeffreys para la familia param\'etrica Bernoulli: 
    $$\pi(\theta)\propto\theta^{-1/2}(1-\theta)^{-1/2}$$
esto es, el \textit{kernel} de $\pi(\theta)$ corresponde a una distribuci\'on beta por lo que en este caso $\pi(\theta)=\mbox{Beta}(\theta\,|\,\frac{1}{2},\frac{1}{2})$
y como la distribuci\'on beta es conjugada de la familia Bernoulli, del Ejercicio 1 del Cap\'itulo 2 tenemos que la distribuci\'on a posteriori de $\theta$ es $\pi(\theta\,|\,(x_1,\ldots,x_n))=\mbox{Beta}(\theta\,|\, \frac{1}{2}+r, \frac{1}{2}+n-r)$ donde $r:=\sum x_j\,.\quad\diamond$
  \end{ejem}
  
  \medskip
  
  El siguiente es un ejemplo en el que la distribuci\'on a priori de Jeffreys es impropia; sin embargo, esto no es del todo relevante ya que lo que se busca es que las inferencias se apoyen exclusivamente en la informaci\'on muestral que se obtenga por lo que lo importante ser\'a que las distribuciones a posteriori sean \textit{propias} (i.e. que integren a 1):\par
  
  \begin{ejem}\label{ejem a priori de jeffreys para la poisson}
    Consideremos la familia param\'etrica $\mbox{Poisson}(x\,|\,\lambda)$, en donde $\lambda\in\mathbb{R}^{+}$. Es inmediato verificar que la informaci\'on de Fisher est\'a dada por:
    $$I(\lambda)=\frac{1}{\lambda}$$
y por lo tanto la distribuci\'on a priori de Jeffreys es $\pi(\lambda)\propto \lambda^{-1/2}$ la cual resulta ser impropia. Sin embargo, la distribuci\'on a posteriori de $\lambda$ :
   $$\pi(\lambda\,|\,\mathbf{x})\propto p(\mathbf{x}\,|\,\lambda)\,\pi(\theta) \propto e^{-n\lambda}\lambda^{\sum x_j - 1/2}$$
Lo anterior es el kernel de la distribuci\'on gamma por lo que $\pi(\lambda\,|\,(x_1,\ldots,x_n))=\mbox{Gamma}(\lambda\,|\,\sum x_j + 1/2,n)\,.\quad\diamond$
  \end{ejem}
  
  \bigskip
  
  Adem\'as de la Regla de Jeffreys existen otras propuestas para la construcci\'on de distribuciones no informativas, entre las que destacan las \textit{distribuciones de referencia} de Bernardo (1979), ver tambi\'en Bernardo y Smith (1994).
  
  \bigskip
  \bigskip
  
  $\S$ \textbf{EJERCICIOS}
  \bigskip
  \begin{enumerate}
    \item Verifique las familias conjugadas del Cuadro 3.1.
    \item Un problema que interesa en riesgo de cr\'edito es la posibilidad de que una empresa que emiti\'o t\'itulos de deuda (pagar\'es, bonos, etc.) para financiarse, incumpla en el pago de dicha obligaci\'on al vencimiento de dichos t\'itulos. En primer t\'ermino, existe incertidumbre en cuanto ha si ser\'a o no solvente para regresar el dinero que obtuvo en pr\'estamo en la fecha fijada. En segundo t\'ermino y en caso de que incurra en incumplimiento, existe tambi\'en incertidumbre en cuanto al porcentaje de incumplimiento, esto es, puede ocurrir que no pueda cumplir al 100\% con el reembolso pero quiz\'as pueda hacer un pago parcial del $c\%$ de la obligaci\'on y en tal caso diremos que el incumplimiento fue del $(100-c)\%$, con $0\leq c\leq 100$. Por lo general cada empresa tiene un perfil particular y no es comparable con otras, y de hecho ni con su propio historial crediticio ya que las condiciones del pasado para una misma empresa suelen ser muy diferentes a las del presente, por lo que se pretende modelar el porcentaje de incumplimiento \'unicamente con base en la informaci\'on a priori que proporcione un analista o grupo de analistas de cr\'edito. Supongamos que se cuenta con la siguiente informaci\'on a priori: los analistas de cr\'edito estiman que la probabilidad de incumplimiento se ubica entre 5 y 15\% y que en caso de que se de el incumplimiento el porcentaje de incumplimiento se ubica entre 60 y 100\%. Proponga los modelos adecuados, obtenga la distribuci\'on predictiva a priori del porcentaje de incumplimiento y calcule el porcentaje esperado de incumplimiento.
    \item Supongamos que la llegada de autos a la caseta de cobro de una autopista los d\'ias viernes de 5 a 8 p.m. se puede modelar mediante la familia param\'etrica Poisson. Hacemos dos preguntas al encargado de la caseta: ?`Como cu\'antos autos llegan en promedio por minuto a la caseta? A lo cual nos responde que 5. Tomando en cuenta que el dato anterior es una apreciaci\'on subjetiva ?`Cu\'al cree usted que ser\'ia en el mayor de los casos el n\'umero promedio de autos por minuto? A lo cual nos responde que 12.
      \begin{enumerate}
        \item Utilizando una distribuci\'on conjugada especifique la distribuci\'on a priori del par\'ametro con base en la informaci\'on que se tiene. Calcule el valor esperado del par\'ametro as\'i como la probabilidad de que dicho par\'ametro sea mayor a 8.
        \item Supongamos ahora que procedemos a tomar una muestra aleatoria y obtenemos $\mathbf{x}=(679,703,748,739,693)$. Obtenga la distribuci\'on a posteriori del par\'ametro y calcule el valor esperado del par\'ametro as\'i como la probabilidad de que dicho par\'ametro sea mayor a 8. Compare con el inciso a). Grafique en una misma hoja la distribuci\'on a priori, la distribuci\'on a posteriori con el primer dato, con los primeros dos y as\'i sucesivamente hasta la a posteriori con los cinco datos.
        \item Utilizando la Regla de Jeffreys y la informaci\'on del inciso anterior obtenga la distribuci\'on a posteriori del par\'ametro y calcule el valor esperado del par\'ametro as\'i como la probabilidad de que dicho par\'ametro sea mayor a 8. Compare con el inciso b). Grafique lo an\'alogo al inciso anterior. ?`Qu\'e se puede concluir sobre la informaci\'on a priori proveniente del encargado de la caseta?
      \end{enumerate}
    \item Utilizando la Regla de Jeffreys obtenga las distribuciones a posteriori de las siguientes distribuciones uniparam\'etricas univariadas:
      \begin{enumerate}
        \item Geom\'etrica
        \item Exponencial
        \item Normal (con precisi\'on conocida)
        \item Normal (con media conocida)
      \end{enumerate}
    \item Verifique si las siguientes distribuciones a priori impropias son no informativas ya que producen distribuciones a posteriori que dependen \'unicamente de la informaci\'on muestral y son invariantes ante transformaciones uno a uno:
      \begin{enumerate}
        \item $p(\theta)\propto\theta^{-1}$ para el modelo continuo Uniforme $(0,\theta)$, $\theta\in\mathbb{R}^{+}$.
        \item $p(\theta)\propto k$, $k$ una constante, para el modelo Normal con precisi\'on conocida.
      \end{enumerate}
  \end{enumerate}

  \chapter{Elementos de la teor\'ia de la decisi\'on}

  Revisaremos algunos resultados de la teor\'ia de la decisi\'on que son \'utiles para hacer inferencias pero no daremos aqu\'i ni la construcci\'on axiom\'atica ni la mayor\'ia de las demostraciones de los resultados que al respecto se utilizar\'an. Para detalles sobre esto se recomienda ampliamemte el libro de Bernardo y Smith (1994).\par
  
  \smallskip
  
  Uno de los principales objetivos de la teor\'ia de la decisi\'on es el desarrollo de procesos l\'ogicos para la toma de decisiones bajo condiciones de incertidumbre. La idea es plantear los problemas de inferencia estad\'istica como problemas de decisi\'on, y aprovechar por tanto los resultados que ya se tienen respecto a esto \'ultimo.\par

 \bigskip
 
 \section{Representaci\'on formal}
  
  \begin{enun}
    \begin{defn}\label{def problema de decision}
     Un \textit{problema de decisi\'on} est\'a definido conjuntamente por los elementos $(\mathcal{E},\mathcal{C},\mathcal{A},\preceq)$ en donde:
      \begin{enumerate}
        \item $\mathcal{E}$ es un \'algebra de eventos relevantes que denotaremos mediante $E_j$ ,
        \item $\mathcal{A}$ es un conjunto de opciones o acciones potenciales, cuyos elementos denotaremos mediante $a_i$ ,
        \item $\mathcal{C}$ es el conjunto de consecuencias posibles y mediante $c_{ij}$ denotaremos la consecuencia de haber elegido la acci\'on $a_i\in\mathcal{A}$ bajo la ocurrencia de el evento $E_j\in\mathcal{E}$ ,
        \item $\preceq$ es una relaci\'on (binaria) de \textit{preferencia} para algunos de los elementos de $\mathcal{A}$.
      \end{enumerate}
    \end{defn}
  \end{enun}
  
 \medskip
  
 \begin{ejem}\label{ejem paraguas}
   Supongamos que nos enfrentamos al trivial problema de decidir si salimos a la calle con o sin paraguas. Como conjunto de acciones posibles tenemos entonces que $\mathcal{A}:=\{a_1,a_2\}$ en donde $a_1$ puede representar la acci\'on de llevar paraguas y $a_2$ la acci\'on de no llevarlo. Son muchos los eventos que al respecto podr\'iamos considerar, pero por el momento aceptemos la idea intuitiva de que esencialmente tenemos dos eventos relevantes (para lo que se pretende decidir) : llueve $(E_1)$ o no llueve $(E_2)$. Con lo anterior podemos entonces determinar el conjunto de consecuencias posibles:
   $$\mathcal{C}=\{c_{ij}=(a,E)\,:\,a\in\mathcal{A}\,,\,E\in\{E_1,E_2\}\}$$
As\'i por ejemplo $c_{22}$ es la consecuencia de haber decidido no llevar paraguas y que efectivamente no haya llovido; $c_{21}$ es la consecuencia de haber decido no llevar paraguas y que haya llovido. Intuitivamente podr\'iamos decir que nos gusta m\'as la consecuencia $c_{22}$ que la $c_{21}$ (esto en la mayor\'ia de los casos quiz\'as, porque hay quienes disfutan mojarse).$\quad\diamond$ 
 \end{ejem}
 
 \medskip
  
 \textit{Resolver un problema de decisi\'on} significa determinar $\preceq$ , esto es, definir un criterio para decidir qu\'e acciones son preferibles a otras. Hay que notar que $\mathcal{A}$ representa la parte del problema de decisi\'on que controlamos, que est\'a en nuestras manos en un momento dado. $\mathcal{E}$ representa la parte que no controlamos pues se refiere a eventos cuya ocurrencia no depende de nosotros.\par
 
 \smallskip
 
 En la Definici\'on \ref{def problema de decision} se defini\'o a $\mathcal{E}$ como un \textit{\'algebra}. En un problema de decisi\'on tenemos involucrado un fen\'omeno o experimento aleatorio para la parte que no controlamos. En probabilidad, se denota por $\Omega$ el \textit{espacio muestral}, esto es, el conjunto de resultados posibles del fen\'omeno o experimento aleatorio. Los distintos eventos relacionados a este experimento se pueden identificar como subconjuntos de $\Omega$. Se define un \textit{espacio de eventos} como un conjunto de eventos asociados a un experimento o fen\'omeno aleatorio, esto es, un espacio de eventos es, en principio, un conjunto de subconjuntos de $\Omega$. Sea $\mathcal{E}$ justamente ese espacio de eventos. Es cuesti\'on de analizar algunos ejemplos sencillos de probabilidad para motivar algunas propiedades que debe tener dicho espacio de eventos:
 
 \begin{enun}
   \begin{defn}\label{def algebra}
     Sea $\Omega$ un conjunto arbitrario y sea $\mathcal{E}$ un conjunto de subconjuntos de $\Omega$. Se dice que $\mathcal{E}$ es un \textit{\'algebra} si:
       \begin{enumerate}
         \item $\Omega\in\mathcal{E}$\,,
         \item Si $E\in\mathcal{E}$ entonces $E^{c}\in\mathcal{E}$\,,
         \item Si $E_1,\ldots,E_n\,\in\mathcal{E}$ entonces $\displaystyle{\mathop{\cup}_{j=1}^n E_j\,\in\mathcal{E}}$\,.
       \end{enumerate}
   \end{defn}
 \end{enun}
 
 Es f\'acil verificar que consecuencia de lo anterior es que el conjunto vac\'io $\varnothing\in\mathcal{E}$ y que si $E_1,\ldots,E_n\,\in\mathcal{E}$ entonces $\displaystyle{\mathop{\cap}_{j=1}^n E_j\,\in\mathcal{E}}$. Pero en probabilidad esto no es suficiente, se pide adem\'as que la uni\'on (infinito) numerable de eventos tambi\'en est\'e en el espacio de eventos, en cuyo caso se le denomina $\sigma$-\textit{\'algebra}. Sin embargo, m\'as adelante mencionaremos el por qu\'e se asigna a $\mathcal{E}$ una estructura de \'algebra y no de $\sigma$-\'algebra.\par
 
 \smallskip
 
 En un problema de decisi\'on no trabajamos directamente con todo $\mathcal{E}$ sino con algunos de sus elementos que llamamos \textit{eventos relevantes}, relevantes respecto a lo que se quiere decidir. Por el momento estableceremos que dicho conjunto de eventos relevantes sea finito. Tambi\'en pediremos que sea una partici\'on de $\Omega$. Igualmente pediremos que $\mathcal{A}$ sea finito.\par
 
 \medskip
 
 \hyphenation{va-lo-res}
 
 Mencionamos ya que el conjunto de eventos relevantes es la parte que no controlamos del problema de decisi\'on, esto es, tenemos \textit{incertidumbre} \linebreak respecto a cu\'al de los eventos relevantes ocurrir\'a, pero esto no nos impide estudiar cient\'ificamente el fen\'omeno o experimento aleatorio asociado a \linebreak ellos e intentar reunir informaci\'on que nos de idea acerca de la posibilidad de ocurrencia de cada uno de los eventos. Al respecto Lindley (2000) menciona que ``un enfoque cient\'ifico [sobre este problema] implica la medici\'on de la \linebreak incertidumbre ya que, citando a Kelvin, es s\'olo asociando n\'umeros a cualquier concepto cient\'ifico como puede ser adecuadamente entendido. La raz\'on de medir no es s\'olo para ser m\'as precisos respecto a la noci\'on de que tenemos m\'as incertidumbre acerca de lo que suceder\'a ma\~nana en el mercado de valores en comparaci\'on con que salga el sol, sino tambi\'en para poder combinar incertidumbres''. Lindley (2000) argumenta el por qu\'e la incertidumbre debe ser medida con probabilidad y Bernardo y Smith (1994) hacen una fundamentaci\'on rigurosa de ello y de c\'omo tomar decisiones bajo condiciones de incertidumbre. Esto \'ultimo se traduce en poder determinar una relaci\'on de preferencia $\preceq$ sobre $\mathcal{A}$ y escoger la acci\'on \'optima.\par
 
 \medskip
 \hyphenation{su-po-nien-do}
 \hyphenation{me-dian-te}
 
 Al hablar de una relaci\'on (binaria) de preferencia  $\preceq$  no estamos suponiendo que cualquier par de acciones $(a_1,a_2)\in\mathcal{A}\times\mathcal{A}$ est\'a necesariamente relacionado mediante $\preceq$ . En caso de que dicha relaci\'on sea aplicable, mediante  $a_1\preceq a_2$ entenderemos que, bajo alg\'un criterio que se defina, $a_1$ no es m\'as preferible que $a_2$.\par 
 
 \begin{enun}
   \begin{defn}\label{def relaciones binarias inducidas}
     La relaci\'on de preferencia  $\,\preceq\,$  induce las siguientes relaciones binarias para elementos $a_1,a_2\in\mathcal{A}$ :
     \begin{enumerate}
       \item $a_1\sim a_2$ si y s\'olo si $a_1\preceq a_2$ y $a_2\preceq a_1$ (indiferencia),
       \item $a_1\prec a_2$ si y s\'olo si $a_1\preceq a_2$ pero no se cumple que $a_2\preceq a_1$ (preferencia estricta),
       \item $a_1\succeq a_2$ si y s\'olo si $a_2\preceq a_1$\,,
       \item $a_1\succ a_2$ si y s\'olo si $a_2\prec a_1$\,.
     \end{enumerate}
   \end{defn}
 \end{enun}
 
 \medskip
 
 Y as\'i como resulta necesario cuantificar la incertidumbre de alg\'un modo, que en nuestro caso ser\'a por medio de probabilidad, tambi\'en es necesario cuantificar las consecuencias. En el Ejemplo \ref{ejem paraguas} result\'o (quiz\'as) intuitivamente claro que la consecuencia $c_{22}$ es preferible a la consecuencia $c_{21}$ pero si nos preguntamos lo mismo respecto a $c_{12}$ y $c_{21}$ posiblemente no resulte tan contundente la respuesta, o al menos no con la intensidad del otro caso. N\'otese que de inicio evitamos escribir, por ejemplo, que $c_{22}\succ c_{21}$ porque hemos definido las relaciones de preferencia para elementos de $\mathcal{A}$ y no de $\mathcal{C}$. Claro que podr\'iamos definir relaciones de preferencia an\'alogas para $\mathcal{C}$ y tener cuidado en utilizar una simbolog\'ia diferente, lo cual no ser\'ia pr\'actico, as\'i que utilizaremos los mismos s\'imbolos pero conscientes de que las relaciones de preferencia de $\mathcal{A}$ son distintas a las de $\mathcal{C}$. \par
 
 \begin{enun}
   \begin{defn} \label{def espacio de estados}
     Entenderemos por \textit{espacio de estados}, y lo denotaremos $\Theta$, a una partici\'on de $\Omega$ en eventos relevantes.
   \end{defn}
 \end{enun}
 
 Como la incertidumbre sobre los eventos relevantes la mediremos con probabilidad, si se tiene una medida de probabilidad $\mathbf{P}:\mathcal{E}\rightarrow [\,0,1\,]$ entonces tenemos una funci\'on de probabilidad $\prob:\Theta\rightarrow [\,0,1\,]$. N\'otese adem\'as que $\mathcal{C}=\mathcal{A}\times\Theta$ .\par
 
 \begin{enun}
   \begin{defn}\label{def funcion de utilidad}
     Una \textit{funci\'on de utilidad} es una funci\'on $u:\mathcal{C}\rightarrow\mathbb{R}$ .
   \end{defn}
 \end{enun}
  
 \medskip
 
 El poder cuantificar de alg\'un modo las distintas consecuencias nos provee de un criterio inmediato para determinar las relaciones de preferencia en $\mathcal{C}$ :\par
 
 \begin{enun}
   \begin{defn} \label{def preferencia de consecuencias}
     $c_{ij}\preceq c_{kl}$ si y s\'olo si $u(c_{ij})\leq u(c_{kl})$ .
   \end{defn}
 \end{enun}
 
 La definici\'on anteior induce las siguientes relaciones binarias:
    $$c_{ij}\sim c_{kl}\,\Leftrightarrow\,c_{ij}\preceq c_{kl}\,\,\mbox{y}\,\, c_{kl}\preceq c_{ij}\,,$$
    $$c_{ij}\prec c_{kl}\,\Leftrightarrow\,c_{ij}\preceq c_{kl}\,\,\mbox{pero no se cumple que}\,\, c_{kl}\preceq c_{ij}\,.$$

 \section{Soluci\'on de un problema de decisi\'on}
  
 Dijimos ya que resolver un problema de decisi\'on $(\mathcal{E},\mathcal{C},\mathcal{A},\preceq)$ consiste en determinar $\,\preceq\,$, es decir, definir una relaci\'on de preferencia entre los elementos de $\mathcal{A}$ y escoger la acci\'on \'optima (la m\'as preferible). Existen diversas formas de hacerlo, pero aqu\'i trataremos exclusivamente la forma que nos interesa para hacer inferencias desde el enfoque bayesiano y para ello requerimos tener identificado lo siguiente:\par
 
   \begin{itemize}
     \item El espacio de estados $\Theta$,
     \item una funci\'on de probabilidad $\prob$ sobre los elementos de $\Theta$,
     \item una funci\'on de utilidad $u$ sobre $\mathcal{C}$.
   \end{itemize}
     
 La funci\'on de probabilidad $\prob:\Theta\rightarrow [\,0,1\,]$ puede ser \textit{a priori} o \textit{a posteriori}, en el sentido en que se trat\'o en el Cap\'itulo 2. El c\'omo establecer o construir una funci\'on de utilidad depender\'a de cada problema particular. Para mayor detalle acerca de algunas formas generales de funciones de utilidad nuevamente insistimos en consultar el libro de Bernardo y Smith (1994). Aqu\'i nos limitaremos a ilustrar lo anterior mediante el siguiente:\par
 
 \begin{ejem} \label{ejemplo 1 revisited}
   Retomando el Ejemplo \ref{e1} en la parte referente a la apuesta, podemos plantearlo como un problema de decisi\'on. El fen\'omeno aleatorio involucrado es el lanzamiento de una moneda por lo que su espacio muestral es $\Omega=\{\mbox{\'aguila},\mbox{sol}\}$ y su \'algebra de eventos es $\mathcal{E}=\{\Omega,\varnothing,\{\mbox{\'aguila}\},\{\mbox{sol}\}\}$. El espacio de estados $\Theta=\{E_1,E_2\}$ donde $E_1:=\{\mbox{\'aguila}\}$ y $E_2:=\{\mbox{sol}\}$. N\'otese que $\Theta$ es partici\'on de $\Omega$. El conjunto de acciones es $\mathcal{A}=\{a_1,a_2\}$ donde $a_1$ representa la acci\'on de apostar en favor de que salga \'aguila y $a_2$ en favor de sol. Si el esquema de apuesta consiste en que quien apueste a favor de \'aguila arriesgue $b$ pesos y quien lo haga en favor de sol arriesgue $a$ pesos entonces la funci\'on de utilidad queda como sigue:
   $$u(c_{11})=a\qquad u(c_{12})=-b\qquad u(c_{21})=-a\qquad u(c_{22})=b$$
De acuerdo al Ejemplo \ref{e1} n\'otese que $E_1\equiv\{X=1\}$ y $E_2\equiv\{X=0\}$ por lo que:
   $$\prob(E_1)=\prob(X=1)\quad\mbox{y}\quad\prob(E_2)=\prob(X=0)$$
para lo cual podemos utilizar la distribuci\'on predictiva a priori o a posteriori, seg\'un sea el caso, es decir, $\prob(X=x)=p(x)$ o bien $\prob(X=x)=p(x\,|\,\mathbf{x})$, y con esto queda definida una funci\'on de probabilidad $\prob$ sobre el espacio de estados (relevantes) $\Theta$. De manera tabular podemos resumir lo anterior como sigue:
   $$\begin{array}{c|c|c}
        \prob(E_j)   & \prob(E_1)   &   \prob(E_2) \\ \hline
        u(a_i,E_j)   &     E_1      &     E_2      \\ \hline
             a_1     &      a       &      -b      \\ \hline
             a_2     &     -a       &       b      \\ \hline
     \end{array}$$
 \end{ejem}
En el Ejemplo \ref{e1} se obtuvo la relaci\'on que debe existir entre los montos de apuesta $a$ y $b$ pesos para tener una \textit{apuesta justa} y dicha relaci\'on se obtuvo a partir de la ecuaci\'on:
   $$a\prob(E_1)-b\prob(E_2)=0$$
El tener una apuesta o juego justo implica que seamos indiferentes respecto a apostar en favor de cualquiera de las opciones disponibles, que en t\'erminos de este problema de decisi\'on lo escribimos como $a_1\sim a_2$. Pero eso es tan s\'olo un caso particular. De forma m\'as general podemos definir las variables aleatorias:
   $$U_1:=\;a\indic_{E_1}-b\indic_{E_2}$$
   $$U_2:=-a\indic_{E_1}+b\indic_{E_2}$$
esto es, $U_1$ representa la ganancia/p\'erdida que obtendr\'a quien decida tomar la acci\'on $a_1$ y $U_2$ lo an\'alogo para la acci\'on $a_2$. Calculando sus esperanzas:
   $$\esper(U_1):=\;a\prob(E_1)-b\prob(E_2)$$
   $$\esper(U_2):=-a\prob(E_1)+b\prob(E_2)$$
Entonces para tener un juego justo se requiere que $\esper(U_1)=0$ y que $\esper(U_2)=0$, que de hecho tienen la misma soluci\'on, por lo que tendremos que $a_1\sim a_2$ si $\esper(U_1)=\esper(U_2)$. Si por el contrario ocurriera que $\esper(U_1)>\esper(U_2)$ entonces si nos dan a escoger preferimos la acci\'on $a_1$, esto es, $a_1\succ a_2$. $\esper(U_i)$ es lo que se conoce como la \textit{utilidad esperada de la acci\'on} $a_i$ y es justamente lo que nos servir\'a como criterio para definir una relaci\'on de preferencia $\,\preceq\,$ sobre $\mathcal{A}$ y poder as\'i elegir la acci\'on \'optima. $\quad\diamond$\par

 \medskip
 
 \begin{enun}
   \begin{defn} \label{def utilidad esperada de una accion}
     En un problema de decisi\'on $(\mathcal{E},\mathcal{C},\mathcal{A},\preceq)$ con espacio de estados (relevantes) $\Theta=\{E_1,\ldots,E_m\}\subset\mathcal{E}$, funci\'on de probabilidad $\prob$ sobre $\Theta$ y funci\'on de utilidad $u$ sobre $\mathcal{C}$, la \textit{utilidad esperada de la acci\'on} $a_i\in\mathcal{A}=\{a_1,\ldots,a_k\}$ se denota $\overline{u}(a_i)$ y se define como:
     $$\overline{u}(a_i):=\sum_{j=1}^m u(a_i,E_j)\,\prob(E_j)\qquad i=1,\ldots,k$$
   \end{defn}
 \end{enun}
 
 \begin{enun}
   \begin{defn} \label{def criterio general de decision}
     (Criterio general de decisi\'on). En un problema de decisi\'on como el de la Definici\'on \ref{def utilidad esperada de una accion}, la relaci\'on de preferencia $\,\preceq\,$ sobre $\mathcal{A}$ queda definida por:
     $$a_1\preceq a_2\Leftrightarrow\overline{u}(a_1)\leq \overline{u}(a_2)$$
   \end{defn}
 \end{enun}

 Estrictamente hablando, lo anterior no es una definici\'on sino una proposici\'on que se demuestra despu\'es de una rigurosa axiomatizaci\'on de lo que hemos visto hasta el momento como lo desarrollan Bernardo y Smith (1994), pero como aqu\'i nos limitamos a dar la motivaci\'on intuitiva para establecer dicho criterio, no qued\'o m\'as remedio que definirlo as\'i.\par
 
 \smallskip
 
 A partir de la Definici\'on \ref{def criterio general de decision} es inmediato que:
   $$a_1\sim a_2 \Leftrightarrow \overline{u}(a_1)=\overline{u}(a_2)$$
   $$a_1\prec a_2 \Leftrightarrow \overline{u}(a_1)<\overline{u}(a_2)$$
   
 \smallskip
 
 Finalmente, el criterio que utilizaremos para elegir la acci\'on \'optima de $\mathcal{A}$, misma que denotaremos $a_{*}$, ser\'a aqu\'ella que satisfaga:
   $$\overline{u}(a_{*})=\max_{i}\,\overline{u}(a_i)$$
Puede ocurrir que $a_{*}$ no sea \'unica. En tal caso hablar\'iamos entonces de el \textit{conjunto de acciones \'optimas} $\mathcal{A}^{*}\subset\mathcal{A}$ y en tal caso diremos que somos indiferentes ante llevar acabo cualquiera de las acciones de $\mathcal{A}^{*}$. Es en este punto donde podemos retomar el por qu\'e pedimos que tanto $\mathcal{A}$ como $\mathcal{E}$ sean finitos, pues de ser as\'i, los resultados que desarrollan Bernardo y Smith (1994) garantizan que $a_{*}$ existe, de otro modo puede o no ocurrir as\'i. Necesitamos un par de definiciones m\'as para ilustrarlo.\par

 \medskip

 \begin{enun}
   \begin{defn} \label{def accion dominada}
     Una acci\'on $a_{i_1}$ est\'a \textit{dominada} por otra acci\'on $a_{i_2}$ si para todo $j$ tenemos que $u(a_{i_1},E_j)\leq u(a_{i_2},E_j)$ y adem\'as existe un $j_0$ tal que $u(a_{i_1},E_{j_0})< u(a_{i_2},E_{j_0})$.
   \end{defn}
 \end{enun}

 \begin{enun}
   \begin{defn} \label{def accion admisible}
     Una acci\'on es \textit{admisible} si no existe otra acci\'on que la domine. Una acci\'on es \textit{inadmisible} si existe al menos otra que la domine.
   \end{defn}
 \end{enun}
 
 Lo anterior nos dice en pocas palabras que es (quiz\'as) posible depurar el espacio de acciones, esto es, habr\'ia que eliminar de $\mathcal{A}$ las acciones inadmisibles, llamadas as\'i porque, independientemente del evento que ocurra, siempre existe una mejor opci\'on.
 
 \medskip
 
 \hyphenation{di-fe-ren-tes}
 
 \begin{ejem} \label{ejem portafolios de inversion}
   Supongamos que una operadora de fondos de inversi\'on nos ofrece cuatro tipos diferentes de sociedades de inversi\'on, esto es, cuatro diferentes estrategias para invertir nuestro dinero. Por simplicidad supongamos que los cuatro portafolios de inversi\'on de dichas sociedades invierten en dos opciones: acciones de la empresa ABC que cotiza en bolsa y en t\'itulos que pagan un rendimiento fijo de 6\%. Lo que distingue a cada portafolios es el porcentaje destinado a una y otra opci\'on de inversi\'on:
   $$\begin{array}{l|c|c}
       \mbox{portafolios} & \mbox{\% en ABC} & \mbox{\% a tasa fija}\\ \hline
       \mbox{agresivo}    & 80 & 20 \\
       \mbox{moderado}    & 50 & 50 \\
       \mbox{conservador} & 20 & 80 \\
       \mbox{sin riesgo}  &  0 & 100 
     \end{array}$$
  De acuerdo a lo anterior, el rendimiento de cada portafolios se desconoce (a excepci\'on del \'ultimo) ya que depende del rendimiento que tenga la empresa ABC y \'este resulta incierto; sin embargo, podemos modelar nuestra incertidumbre respecto al rendimiento de ABC consultando a un analista financiero y pidi\'endole (por ejemplo) nos de los escenarios posibles acerca del rendimiento que ABC tendr\'a de acuerdo a la informaci\'on que \'el maneja. Por simplicidad supongamos que nos plantea los siguientes escenarios con sus respectivas probabilidades:
  $$\begin{array}{l|c|c}
      \mbox{Escenario} & \mbox{Rendimiento} & \mbox{Probabilidad} \\ \hline
      \mbox{pesimista} & -5\% & 0{.}20 \\
      \mbox{realista}  & +15\%& 0{.}60 \\
      \mbox{optimista} & +25\%& 0{.}20
    \end{array}$$
Las probabilidades asignadas constituyen lo que ya hemos definido como \textit{probabilidad a priori}. De acuerdo a lo anterior podemos pensar en una variable aleatoria $X$ que represente el rendimiento de ABC y por tanto $\mbox{Ran}\,X=\{-5\%,+15\%,+25\%\}$ con las probabilidades arriba se\~naladas.\par
  El problema de decisi\'on consiste justamente en elegir de entre los cuatro portafolios el \'optimo de acuerdo a la informaci\'on con que se cuenta. Sea el conjunto de acciones $\mathcal{A}:=\{a_1,a_2,a_3,a_4\}$ y el espacio de estados (relevantes) $\Theta:=\{E_1,E_2,E_3\}$ de modo que:
  \begin{equation*}
    \begin{split}
      a_1 & \equiv\,\,\mbox{invertir en el portafolios agresivo} \\
      a_2 & \equiv\,\,\mbox{invertir en el portafolios moderado} \\
      a_3 & \equiv\,\,\mbox{invertir en el portafolios conservador} \\
      a_4 & \equiv\,\,\mbox{invertir en el portafolios sin riesgo} \\
      { } & { } \\
      E_1 & \equiv\,\{X=-5\%\} \\
      E_2 & \equiv\,\{X=+15\%\} \\
      E_3 & \equiv\,\{X=+25\%\} \\
    \end{split}
  \end{equation*}

 Para poder resolver este problema de decisi\'on s\'olo nos falta especificar una funci\'on de utilidad para las distintas consecuencias. Por el momento consideremos una funci\'on de utilidad igual al rendimiento que puede obtener cada portafolios bajo cada escenario:
   $$u(a_i,E_j):=\alpha_i x_j+(1-\alpha_i)r$$
en donde $\alpha_i$ representa el porcentaje de inversi\'on bajo la acci\'on $a_i$ y $x_j$ representa el rendimiento de ABC bajo el escenario $E_j$ y $r$ la tasa fija del 6\% en este caso. Utilizando el criterio de la utilidad esperada m\'axima:
   $$\begin{array}{c|c|c|c|c}
      \prob(E)& 0{.}20 & 0{.}60 & 0{.}20 & { } \\ \hline
      u(a,E)  &   E_1  &   E_2  &   E_3  & \overline{u}(a_i) \\ \hline
        a_1   & -2{.}8\% & 13{.}2\% & 21{.}2\% & 11{.}6\% \\
        a_2   & 0{.}5\% & 10{.}5\% & 15{.}5\% & 9{.}5\% \\
        a_3   & 3{.}8\% & 7{.}8\% & 9{.}8\% & 7{.}4\% \\
        a_4   & 6{.}0\% & 6{.}0\% & 6{.}0\% & 6{.}0\% 
      \end{array}$$
Claramente la acci\'on \'optima es $a_*=a_1$. Cabe aclarar que $a_1$ es la acci\'on \'optima si realmente la funci\'on de utilidad propuesta corresponde a nuestras preocupaciones como inversionistas. De hecho, veremos r\'apidamente que, salvo casos extremos, la funci\'on de utilidad propuesta no es una buena elecci\'on para este problema espec\'ifico. Suponiendo que tenemos la libertad de elegir libremente los porcentajes de inversi\'on en las opciones ya mencionadas tendremos entonces que el conjunto de acciones es $\mathcal{A}=[\,0,100\%\,]$, esto es, existe una infinidad no numerable de porcentajes distintos que podemos asignar a cada opci\'on inversi\'on (con la condici\'on de que sumen 100\%) y en tal caso $a\in\mathcal{A}$ representa la acci\'on de invertir $a$\% en ABC y el resto a tasa fija por lo que:
   $$\overline{u}(a)=a\esper(X)+(1-a)r\quad,\,a\in\mathcal{A}$$
y como $\esper(X)=13\%$ entonces reescribimos:
   $$\overline{u}(a)=(7\%)a+6\%$$
es decir, $\overline{u}(a)$ es la ecuaci\'on de una recta con pendiente positiva y alcanza su m\'aximo en $a=100\%$ por lo que la acci\'on \'optima ser\'ia en este caso $a_*=100\%$ con una utilidad (rendimiento esperado en este caso) $\overline{u}(a_*)=13\%$. Entonces, con la funci\'on de utilidad propuesta, la acci\'on \'optima es tomar el mayor riesgo posible: invertir el 100\% en ABC. Aqu\'i es donde un inversionista m\'inimamente informado protestar\'ia con semejante decisi\'on. ?`Qu\'e sucede? Pues que normalmente un inversionista considera, al menos, dos aspectos: rendimiento y \textit{riesgo} de la inversi\'on. Normalmente un inversionista busca altos rendimientos pero con el menor riesgo posible y de hecho la decisi\'on de inversi\'on bajo esta doble consideraci\'on implica balancear entre el rendimiento que quiere el inversionista y el riesgo que est\'a dispuesto a tomar porque las inversiones de poco riesgo van acompa\~nadas de rendimientos moderados y las inversiones que tienen la posibilidad de otorgar altos rendimientos van acompa\~nadas de mayor riesgo. As\'i que no quiere decir esto que est\'e mal la teor\'ia, simplemente que hay que tener cuidado con la elecci\'on de una funci\'on de utilidad que refleje todo aquello que nos preocupe o interese sea tomado en cuenta. La funci\'on de utilidad que se propuso \'unicamente toma en cuenta rendimientos mas no riesgo.\par
  Construiremos pues una funci\'on de utilidad (entre muchas que se podr\'ian definir) que de alg\'un modo refleje preocupaci\'on tanto en rendimientos altos como en controlar la cantidad de riesgo que se toma. Sea $u_{ij}:=u(a_i,E_j)$ como se defini\'o anteriormente y sea:
    $$u_{i\bullet}:=\frac{1}{m}\sum_{j=1}^m u_{ij}$$
Definimos la siguiente funci\'on de utilidad $w$ :
   \begin{eqnarray*}
     w(a_i,E_j) &:=& u_{ij}-A(u_{i\bullet}-u_{ij})^2 \\
                &=&  u_{ij}-\alpha_i^2 A(x_{\bullet}-x_j)^2
   \end{eqnarray*}
en donde $x_{\bullet}:=\frac{1}{m}\sum_{j=1}^m x_j$ y en donde $A\geq 0$ es lo que en ocasiones se denomina un \textit{coeficiente de aversi\'on al riesgo}. N\'otese que si $A=0$ entonces $w(a_i,E_j)=u(a_i,E_j)$ por lo que nos ocuparemos s\'olo del caso en que $A>0$. La utilidad esperada para cada acci\'on $a_i\in\mathcal{A}$ es:
   \begin{eqnarray*}
     \overline{w}(a_i) &=& \overline{u}(a_i)- \alpha_i^2 A \sum_{j=1}^m (x_{\bullet}-x_j)^2\,\prob(E_j) \\
       &=& \alpha_i\big[\esper(X)-r\big]+r-\alpha_i^2 A\big[\vari(X) + (x_{\bullet}-\esper(X))^2\big]
   \end{eqnarray*}

N\'otese c\'omo ahora la f\'ormula general para la utilidad esperada de una acci\'on est\'a tanto en t\'erminos de el valor esperado del rendimiento de ABC, esto es $\esper(X)$, como del riesgo o dispersi\'on de dicho rendimiento, es decir, $\vari(X)$ en este caso. Y de hecho, es inmediato a partir de lo anterior obtener la f\'ormula general de la utilidad esperada por acci\'on para el caso en que $a\in\mathcal{A}=[\,0,100\%\,]$ :
    $$\overline{w}(a)=a\big[\esper(X)-r\big]+r-a^2 A\big[\vari(X) + (x_{\bullet}-\esper(X))^2\big]$$
y para encontrar el valor de $a$ que maximiza $\overline{w}(a)$ resolvemos:
    $$\overline{w}\,'(a)=\esper(X)-r-2aA\big[\vari(X) + (x_{\bullet}-\esper(X))^2\big]=0$$
y como $\overline{w}\,''(a)<0$ entonces el valor de $a$ que maximiza $\overline{w}(a)$, esto es el porcentaje \'optimo de inversi\'on en ABC (invirtiendo el resto a tasa fija):
  $$a_*=\frac{\esper(X)-r}{2A\big[\vari(X) + (x_{\bullet}-\esper(X))^2\big]}$$
 Para analizar el resultado anterior definamos $\rho_X:=\esper(X)-r$ y $\delta_X:=\vari(X)+(x_{\bullet}-\esper(X))^2$, y entonces:
   $$a_*=\frac{\rho_X}{2A\delta_X}$$

 En la expresi\'on anterior $\rho_X$ representa el rendimiento esperado de ABC por encima de lo que se obtiene a tasa fija $r$. Normalmente tendremos que $\rho_X>0$, esto quiere decir que por lo general si vamos a considerar invertir en una opci\'on con riesgo pedimos que al menos su rendimiento esperado sea mayor al de una opci\'on sin riesgo. Por otro lado, $\delta_X$ representa una medida de riesgo y tiene dos componentes: la varianza del rendimiento de ABC as\'i como una consideraci\'on respecto a asimetr\'ia en las probabilidades asignadas a los escenarios ya que $x_{\bullet}=\esper(X)$ si, por ejemplo, $X$ tiene distribuci\'on uniforme discreta. Y es aqu\'i donde se ve que la elecci\'on \'optima est\'a considerando tanto rendimiento como riesgo ya que a mayor rendimiento de ABC se tendr\'a un mayor valor de $a_*$ y a mayor riesgo (varianza y asimetr\'ia) de ABC se tendr\'a un menor valor de $a_*$. Con la soluci\'on \'optima anterior se obtiene la siguiente utilidad \'optima:
    $$\overline{w}(a_*)=\frac{\rho_X^2}{4A\delta_X} + r$$
    
  Aqu\'i es interesante notar que a\'un cuando $\rho_X<0$ se tendr\'a una utilidad (rendimiento esperado) por encima de la tasa fija $r$ pero el que $\rho_X<0$ implica que $a_*<0$, y a primera vista parece un sin sentido un porcentaje negativo de inversi\'on, pero quienes tengan un poco de conocimientos burs\'atiles saben que esto corresponde a lo que se conoce como \textit{ventas en corto}, concepto que no discutiremos aqu\'i pero baste mencionarlo para que nos quedemos tranquilos de que estamos obteniendo resultados coherentes. Tambi\'en puede ocurrir que $a_*>100\%$ en cuyo caso tendremos un porcentaje negativo de inversi\'on a tasa fija, lo cual tambi\'en es posible pues esto quiere decir que, adem\'as del dinero propio, habr\'ia que pedir prestado m\'as dinero (a dicha tasa fija del 6\%) para invertirlo tambi\'en en ABC. Con los datos espec\'ificos de este ejemplo obtenemos:
  $$a_*=\frac{357{.}95\%}{A}\qquad\qquad\overline{w}(a_*)=\frac{12{.}53\%}{A}+6\%$$
  
 Para distintos valores de $A$ tenemos las siguientes acciones \'optimas:
    $$\begin{array}{c|l}
       A & a_* \\ \hline
       [\,0,3{.}57\,] & a_1\quad\mbox{(pidiendo prestado)} \\
       3{.}58 & a_1\quad\mbox{(con inversi\'on al 100\% en ABC)} \\
       \mbox{$[\,3{.}58,5{.}48\,]$} & a_1 \\
       \mbox{$[\,5{.}49,5{.}52\,]$} & a_1\sim a_2 \\
       \mbox{$[\,5{.}53,10{.}20\,]$} & a_2 \\
       \mbox{$[\,10{.}21,10{.}24\,]$} & a_2\sim a_3 \\
       \mbox{$[\,10{.}25,\,35{.}6\,]$} & a_3 \\
       \mbox{$[\,35{.}7,35{.}9\,]$} & a_3\sim a_4 \\
       \mbox{$[\,36,62{.}2\,]$} & a_4 \\
       \mbox{$[\,62{.}3,115{.}3\,]$} & a_4\quad\mbox{(con $a_1$ inadmisible)} \\
      \mbox{$[\,115{.}4,403{.}8\,]$} & a_4\quad\mbox{(con $a_1,a_2$ inadmisibles)} \\
      \mbox{$[\,403{.}9,\infty\,[$} &a_4 \quad\mbox{(con $a_1,a_2,a_3$ inadmisibles)}\\
     \end{array}$$
?`Qu\'e valor de $A$ se debe usar? Esto depender\'a de qu\'e tanto riesgo se est\'e dispuesto a tomar. N\'otese que si no existe preocupaci\'on alguna por el riesgo ($A=0$) entonces $w(a_i,E_j)=u(a_i,E_j)$. En cambio una gran preocupaci\'on por el riesgo se refleja en valores grandes para $A$. El c\'omo traducir el nivel de aversi\'on al riesgo de un determinado inversionista en un valor para $A$ es materia ya de un estudio m\'as profundo que, como elegantemente dicen muchos libros, \textit{is beyond the scope of this book}. Pero para no dejarlo as\'i, una manera simplista de averiguar el coeficiente de aversi\'on al riesgo $A$ de un determinado inversionista ser\'ia, por ejemplo, pregunt\'andole como qu\'e rendimiento anda buscando. Supongamos que busca 1.5\% por arriba de la tasa fija del 6\%, entonces:
  $$\overline{u}(a_*)=a_*[\esper(X)-r]+r=r+1{.}5\%$$
despejando $a_*$ e igual\'andola con la f\'ormula de la acci\'on \'optima obtenemos $A=16{.}7$ de donde obtenemos $a_*=a_3$. M\'as a\'un, si en lugar de los cuatro portafolios que se mencionaron existe libertad para decidir los porcentajes como se quiera, la inversi\'on \'optima ser\'ia en este caso invertir 21.43\% en ABC y el resto a tasa fija. $\quad\diamond$ 
 \end{ejem}
 
 \medskip
 
 Volviendo a la restricci\'on de que $\mathcal{A}$ y $\Theta$ sean finitos, si alguno de ellos no lo fuera, el problema de encontrar la acci\'on \'optima puede a\'un as\'i tener soluci\'on (como en el ejemplo anterior con $\mathcal{A}:=[\,0,100\%\,]$) o bien no tenerla, como se ilustra en el Ejercicio 3, al final de este cap\'itulo.

 \section{Problemas de decisi\'on secuencial}
 
 Hasta el momento hemos hablado de problemas de decisi\'on en donde el espacio de estados o eventos relevantes $\Theta$ es el mismo bajo cualquier acci\'on $a_i\in\mathcal{A}$, pero esto no tiene que ser necesariamente as\'i, bien puede ocurrir que dependiendo de la acci\'on que se tome se tenga un conjunto de eventos relevantes diferente, es decir, bajo una acci\'on $a_i\in\mathcal{A}$ se tiene un conjunto particular de $m_i$ eventos o estados relevantes $\Theta_i:=\{E_{i1},E_{i2},\ldots,E_{im_i}\}$ dando lugar a los conjuntos de consecuencias $\mathcal{C}_i:=\{c_{i1},c_{i2},\ldots,c_{im_i}\}$. Y al igual que en la secci\'on anterior, si se tiene una funci\'on de utilidad definida para el conjunto de consecuencias $\bigcup\mathcal{C}_i$ y funciones de probabilidad $\prob_i$ para los espacios de estados $\Theta_i$ entonces nuevamente la acci\'on \'optima ser\'a aqu\'ella $a_{*}\in\mathcal{A}$ que satisfaga:
   $$\overline{u}(a_{*})=\max_i \overline{u}(a_i)$$
en donde:
   $$\overline{u}(a_i):=\sum_{j=1}^{m_i} u(a_i,E_{ij})\,\prob_i(E_{ij})$$
   
   De manera esquem\'atica:

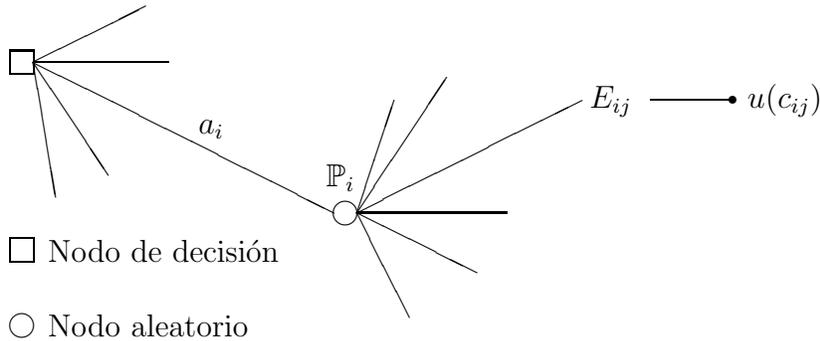
 \setlength{\unitlength}{1cm}   
 \begin{picture}(14,6)(0,4)
   \put(0,8){\framebox(0.3,0.3){}}
   \put(0.3,8.15){\line(2,1){1.5}}
   \put(0.3,8.15){\line(1,0){1.8}}
   \put(0.3,8.15){\line(2,-1){4}}
   \put(0.3,8.15){\line(2,-3){1}}
   \put(0.3,8.15){\line(1,-6){0.3}}
   \put(2.5,7.2){$a_i$}
   \put(4.45,6.15){\circle{0.3}}
   \put(4.2,6.5){$\prob_i$}
   \put(4.6,6.15){\line(1,3){0.5}}
   \put(4.6,6.15){\line(2,3){1.2}}
   \put(4.6,6.15){\line(2,1){3}}
   \put(4.6,6.15){\line(1,0){2}}
   \put(4.6,6.15){\line(2,-1){1.6}}
   \put(4.6,6.15){\line(1,-2){0.7}}
   \put(7.7,7.55){$E_{ij}$}
   \put(8.5,7.65){\line(1,0){1.1}}
   \put(9.6,7.65){\circle*{0.1}}
   \put(9.8,7.55){$u(c_{ij})$}
   \put(0,5.5){\framebox(0.3,0.3){}}
   \put(0.5,5.5){Nodo de decisi\'on}
   \put(0.15,4.65){\circle{0.3}}
   \put(0.5,4.5){Nodo aleatorio}
 \end{picture}
 
 Aunque \'esta es una forma m\'as general de un problema de decisi\'on que el inicialmente presentado sigue siendo un \textit{problema de decisi\'on simple o de una sola etapa} en el sentido de que de que se toma una sola decisi\'on, pero es posible tener un \textit{problema de decisi\'on secuencial} que es una concatenaci\'on de problemas de decisi\'on simples, en donde algunas o todas las consecuencias consisten en tener que resolver nuevos problemas de decisi\'on.\par
 \smallskip
 En un problema de decisi\'on secuencial la decisi\'on \'optima en la primera etapa depende de las elecciones \'optimas en las etapas subsecuentes. En el caso general de un problema de decisi\'on con $n$ etapas, la soluci\'on puede obtenerse de la siguiente manera:\par
 \hyphenation{uti-li-da-des}
 \begin{enumerate}
   \item Se resuelve primero la $n$-\'esima etapa (la \'ultima) maximizando las utilidades esperadas apropiadas,
   \item se resuelve la $(n-1)$-\'esima etapa maximizando las correspondientes utilidades esperadas condicionalmente es las elecciones \'optimas de la $n$-\'esima etapa,
   \item se contin\'ua de esta manera siempre trabajando hacia atr\'as hasta que se obtenga la elecci\'on \'optima en la primera etapa.
 \end{enumerate}
  
 En el caso de un problema de decisi\'on secuencial de $n=2$ etapas:
 
 \setlength{\unitlength}{1cm}   
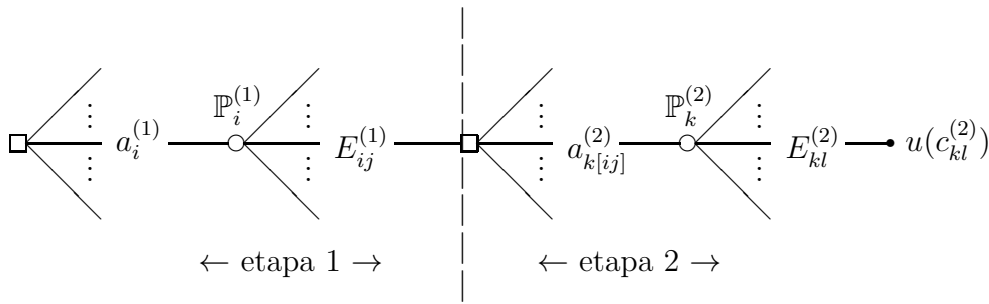
 \begin{picture}(15,4)(0,-2)
   \put(0,-0.1){\framebox(0.2,0.2){}}
   \put(0.2,0){\line(1,1){1}}
   \put(1,0.2){$\vdots$}
   \put(0.2,0){\line(1,0){1}}
   \put(1,-0.5){$\vdots$}
   \put(0.2,0){\line(1,-1){1}}
   \put(1.4,-0.1){$a_i^{(1)}$}
   \put(2.1,0){\line(1,0){0.8}}
   \put(3,0){\circle{0.2}}
   \put(2.7,0.35){$\prob_i^{(1)}$}
   \put(3.1,0){\line(1,1){1}}
   \put(3.9,0.2){$\vdots$}
   \put(3.1,0){\line(1,0){1}}
   \put(3.9,-0.5){$\vdots$}
   \put(3.1,0){\line(1,-1){1}}
   \put(4.3,-0.2){$E_{ij}^{(1)}$}
   \put(5.1,0){\line(1,0){0.9}}
   %segundo nivel
   \put(6,-0.1){\framebox(0.2,0.2){}}
   \put(6.2,0){\line(1,1){1}}
   \put(7,0.2){$\vdots$}
   \put(6.2,0){\line(1,0){1}}
   \put(7,-0.5){$\vdots$}
   \put(6.2,0){\line(1,-1){1}}
   \put(7.4,-0.2){$a_{k[ij]}^{(2)}$}
   \put(8.1,0){\line(1,0){0.8}}
   \put(9,0){\circle{0.2}}
   \put(8.7,0.35){$\prob_k^{(2)}$}
   \put(9.1,0){\line(1,1){1}}
   \put(9.9,0.2){$\vdots$}
   \put(9.1,0){\line(1,0){1}}
   \put(9.9,-0.5){$\vdots$}
   \put(9.1,0){\line(1,-1){1}}
   \put(10.3,-0.2){$E_{kl}^{(2)}$}
   \put(11.1,0){\line(1,0){0.6}}
   \put(11.7,0){\circle*{0.1}}
   \put(11.9,-0.1){$u(c_{kl}^{(2)})$}
   % division
   \multiput(5.95,-2)(0,0.5){8}{$|$}
   \put(2.5,-1.7){$\leftarrow$ etapa 1 $\rightarrow$}
   \put(7,-1.7){$\leftarrow$ etapa 2 $\rightarrow$}
  \end{picture}
  
  \medskip
  
  Resolviendo la \'ultima etapa:
  
 \setlength{\unitlength}{1cm}   
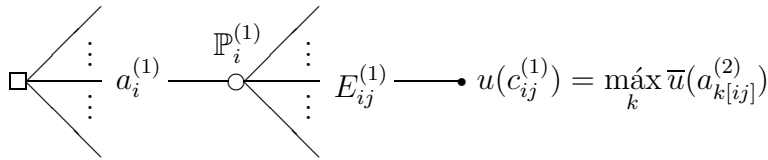
 \begin{picture}(15,4)(0,-2)
   \put(0,-0.1){\framebox(0.2,0.2){}}
   \put(0.2,0){\line(1,1){1}}
   \put(1,0.2){$\vdots$}
   \put(0.2,0){\line(1,0){1}}
   \put(1,-0.5){$\vdots$}
   \put(0.2,0){\line(1,-1){1}}
   \put(1.4,-0.1){$a_i^{(1)}$}
   \put(2.1,0){\line(1,0){0.8}}
   \put(3,0){\circle{0.2}}
   \put(2.7,0.35){$\prob_i^{(1)}$}
   \put(3.1,0){\line(1,1){1}}
   \put(3.9,0.2){$\vdots$}
   \put(3.1,0){\line(1,0){1}}
   \put(3.9,-0.5){$\vdots$}
   \put(3.1,0){\line(1,-1){1}}
   \put(4.3,-0.2){$E_{ij}^{(1)}$}
   \put(5.1,0){\line(1,0){0.9}}
   \put(6,0){\circle*{0.1}}
   \put(6.2,-0.1){$\displaystyle{u(c_{ij}^{(1)})= \max_{k}\overline{u}(a_{k[ij]}^{(2)})}$}
 \end{picture}
 
 en donde:
    $$\overline{u}(a_{k[ij]}^{(2)})=\sum_l u(c_{kl}^{(2)})\,\prob_k^{(2)}(E_{kl}^{(2)})$$
 
 \bigskip
 \begin{ejem}\label{ejem empresa farmaceutica}
   Una empresa farmac\'eutica se plantea la posibilidad de lanzar al mercado un nuevo antigripal. Un despacho de actuarios le ofrece la realizaci\'on de un estudio de mercado para reducir la incertidumbre sobre la proporci\'on de m\'edicos que lo recetar\'ian. Sean los eventos:
   \begin{equation*}
     \begin{split}
       & E_1:=\,\mbox{una proporci\'on alta de m\'edicos lo recetar\'an} \\
       & E_2:=\,\mbox{una proporci\'on moderada de m\'edicos lo recetar\'an} \\
       & E_3:=\,\mbox{una proporci\'on baja de m\'edicos lo recetar\'an}
     \end{split}
   \end{equation*}
A priori la compa\~n\'ia estima que $\prob(E_1)=0{.}2$, $\prob(E_2)=0{.}5$ y por tanto $\prob(E_3)=0{.}3$ y las ganancias que obtendr\'ia la empresa si lanza el producto bajo cada escenario ser\'ian +\$5, +\$1 y -\$3 millones de pesos, respectivamente. El estudio propuesto puede aconsejar la producci\'on ($X=1$) o desaconsejarla ($X=0$) y las probabilidades de que el resultado del estudio sea aconsejar la producci\'on dada la proporci\'on de m\'edicos que recetar\'ian el antigripal son:
   $$\prob(X=1\,|\,E_1)=0{.}9\qquad\prob(X=1\,|\,E_2)=0{.}5\qquad \prob(X=1\,|\,E_3)=0{.}2$$
?`Cu\'al es el precio m\'aximo que la empresa farmac\'eutica debe pagar por el estudio? Sea $c$ el costo de dicho estudio. Tenemos entonces:

 \newpage

 \setlength{\unitlength}{1cm}   
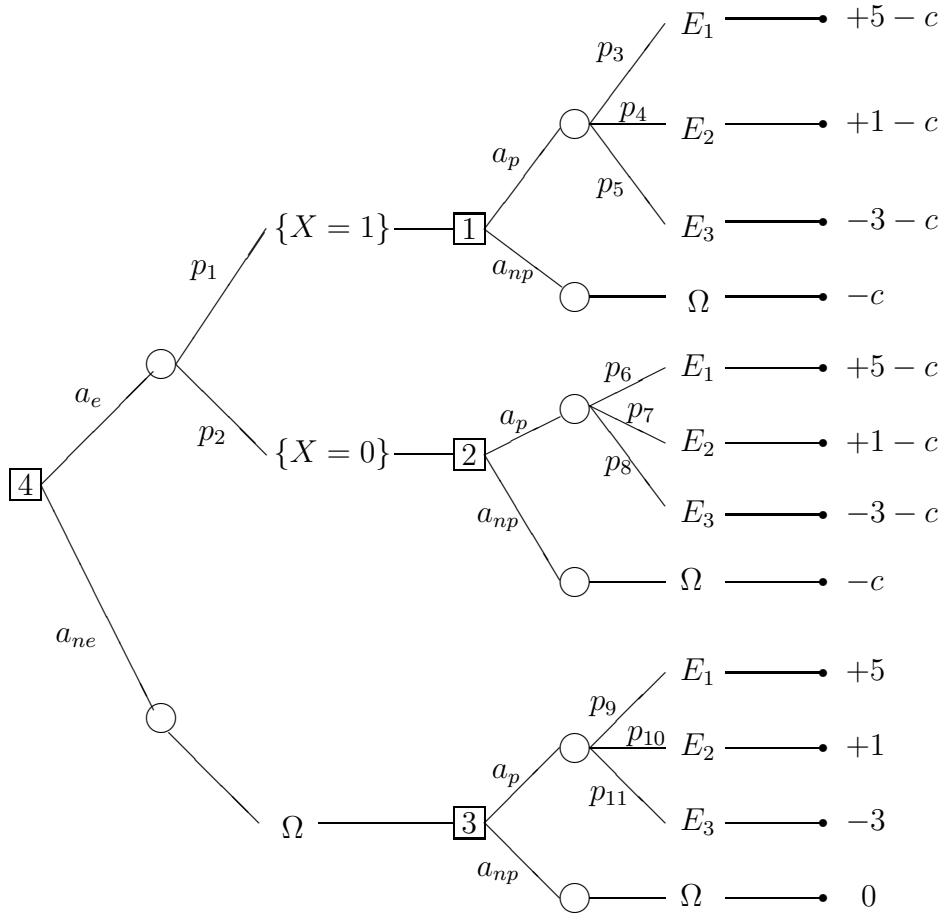
 \begin{picture}(14,15)
   % ETAPA 1
   % Nodo de decisión 4
   \put(0,7.3){\framebox(0.4,0.4){4}}
   \put(0.4,7.5){\line(1,1){1.5}}
   \put(0.85,8.6){$a_e$}
   \put(0.4,7.5){\line(1,-2){1.5}}
   \put(0.6,5.4){$a_{ne}$}
   \put(2,9.1){\circle{0.4}}
   \put(2,4.4){\circle{0.4}}
   % eventos etapa 1
   \put(2.2,9.1){\line(2,3){1.2}}
   \put(2.4,10.3){$p_1$}
   \put(3.5,10.8){$\{X=1\}$}
   \put(5.1,10.9){\line(1,0){0.8}}
   \put(2.2,9.1){\line(1,-1){1.2}}
   \put(2.5,8.1){$p_2$}
   \put(3.5,7.8){$\{X=0\}$}
   \put(5.1,7.9){\line(1,0){0.8}}
   \put(2.1,4.2){\line(1,-1){1.2}}
   \put(3.6,2.8){$\Omega$}
   \put(4.1,3){\line(1,0){1.8}}
   % ETAPA 2
   % Nodo de decisión 1
   \put(5.9,10.7){\framebox(0.4,0.4){1}}
   \put(6.3,10.9){\line(3,4){1}}
   \put(6.4,11.8){$a_p$}
   \put(7.5,12.3){\circle{0.4}}
   \put(7.7,12.3){\line(3,4){1}}
   \put(7.7,12.3){\line(1,0){1}}
   \put(7.7,12.3){\line(3,-4){1}}
   \put(7.8,13.2){$p_3$}
   \put(8.1,12.4){$p_4$}
   \put(7.8,11.4){$p_5$}
   \put(8.9,13.5){$E_1$}
   \put(9.5,13.7){\line(1,0){1.3}}
   \put(10.8,13.7){\circle*{0.1}}
   \put(11.1,13.6){$+5-c$}
   \put(8.9,12.1){$E_2$}
   \put(9.5,12.3){\line(1,0){1.3}}
   \put(10.8,12.3){\circle*{0.1}}
   \put(11.1,12.2){$+1-c$}
   \put(8.9,10.8){$E_3$}
   \put(9.5,11){\line(1,0){1.3}}
   \put(10.8,11){\circle*{0.1}}
   \put(11.1,10.9){$-3-c$}
   
   \put(6.3,10.9){\line(4,-3){1.03}}
   \put(6.4,10.3){$a_{np}$}
   \put(7.5,10){\circle{0.4}}
   \put(7.7,10){\line(1,0){1}}
   \put(9,9.8){$\Omega$}
   \put(9.5,10){\line(1,0){1.3}}
   \put(10.8,10){\circle*{0.1}}
   \put(11.1,9.9){$-c$}
   % Nodo de decisión 2
   \put(5.9,7.7){\framebox(0.4,0.4){2}}
   \put(6.3,7.9){\line(2,1){1.01}}
   \put(6.5,8.3){$a_p$}
   \put(7.5,8.5){\circle{0.4}}
   \put(7.7,8.55){\line(2,1){1}}
   \put(7.9,8.95){$p_6$}
   \put(7.7,8.55){\line(2,-1){1}}
   \put(8.2,8.4){$p_7$}
   \put(7.7,8.55){\line(3,-4){1}}
   \put(7.9,7.7){$p_8$}
   \put(8.9,8.95){$E_1$}
   \put(9.5,9.05){\line(1,0){1.3}}
   \put(10.8,9.05){\circle*{0.1}}
   \put(11.1,8.95){$+5-c$}
   \put(8.9,7.95){$E_2$}
   \put(9.5,8.05){\line(1,0){1.3}}
   \put(10.8,8.05){\circle*{0.1}}
   \put(11.1,7.95){$+1-c$}
   \put(8.9,7){$E_3$}
   \put(9.5,7.1){\line(1,0){1.3}}
   \put(10.8,7.1){\circle*{0.1}}
   \put(11.1,7){$-3-c$}
   
   \put(6.3,7.9){\line(3,-5){1}}
   \put(6.2,7){$a_{np}$}
   \put(7.5,6.2){\circle{0.4}}
   \put(7.7,6.2){\line(1,0){1}}
   \put(8.9,6.1){$\Omega$}
   \put(9.5,6.2){\line(1,0){1.3}}
   \put(10.8,6.2){\circle*{0.1}}
   \put(11.1,6.1){$-c$}
   % Nodo de decisión 3
   \put(5.9,2.8){\framebox(0.4,0.4){3}}
   \put(6.3,3){\line(1,1){1}}
   \put(6.4,3.6){$a_p$}
   \put(7.5,4){\circle{0.4}}
   \put(7.7,4){\line(1,1){1}}
   \put(7.7,4){\line(1,0){1}}
   \put(7.7,4){\line(1,-1){1}}
   \put(7.7,4.5){$p_9$}
   \put(8.2,4.1){$p_{10}$}
   \put(7.7,3.3){$p_{11}$}
   \put(8.9,4.9){$E_1$}
   \put(9.5,5){\line(1,0){1.3}}
   \put(10.8,5){\circle*{0.1}}
   \put(11.1,4.9){$+5$}
   \put(8.9,3.9){$E_2$}
   \put(9.5,4){\line(1,0){1.3}}
   \put(10.8,4){\circle*{0.1}}
   \put(11.1,3.9){$+1$}
   \put(8.9,2.9){$E_3$}
   \put(9.5,3){\line(1,0){1.3}}
   \put(10.8,3){\circle*{0.1}}
   \put(11.1,2.9){$-3$}
      
   \put(6.3,3){\line(1,-1){1}}
   \put(6.2,2.3){$a_{np}$}
   \put(7.5,2){\circle{0.4}}
   \put(7.7,2){\line(1,0){1}}
   \put(8.9,1.9){$\Omega$}
   \put(9.5,2){\line(1,0){1.3}}
   \put(10.8,2){\circle*{0.1}}
   \put(11.1,1.9){$\,\,\,0$}
 \end{picture}
 
 El nodo de decisi\'on 4 corresponde a la primera etapa y los nodos de decisi\'on 1,2 y 3 corresponden a la segunda. En la primera etapa el conjunto de acciones es $\mathcal{A}^{(1)}:=\{a_e,a_{ne}\}$ en donde:
   \begin{equation*}
     \begin{split}
       & \,\,a_e:=\,\mbox{hacer el estudio} \\
       & a_{ne}:=\,\mbox{no hacer el estudio}
     \end{split}
   \end{equation*}

 En la segunda etapa los nodos de decisi\'on tienen el mismo conjunto de acciones $\mathcal{A}^{(2)}:=\{a_p,a_{np}\}$ en donde:
   \begin{equation*}
     \begin{split}
       & \,\,a_p:=\,\mbox{producir el antigripal} \\
       & a_{np}:=\,\mbox{no producir el antigripal}
     \end{split}
   \end{equation*} 
   
 La medida de probabilidad para el espacio de estados $\{E_1,E_2,E_3\}$ var\'ia seg\'un el nodo aleatorio. Para el nodo aleatorio correspondiente al nodo de decisi\'on 1 tenemos, utilizando la regla de Bayes:
   \begin{eqnarray*}
     p_3 & = & \prob(E_1\,|\,X=1) \\
         & = & \frac{\prob(X=1\,|\,E_1)\prob(E_1)}{\prob(X=1)\,|\,E_1)
               \prob(E_1)+\prob(X=1)\,|\,E_1^{\,c})\prob(E_1^{\,c})} \\
         & = & 0{.}367
   \end{eqnarray*}
y de manera an\'aloga $p_4=0{.}510,p_5=0{.}123,p_6=0{.}039,p_7=0{.}490,p_8=0{.}471,p_9=0{.}2, p_{10}=0{.}5,p_{11}=0{.}3,p_1=0{.}49,p_2=0{.}51$. Resolviendo los nodos de decisi\'on de la segunda etapa obtenemos:

 \setlength{\unitlength}{1cm}   
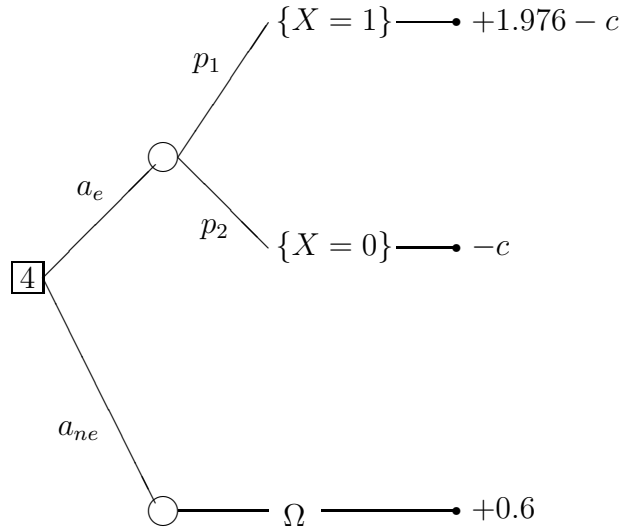
 \begin{picture}(14,7.5)(0,4)
   % ETAPA 1
   % Nodo de decisión 4
   \put(0,7.3){\framebox(0.4,0.4){4}}
   \put(0.4,7.5){\line(1,1){1.5}}
   \put(0.85,8.6){$a_e$}
   \put(0.4,7.5){\line(1,-2){1.5}}
   \put(0.6,5.4){$a_{ne}$}
   \put(2,9.1){\circle{0.4}}
   \put(2,4.4){\circle{0.4}}
   % eventos etapa 1
   \put(2.2,9.1){\line(2,3){1.2}}
   \put(2.4,10.3){$p_1$}
   \put(3.5,10.8){$\{X=1\}$}
   \put(5.1,10.9){\line(1,0){0.8}}
   \put(5.9,10.9){\circle*{0.1}}
   \put(6.1,10.8){$+1{.}976-c$}
   
   \put(2.2,9.1){\line(1,-1){1.2}}
   \put(2.5,8.1){$p_2$}
   \put(3.5,7.8){$\{X=0\}$}
   \put(5.1,7.9){\line(1,0){0.8}}
   \put(5.9,7.9){\circle*{0.1}}
   \put(6.1,7.8){$-c$}
   
   \put(2.2,4.4){\line(1,0){1.2}}
   \put(3.6,4.2){$\Omega$}
   \put(4.1,4.4){\line(1,0){1.8}}
   \put(5.9,4.4){\circle*{0.1}}
   \put(6.1,4.3){$+0{.}6$}
 \end{picture}
   
   Resulta entonces preferible hacer el estudio a no hacerlo ($a_e\succ a_{ne}$) siempre y cuando se cumpla $\overline{u}(a_e)>\overline{u}(a_{ne})$ lo cual ocurre si y s\'olo si $c<0{.}368$ as\'i que para la empresa farmac\'eutica resulta conveniente pagar porque se haga el estudio siempre y cuando el costo de \'este no exceda \$368,000. $\quad\diamond$
 \end{ejem}  
 
 \bigskip

 \hyphenation{par-ti-cu-lar}
    
 Un problema de decisi\'on secuencial que nos interesa de manera particular es aqu\'el en que se tiene que decidir llevar a cabo un experimento de entre varios posibles, y una vez escogido el experimento \'este es utilizado en un problema de decisi\'on subsecuente, y se desea escoger el experimento \'optimo. Este problema particular se conoce como \textit{dise\~no experimental}. Esquem\'aticamente:
 
 \setlength{\unitlength}{1cm}   
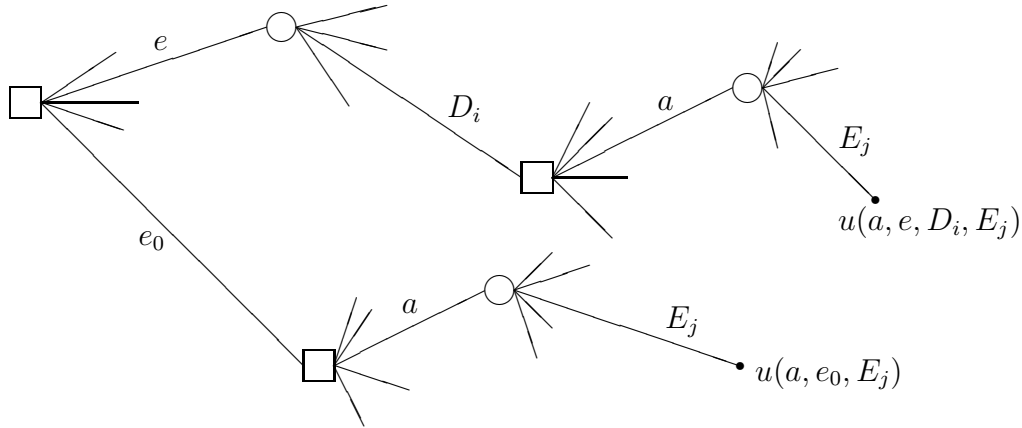
 \begin{picture}(14,6)
   \put(0,4){\framebox(0.4,0.4)}
   \put(0.4,4.2){\line(3,2){1}}
   \put(0.4,4.2){\line(3,1){3}}
   \put(1.9,4.9){$e$}
   \put(0.4,4.2){\line(1,0){1.3}}
   \put(0.4,4.2){\line(3,-1){1.1}}
   \put(3.6,5.2){\circle{0.4}}
   \put(3.8,5.2){\line(4,1){1.2}}
   \put(3.8,5.2){\line(4,-1){1.2}}
   \put(3.8,5.2){\line(3,-2){3}}
   \put(3.8,5.2){\line(2,-3){0.7}}
   \put(5.8,4){$D_i$}
   \put(6.8,3){\framebox(0.4,0.4)}
   \put(7.2,3.2){\line(1,2){0.5}}
   \put(7.2,3.2){\line(1,1){0.8}}
   \put(7.2,3.2){\line(2,1){2.4}}
   \put(7.2,3.2){\line(1,0){1}}
   \put(7.2,3.2){\line(1,-1){0.8}}
   \put(8.6,4.1){$a$}
   \put(9.8,4.4){\circle{0.4}}
   \put(10,4.4){\line(1,3){0.2}}
   \put(10,4.4){\line(1,1){0.5}}
   \put(10,4.4){\line(4,1){1}}
   \put(10,4.4){\line(1,-1){1.5}}
   \put(10,4.4){\line(1,-4){0.2}}
   \put(11,3.6){$E_j$}
   \put(11.5,2.9){\circle*{0.1}}
   \put(11,2.5){$u(a,e,D_i,E_j)$}
      
   \put(0.4,4.2){\line(1,-1){3.5}}
   \put(1.7,2.3){$e_0$}
   \put(3.9,0.5){\framebox(0.4,0.4)}
   \put(4.3,0.7){\line(1,3){0.3}}
   \put(4.3,0.7){\line(2,3){0.5}}
   \put(4.3,0.7){\line(2,1){2}}
   \put(4.3,0.7){\line(3,-1){1}}
   \put(4.3,0.7){\line(1,-2){0.4}}
   \put(5.2,1.4){$a$}
   \put(6.5,1.7){\circle{0.4}}
   \put(6.7,1.7){\line(1,1){0.5}}
   \put(6.7,1.7){\line(3,1){1}}
   \put(6.7,1.7){\line(3,-1){3}}
   \put(6.7,1.7){\line(1,-1){0.5}}
   \put(6.7,1.7){\line(1,-3){0.3}}
   \put(8.7,1.2){$E_j$}
   \put(9.7,0.7){\circle*{0.1}}
   \put(9.9,0.5){$u(a,e_0,E_j)$}
 \end{picture}
   
   Primero escogemos un experimento $e$ y de acuerdo a los datos obtenidos $D$ tomamos la acci\'on $a$, despu\'es de la cual y ante la ocurrencia del evento $E$ produce una consecuencia cuya utilidad denotaremos $u(a,e,D,E)$. Entre los posibles experimentos a escoger incluimos un \textit{experimento nulo} $e_0$ que representa el caso en que decidamos irnos directo a las acciones posibles sin llevar a cabo experimento alguno.\par
   
 \medskip
   
 Resolviendo primero los nodos de decisi\'on de la segunda etapa tendremos que calcular la utilidad esperada de las acciones:
   $$\overline{u}(a,e,D_i)=\sum_j u(a,e,D_i,E_j)\,\prob(E_j\,|\,e,D_i,a)$$ 
   
 Con lo anterior y para cada par $(e,D_i)$ podemos escoger la acci\'on \'optima a seguir en cada caso, esto es, una acci\'on $a_i^*$ que maximice la expresi\'on anterior. De este modo, la utilidad de la consecuencia $(e,D_i)$ estar\'a dada por:
   $$u(e,D_i)=\overline{u}(a_i^*,e,D_i)=\max_a\overline{u}(a,e,D_i)$$
   
 Ahora s\'olo queda resolver la primera etapa, es decir, determinar cu\'al es el experimento \'optimo y para ello calculamos la utilidad esperada de cada experimento:
   $$\overline{u}(e)=\sum_i \overline{u}(a_i^*,e,D_i)\,\prob(D_i\,|\,e)$$

En el caso particular del experimento nulo tenemos:
   $$\overline{u}(e_0)=\overline{u}(a_0^*,e_0)=\max_a \sum_j u(a,e_0,E_j) \prob(E_j\,|\,e_0,a)$$
por lo que vale la pena llevar a cabo un experimento $e$ siempre y cuando $\overline{u}(e)>\overline{u}(e_0)$ :

 \medskip
 
 \begin{enun}
   \begin{prop}\label{diseno experimental optimo}
     La acci\'on \'optima es llevar a cabo el experimento $e^*$ si \linebreak $\overline{u}(e^*)>\overline{u}(e_0)$ y $\overline{u}(e^*)=\max_e\overline{u}(e)$; de lo contrario, la acci\'on \'optima es no realizar experimento alguno.
   \end{prop}
 \end{enun}
 \begin{proof}
    Es inmediata a partir de la Definici\'on \ref{def criterio general de decision}.
 \end{proof}
 
 \medskip
 
 Con lo anterior tenemos forma de definir un valor para la informaci\'on adicional que se puede obtener en el contexto de un problema de decisi\'on dado. Es posible calcular el valor esperado de la informaci\'on que nos dan los datos como la esperanza (a posteriori) de la diferencia entre las utilidades que corresponden a las acciones \'optimas despu\'es y antes de considerar los datos obtenidos:
 
 \medskip
 
 \begin{enun}
   \begin{defn}\label{valor de info adicional}
     \begin{enumerate}
       \item El \textit{valor esperado de los datos} $D_i$ proveniente de un experimento $e$ est\'a definido por:
  $$v(e,D_i):=\sum_j \big[u(a_i^*,e,D_i,E_j)-u(a_0^*,e_0,E_j)\big]\, \prob(E_j\,|\,e,D_i,a_i^*)$$
donde $a_i^*$ y $a_0^*$ son las acciones \'optimas dados los datos $D_i$ y en ausencia de datos, respectivamente.
       \item El \textit{valor esperado de un experimento} $e$ est\'a dado por:
          $$v(e):=\sum_i v(e,D_i)\,\prob(D_i\,|\,e)$$
     \end{enumerate}
   \end{defn}
 \end{enun}
 
 \medskip
 
 Y para tener una idea de qu\'e tan grande es el valor $v(e)$ de un experimento $e$ es posible calcularle una cota superior. Consideremos las acciones \'optimas que estar\'ian disponibles bajo \textit{informaci\'on perfecta}, esto es, suponiendo que sabemos de antemano que el evento $E_j$ va a ocurrir, y sea  $a_{(j)}^*$ la acci\'on \'optima dado $E_j$, es decir tal que:
   $$u(a_{(j)}^*,e_0,E_j)=\max_a u(a,e_0,E_j)$$
  
  De este modo, dado $E_j$, la p\'erdida que se tiene por escoger cualquier otra acci\'on $a$ ser\'a:
   $$u(a_{(j)}^*,e_0,E_j)-u(a,e_0,E_j)$$
   
  Para $a=a_0^*$ (la acci\'on \'optima a priori) esta diferencia proporciona, condicional en $E_j$, el \textit{valor de informaci\'on perfecta} y, bajo ciertas condiciones, su valor esperado nos dar\'a una cota superior para el incremento en utilidad que nos proporcionar\'ian datos adicionales acerca de los eventos $E_j$ :
 
  \medskip
  
  \begin{enun}
    \begin{defn}\label{valor esperado de info perfecta}
      La \textit{p\'erdida de oportunidad} que se tiene si se toma la acci\'on $a$ y ocurre el evento $E_j$ est\'a dada por:
      $$l(a,E_j):=\max_{a_i}u(a_i,e_0,E_j)-u(a,e_0,E_j)$$
y el \textit{valor esperado de informaci\'on perfecta} est\'a dado por:
      $$v^*(e_0):=\sum_j l(a_0^*,E_j)\,\prob(E_j\,|\,a_0^*)$$
    \end{defn}
  \end{enun}
  
  \smallskip
  
  Es importante no perder de vista que las funciones $v(e,D_i)$ y $v(e)$ as\'i como el n\'umero $v^*(e_0)$ dependen de las distribuciones (a priori) : 
      $$\{\prob(E_j\,|\,a)\,:\,a\in\mathcal{A}\}$$
  
  \medskip
  
  Existen situaciones en las que es posible separar la funci\'on de utilidad $u(a,e,D_i,E_j)$ en dos componentes: el \textit{costo} de llevar a cabo el experimento $e$ para obtener los datos $D_i$ y la \textit{utilidad} que se obtiene cuando se escoge la acci\'on $a$ y ocurre el evento $E_j$. Com\'unmente el componente \textit{utilidad} no depende de $(e,D_i)$ por lo que, suponiendo aditividad de ambos componentes:
    $$u(a,e,D_i,E_j)=u(a,e_0,E_j)-c(e,D_i)\,\,,\quad c(e,D_i)\geq 0$$
    
  M\'as a\'un, las distribuciones de probabilidad sobre los eventos son, por lo general, independientes de las acciones. Bajo las condiciones anteriores es posible calcular una cota superior para el valor esperado de un experimento:
  
 \smallskip
  
 \begin{enun}
   \begin{prop}\label{descomposicion aditiva}
     Si la funci\'on de utilidad es de la forma:
     $$u(a,e,D_i,E_j)=u(a,e_0,E_j)-c(e,D_i)\,\,,\quad c(e,D_i)\geq 0\,,$$
y las distribuciones de probabilidad son tales que
     $$\prob(E_j\,|\,e,D_i,a)=\prob(E_j\,|\,e,D_i)\,,\quad \prob(E_j\,|\,e_0,a)=\prob(E_j\,|\,e_0)\,,$$
entonces, para cualquier experimento disponible $e$, se tiene que
     $$v(e)\leq v^*(e_0)-\overline{c}(e)\,,$$
en donde
     $$\overline{c}(e):=\sum_i c(e,D_i)\,\prob(D_i\,|\,e)$$
es el costo esperado del experimento $e$.
   \end{prop}
 \end{enun}
 \begin{proof}
  Utilizando las definiciones \ref{valor de info adicional} y \ref{valor esperado de info perfecta} podemos reexpresar $v(e)$ como:
   \begin{equation*}
     \begin{split}
     v(e) &= \sum_i\bigg[\sum_j \Big\{ u(a_i^*,e_0,E_j)-c(e,D_i)-u(a_0^*,e_0,E_j)\Big\}\prob(E_j\,|\,e,D_i)\bigg]\prob(D_i\,|\,e) \\
          &= \sum_i\bigg[\max_a \sum_j \Big\{ u(a,e_0,E_j)-u(a_0^*,e_0,E_j) \Big\} \prob(E_j\,|\,e,D_i) \bigg]\prob(D_i\,|\,e) - \overline{c}(e) \\
          &\leq \sum_i\sum_j \Big[ \max_a u(a,e_0,E_j)-u(a_0^*,e_0,E_j) \Big]\prob(E_j\cap D_i\,|\,e)-\overline{c}(e) \\
          &\leq \Big\{ \sum_j l(a_0^*,E_j)\,\prob(E_j\,|\,a_0^*)\Big\}\Big\{ \sum_i \prob(D_i\,|\,E_j,e) \Big\} - \overline{c}(e) \\
          &\leq \Big\{ \sum_j l(a_0^*,E_j)\,\prob(E_j\,|\,a_0^*)\Big\}-\overline{c}(e)\\
          &\leq v^*(e_0)-\overline{c}(e)
     \end{split}
   \end{equation*}
 \end{proof}
 
 \begin{ejem}\label{ejem empresa farmaceutica continuacion}
   Continuando con el Ejemplo \ref{ejem empresa farmaceutica} tendr\'iamos como $e$ el experimento de llevar a cabo el estudio y como $e_0$ el no llevarlo a cabo. Para evitar conflicto de notaci\'on lo que en el ejemplo anterior denotamos como $c$ ahora lo denotaremos $k$. Entonces:
     $$\overline{u}(e)=0{.}968 - k\,\,,\quad \overline{u}(e_0)=0{.}6$$
por lo que vale la pena llevar a cabo el experimento $e$ siempre y cuando $\overline{u}(e)>\overline{u}(e_0)$, es decir, siempre que $k<0{.}368$. Los datos $D_i$ que se pueden obtener est\'an en este caso representados por los eventos $D_1:=\{X=1\}$ y $D_2:=\{X=0\}$ :
     $$v(e,D_1)=-k\,\,,\quad v(e,D_2)=-k-1{.}00776$$
y como $\overline{c}(e)=k$ entonces:
     $$v(e)=-k-1{.}00776\leq 0{.}9-k=v^*(e_0)-\overline{c}(e)\qquad\diamond$$
 \end{ejem}
 
 \bigskip

 \section{Inferencia e informaci\'on}
 
 En las secciones previas hemos visto que para resolver un problema de decisi\'on $(\mathcal{E},\mathcal{C},\mathcal{A},\preceq)$ requerimos de una funci\'on de utilidad $\,u:\mathcal{C}\rightarrow\mathbb{R}\,$ y de una medida de probabilidad $\prob$ sobre $\mathcal{E}$, y aplicamos el criterio de la utilidad esperada m\'axima sobre el conjunto de acciones $\mathcal{A}$. En esta secci\'on nos concentraremos en algunos aspectos sobre la asignaci\'on de dicha medida de probabilidad.\par
 
 \smallskip
 
 La medida de probabilidad $\prob$ se asigna de acuerdo al estado de informaci\'on que en un momento dado tiene un individuo (o grupo de individuos). Dado un estado inicial de informaci\'on $M_0$, las probabilidades que se asignen a los distintos eventos de $\mathcal{E}$ son probabilidades condicionales en dicha informaci\'on inicial (o probabilidades a priori) y en estricto sentido debi\'eramos denotarlas $\prob(\,\,\cdot\,\,|\,M_0)$ aunque por simplicidad se suele omitir el condicionamiento y lo denotamos simplemente $\prob(\,\,\cdot\,\,)$. Despu\'es de ese momento inicial puede el individuo (o grupo de individuos) recibir informaci\'on adicional (por ejemplo, resultados de un nuevo experimento, encuesta, etc.) misma que denotamos como un evento $G$. Esto nos lleva inmediatamente a \textit{actualizar} la medida de probabilidad $\prob(\,\,\cdot\,\,)$ a una medida de probabilidad $\prob_1(\,\,\cdot\,\,):=\prob(\,\,\cdot\,\,|\,G)$. Normalmente dicha informaci\'on adicional $G$ corresponde a datos recolectados en relaci\'on con el fen\'omeno aleatorio de inter\'es y en tal caso denotaremos dicha informaci\'on como $D$ y por tanto la medida de probabilidad inicial quedar\'a actualizada a una medida de probabilidad $\prob(\,\,\cdot\,\,|\,D)$, que equivale a lo que en el Cap\'itulo 2 definimos como probabilidad a posteriori.\par
 
 \smallskip
 
 La utilidad esperada de cada acci\'on (y por tanto la utilidad esperada m\'axima) depende tanto de la funci\'on de utilidad sobre $\mathcal{C}$ como de la medida de probabilidad que se utilice y por ello analizaremos el conjunto de medidas de probabilidad que se pueden asignar y el problema de escoger una en particular como un problema de decisi\'on.\par
 
 \smallskip 
 
 \begin{enun}
   \begin{defn}
      Sea el espacio de estados (relevantes) $\Theta:=\{E_j\,:\,j\in J\}$. Definimos como la \textit{clase de distribuciones condicionales} sobre $\Theta\,$:
 $$\mathcal{Q}:=\{\mathbf{q}\equiv(q_j,j\in J)\,:\,q_j\geq 0, \sum_{j\in J} q_j=1\}\,.$$
   \end{defn}
 \end{enun}
 
 Si supi\'eramos de antemano que el evento $E_{j^*}$ va a ocurrir (informaci\'on perfecta) la distribuci\'on de probabilidad ``ideal'' sobre el espacio de estados $\Theta$ ser\'ia $\mathbf{q}^*\equiv(q_j=\indic_{\{j=j^*\}})$. No siempre es posible tener informaci\'on perfecta ni tampoco garant\'ia de que una determinada distribuci\'on de probabilidad $\mathbf{q}$ que elijamos sea la m\'as adecuada. Supongamos por un momento que la distribuci\'on de probabilidad ``correcta'' sobre $\Theta$ la denotamos por:
 
 $$\mathbf{p}\equiv\big(p_j=\prob(E_j\,|\,D)\,:\,j\in J,p_j>0,\sum_{j\in J}p_j=1\big)$$
 
 N\'otese que en la definici\'on de $\mathbf{p}$ tenemos la condici\'on estricta $p_j>0$. Esto es, pediremos que los elementos del  espacio de eventos relevantes tengan medida de probabilidad distinta de cero.\par
 
  \smallskip
  
 Consideremos el problema de decisi\'on $(\mathcal{E},\mathcal{C},\mathcal{A},\preceq)$ en donde $\mathcal{A}:=\mathcal{Q}$ y el espacio de estados relevantes es $\Theta:=\{E_j\,:\,j\in J\}$ y por tanto el conjunto de consecuencias est\'a dado por $\mathcal{C}=\mathcal{Q}\times\Theta$. S\'olo nos falta definir una funci\'on de utilidad $u$ sobre $\mathcal{C}$ que describa el ``valor'' $u(\mathbf{q},E_j)$ de utilizar la distribuci\'on de probabilidad $\mathbf{q}$ bajo la ocurrencia del evento $E_j$. De c\'omo definir tal funci\'on de utilidad nos ocuparemos a continuaci\'on, y a este tipo particular de funciones se les conoce como \textit{funciones de puntaje} (``score functions'', en ingl\'es).\par
 
 \medskip
 
 \begin{enun}
   \begin{defn}
      Una \textit{funci\'on de puntaje} para una familia de distribuciones de probabilidad $\mathcal{Q}:=\{\mathbf{q}=(q_j\,:\,j\in J)\}$ definidas sobre una partici\'on $\Theta:=\{E_j,j\in J\}$ es una funci\'on $u:\mathcal{Q}\times\Theta\rightarrow\mathbb{R}$. Se dice que esta funci\'on es \textit{suave} si es continuamente diferenciable como funci\'on de cada $q_j$.
   \end{defn}
 \end{enun}
 
 Esta condici\'on de suavidad resulta deseable en tanto que esperar\'iamos que cambios peque\~nos en alg\'un $q_j$ produzca cambios peque\~nos en el puntaje asignado por $u$.\par
 
 \smallskip
 
 Y nuevamente la elecci\'on \'optima de entre los elementos de $\mathcal{Q}$ ser\'a aqu\'ella $\mathbf{q^*}$ tal que:
    $$\overline{u}(\mathbf{q^*})=\max_{\mathbf{q}\in\mathcal{Q}}\overline{u} (\mathbf{q})$$
en donde
 $$\overline{u}(\mathbf{q})=\sum_{j\in J}u(\mathbf{q},E_j)\,\prob(E_j\,|\,D)$$
 
 Una caracter\'istica razonable que debe tener una funci\'on de puntaje $u$ es que $\mathbf{q}^*=\mathbf{p}$ en donde, recordemos, $\mathbf{p}\equiv(p_j\,:\,j\in J)$ tal que $p_j:=\prob(E_j\,|\,D)\,$:
 
 \begin{enun}
   \begin{defn}
     Una funci\'on de puntaje $u$ es \textit{propia} si para cada distribuci\'on de probabilidad $\mathbf{p}\equiv (p_j\,:\,j\in J)$ definida sobre una partici\'on $\Theta:=\{E_j\,:\,j\in J\}$ se cumple:
     $$\sup_{\mathbf{q}\in\mathcal{Q}}\Big\{\sum_{j\in J} u(\mathbf{q},E_j)p_j \Big\} = \sum_{j\in J} u(\mathbf{p},E_j)p_j$$
en donde el supremo se alcanza s\'olo si $\mathbf{q}=\mathbf{p}\,$.
   \end{defn}
 \end{enun}
 
 \smallskip
 
 Entre las aplicaciones que tienen las funciones de puntaje propias se encuentra el pago de premios justos a meteor\'ologos o analistas financieros por sus predicciones, en donde sus predicciones suelen ser asignaciones de probabilidades para distintos escenarios posibles, m\'as que una predicci\'on de qu\'e escenario va a ocurrir exactamente. Tambi\'en est\'a el caso de ex\'amenes de opci\'on m\'ultiple. El m\'etodo tradicional de evaluaci\'on de estos ex\'amenes suele ser binario (acert\'o o no acert\'o); sin embargo, ante un acierto existe la posibilidad de que le haya atinado a pesar de desconcer la respuesta correcta, y tambi\'en el que no haya acertado no implica que el conocimiento respecto a dicha pregunta sea completamente nulo. Utilizando funciones de puntaje se puede solicitar a quien responde el examen que, en vez de escoger una de las opciones, asigne una distribuci\'on de probabilidad para cada una de las opciones. Si est\'a completamente seguro acerca de la respuesta puede asignar probabilidad 1 a la misma y cero al resto de las opciones; si est\'a indeciso entre dos de las opciones puede asignar toda la masa de probabilidades en ellas para expresarlo. Esto nos da un panorama m\'as amplio acerca de lo que sabe quien presenta un examen de opci\'on m\'ultiple. La funci\'on de puntaje se dise\~na de modo que quien no tenga idea de la respuesta le resulte m\'as conveniente confesarlo v\'ia la asignaci\'on de una distribuci\'on de probabilidad uniforme discreta que intentar simular que sabe.  Un ejemplo de funci\'on de puntaje propia es el siguiente:\par
 \smallskip
 
 \begin{enun}
   \begin{defn}
     Una \textit{funci\'on de puntaje cuadr\'atica} para las distribuciones $\mathbf{q}\equiv(q_j\,:\,j\in J)$ definidas sobre una partici\'on $\{E_j\,:\,j\in J\}$ es cualquier funci\'on de la forma
 $$u(\mathbf{q},E_j)=A\Big\{2q_j-\sum_{i\in J}q_i^2\Big\} + B_j\,\,,\quad A>0$$
o alternativamente
 $$u(\mathbf{q},E_j)=A\Big\{1-\sum_{i\in J}(q_i-\indic_{\{i=j\}})^2 + B_j\Big\}\,\,,\quad A>0$$
en donde $(q_i-\indic_{\{i=j\}})^2$ representa una penalizaci\'on. Es inmediato verificar que ambas expresiones son equivalentes.
   \end{defn}
 \end{enun}
 
 \begin{enun}
   \begin{prop}
     Una funci\'on de puntaje cuadr\'atica es propia.
   \end{prop}
 \end{enun}
 \begin{proof}
   Sea la funci\'on:
   $$f(\mathbf{q})\equiv f(q_1,\ldots,q_m):=\sum_{j\in J}u(\mathbf{q},E_j)p_j$$
   
   Como $u$ es funci\'on de puntaje cuadr\'atica entonces
   $$f(\mathbf{q})=\sum_{j\in J} \Big\{A\big(2q_j-\sum_{i\in J}q_i^2\big) + B_j\Big\}p_j\,\,,\quad A>0$$
   
   Calculando las parciales de $f\,$:
   
   \begin{equation*}
     \begin{split}
       \frac{\partial}{\partial q_k}f(\mathbf{q}) &= \frac{\partial}{\partial q_k} \bigg[ \Big\{ A\big(2q_k-\sum_{i\in J}q_i^2\big) + B_k \Big\}p_k  +  \sum_{j\in J,j\neq k}\Big\{ A\big(2q_j-\sum_{i\in J} q_i^2\big)+B_j \Big\}p_j \bigg] \\
       &= \big\{A(2-2q_k)\big\}p_k + \sum_{j\in J,j\neq k}\big\{ A(-2q_k) \big\}p_j \\       
       &= 2A(1-q_k)p_k - 2Aq_k\sum_{j\in J,j\neq k}p_j\\
       &= 2Ap_k - 2Aq_kp_k - 2Aq_k\sum_{j\in J,j\neq k}p_j\\
       &= 2Ap_k - 2Aq_k\sum_{j\in J}p_j \\
       &= 2A(p_k-q_k)
     \end{split}
   \end{equation*}
   Por lo que
   $$\frac{\partial}{\partial q_k}f(\mathbf{q})=0 \Leftrightarrow p_k=q_k\,,\, k=1,\ldots,m$$
   Adem\'as
   $$\frac{\partial^2}{\partial q_k^2}f(\mathbf{q})=-2A<0$$
lo que implica que $f(\mathbf{q})$ alcanza un m\'aximo si y s\'olo si $\mathbf{q}=\mathbf{p}$ y por lo tanto la funci\'on de puntaje cuadr\'atica es propia.
 \end{proof}
 
 \medskip
 
 \begin{ejem}\label{ejem examenes de opcion multiple}
   Mencionamos ya que una de las aplicaciones de las funciones de puntaje es el caso de ex\'amenes de opci\'on m\'ultiple. En la forma tradicional de evaluar las respuestas de este tipo de ex\'amenes (correcta o incorrecta) no se puede evitar dos problemas: primero, la deshonestidad de un estudiante que sin saber la respuesta escoge una al azar y le atina; segundo, una respuesta incorrecta no implica ausencia total de conocimiento (por ejemplo, de entre 5 respuestas posibles quiz\'as le quedaba claro que tres de ellas no eran la respuesta correcta). Las funciones de puntaje propias nos permiten forzar al estudiante a ser honesto as\'i como reconocer grados parciales de conocimiento.\par
   Una forma de resolver este problema es pedirle al estudiante que proporcione una distribuci\'on de probabilidad sobre las posibles respuestas que describa lo que piensa acerca de la respuesta correcta. Desde el punto de vista del estudiante, contestar la pregunta es un problema de decisi\'on donde el conjunto de acciones es ahora la clase:
 $$\mathcal{Q}:=\{\mathbf{q}\equiv(q_1,\ldots,q_m)\,:\,q_j\geq 0,\sum q_j=1\}$$
de distribuciones de probabilidad sobre el conjunto $\{E_1,\ldots,E_m\}$ de respuestas posibles. En este caso la utilidad esperada puede ser definida de acuerdo a la calificaci\'on que se espera obtener:
   $$\overline{u}(\mathbf{q})=\sum_{j=1}^m u(\mathbf{q},E_j)\,p_j$$
en donde $u(\mathbf{q},E_j)$ es la calificaci\'on otorgada a un estudiante que reporta la distribuci\'on $\mathbf{q}$ cuando la respuesta correcta es $E_j$ y $p_j$ es la probabilidad personal que el estudiante asigna al evento de que la respuesta correcta sea $E_j$. Notemos que, en principio, no hay raz\'on para suponer que la distribuci\'on $\mathbf{q}$ reportada por el estudiante como su respuesta es la misma que la distribuci\'on $\mathbf{p}$ que describe en realidad su conocimiento (es decir, el estudiante puede ser tambi\'en deshonesto al contestar bajo este esquema).\par
  Por su parte, el maestro est\'a interesado en garantizar la honestidad del estudiante y para ello escoge una funci\'on de utilidad propia de modo que la utilidad esperada del estudiante se maximice si y s\'olo si $\mathbf{q}=\mathbf{p}$, y con ello el estudiante se autoperjudica si es deshonesto. Como ya vimos, la funci\'on de utilidad cuadr\'atica es propia, y para determinar las constantes $A$ y $B_j$ establecemos condiciones adicionales. Por ejemplo, supongamos que se decide otorgar un punto si la distribuci\'on $\mathbf{q}$ asigna probabilidad $1$ a la respuesta correcta, y cero puntos a la distribuci\'on uniforme $q_j=\frac{1}{m}$ (la cual describe ausencia de conocimiento). Esto nos lleva al sistema de ecuaciones:
    \begin{eqnarray*}
      A+B_j &=& 1 \\
      \frac{A}{m}+B_j &=& 0
    \end{eqnarray*}
de donde $A=\frac{m}{m-1}$ y $B_j=\frac{1}{m-1}$ y por lo tanto:
$$u(\mathbf{q},E_j)=\frac{m}{m-1}\big(2q_j-\sum_{i=1}^m q_i^2\big)-\frac{1}{m-1}$$
Notemos adem\'as que esta funci\'on asigna valores negativos a distribuciones que asignen probabilidades altas a respuestas incorrectas. En particular, un estudiante que asigne probabilidad $1$ a una respuesta incorrecta tendr\'a un descuento de $\frac{m+1}{m-1}$ puntos en su calificaci\'on, lo que implica que le resulte mejor, en promedio, en caso de desconocer por completo la respuesta correcta, admitir honestamente su ignorancia que intentar atinarle. Y por otro lado, si tenemos un estudiante que a pesar de no estar seguro de la respuesta correcta le queda claro que $m-2$ de las opciones deben descartarse y asigna probabilidad $\frac{1}{2}$ a dos opciones, una de las cuales es la correcta, entonces obtiene al menos $\frac{m-2}{2(m-1)}$ puntos, cantidad menor a uno pero positiva en caso de que $m>2\,.\quad\diamond$
 \end{ejem}
  
 \bigskip
 
 Para otro tipo de aplicaciones, un conjunto particularmente importante de funciones de puntaje son aqu\'ellas que dependen \'unicamente del evento que finalmente ocurre, esto es, de la probabilidad que la distribuci\'on $\mathbf{q}$ haya asignado a dicho evento, y de ah\'i el adjetivo de \textit{local}:\par
 \medskip
 
 \begin{enun}
   \begin{defn}\label{funcion de puntaje local}
     Una funci\'on de puntaje $u$ es \textit{local} si para cada $\mathbf{q}\in\mathcal{Q}$ definida sobre la partici\'on $\{E_j\,:\,j\in J\}$ existen funciones $\{u_j(\,\cdot\,)\,:\,j\in J\}$ tales que $u(\mathbf{q},E_j)=u_j(q_j)\,$.
   \end{defn}
 \end{enun}
 
 Este caso particular de funciones de puntaje es importante ya que una vez que se observe el evento que ocurre es respecto a \'este que se comparan las inferencias o predicciones acerca de lo que iba a ocurrir, y por ello resultan adecuadas para la inferencia estad\'istica. Lo que sigue es caracterizar este tipo de funciones:\par
 
 \medskip
 
 \begin{enun}
   \begin{prop}\label{caracterizar funciones de puntaje local}
     Si $u$ es una funci\'on de puntaje suave, propia y local para una clase de distribuciones $\mathbf{q}\in\mathcal{Q}$ definidas sobre una partici\'on $\{E_j\,:\,j\in J\}$ que contiene m\'as de dos elementos, entonces tiene que ser de la forma $u(\mathbf{q},E_j)=A\log q_j + B_j$ en donde $A>0$ y donde $\{B_j\,:\,j\in J\}$ son constantes arbitrarias.
   \end{prop}
 \end{enun}
 \begin{proof}
   Como $u(\,\cdot\,)$ es local y propia, entonces para algunas $\{u_j(\,\cdot\,):\,j\in J\}$ se tiene que
   $$\sup_{\mathbf{q}}\sum_{j\in J}u(\mathbf{q},E_j)p_j=\sup_{\mathbf{q}}u_j (q_j)p_j=\sum_{j\in J}u_j(p_j)p_j\,,$$
en donde $p_j>0$, $\sum_j p_j=1$, y el supremo se toma sobre la clase de distribuciones $\mathbf{q}\equiv(q_j\,:\,j\in J)$ tales que $q_j\geq 0$ y $\sum_j q_j=1\,$.\par
  Denotando $\mathbf{p}\equiv(p_1,p_2,\ldots)$ y $\mathbf{q}\equiv(q_1,q_2,\ldots)$ en donde
  $$p_1=1-\sum_{j>1}p_j\,,\quad q_1=1-\sum_{j>1}q_j\,,$$
caracterizamos las funciones $\{u_j(\,\cdot\,):\,j\in J\}$ buscando un punto extremo de:
  $$F(q_2,q_3,\ldots):=\Big(1-\sum_{j>1}p_j\Big)u_1\Big(1-\sum_{j>1}q_j\Big) +\sum_{j>1}p_j u_j(q_j)$$
Para que $F$ sea estacionaria en alg\'un punto $(q_2,q_3,\ldots)$ es necesario (ver Jeffreys y Jeffreys (1946), p.315) que
  $$\frac{\partial}{\partial\alpha}F(q_2+\alpha\varepsilon_2, q_3+\alpha\varepsilon_3,\ldots)\restriction_{\alpha=0}=0$$
para cualquier $\varepsilon:=(\varepsilon_2,\varepsilon_3,\ldots)$ tal que las $\varepsilon_j$ son suficientemente peque\~nas. Calculando dicha derivada obtenemos:
  $$\sum_{j>1}\bigg\{ \Big( 1-\sum_{i>1}p_i \Big)u_1'\Big( 1-\sum_{j>1}q_j \Big) \bigg\}\varepsilon_j=0$$
para $\varepsilon_j$ suficientemente peque\~nas y en donde $u'$ es la derivada de $u$. Como $u$ es propia entonces $(p_2,p_3,\ldots)$ debe ser un punto extremo de $F$ y obtenemos as\'i el sistema de ecuaciones:
   $$p_1u_1'(p_1)=p_ju_j'(p_j)\,,\quad j=1,2,\ldots$$
para todos los valores $p_2,p_3,\ldots$ lo que implica que, para una constante $A\,$:
  $$p\,u_j'(p)=A\,,\quad 0<p\leq 1\,,\quad j=1,2,\ldots$$
de donde $u_j(p)=A\log p + B_j\,$. La condici\'on de que $A>0$ es suficiente para garantizar que tal punto extremo es, en efecto, un m\'aximo.
 \end{proof}
 
 \medskip
 
 \begin{enun}
   \begin{defn}\label{funcion de puntaje logaritmica}
     Una \textit{funci\'on de puntaje logar\'itmica} para distribuciones de probabilidad estrictamente positivas $\mathbf{q}\equiv(q_j\,:\,j\in J)$ definidas sobre una partici\'on $\{E_j\,:\,j\in J\}$ es cualquier funci\'on de la forma:
     $$u(\mathbf{q},E_j)=A\log q_j + B_j\,,\quad A>0\,.$$
   \end{defn}
 \end{enun}
 
 \bigskip
 
 A continuaci\'on veremos la aplicaci\'on de este tipo de funciones de puntaje para aproximar, por ejemplo, una distribuci\'on de probabilidad $\mathbf{p}$ por medio de otra $\mathbf{q}$, o bien medir de alg\'un modo qu\'e tanto se aproxima una a la otra:
 
 \begin{enun}
   \begin{prop}\label{perdida esperada al reportar probabilidades}
     Si nuestras preferencias est\'an descritas por una funci\'on de puntaje logar\'itimica, la p\'erdida esperada de utilidad al usar una distribuci\'on de probabilidad $\mathbf{q}\equiv(q_j\,:\,j\in J)$ definida sobre una partici\'on $\{E_j\,:\,j\in J\}$ en lugar de la distribuci\'on $\mathbf{p}\equiv(p_j\,:\,j\in J)$ que representa lo que realmente creemos, est\'a dada por:
    $$\delta(\mathbf{q}\,|\,\mathbf{p})=A\sum_{j\in J}p_j\log\frac{p_j}{q_j}\,, \quad A>0\,.$$
M\'as a\'un, $\delta(\mathbf{q}\,|\,\mathbf{p})\geq 0$ con igualdad si y s\'olo si $\mathbf{q}=\mathbf{p}\,$.
   \end{prop}
 \end{enun}
 \begin{proof}
   Utilizando la Definici\'on \ref{funcion de puntaje logaritmica} tenemos que la utilidad esperada de usar la distribuci\'on $\mathbf{q}$ cuando $\mathbf{p}$ es la correcta es:
   $$\overline{u}=\sum_{j\in J}\big(A\log p_j + b_j\big)p_j$$
por lo que
  \begin{eqnarray*}
    \delta(\mathbf{q}\,|\,\mathbf{p}) &=& \overline{u}(\mathbf{p}) -\overline{u}(\mathbf{q})\\
    &=& \sum_{j\in J}\Big\{ \big(A\log p_j+B_j\big) - \big(A\log q_j+B_j\big) \Big\}p_j\\
    &=& A\sum_{j\in J}p_j\log\frac{p_j}{q_j}
  \end{eqnarray*}
  Como la funci\'on de puntaje es logar\'itmica y por tanto propia entonces por la Proposici\'on \ref{caracterizar funciones de puntaje local} tenemos que $\overline{u}(\mathbf{p})\geq\overline{u}(\mathbf{q})$ con igualdad si y s\'olo si $\mathbf{p}=\mathbf{q}$. O bien, utilizando el hecho de que $1+x\leq e^x$ de donde $x\leq e^{x-1}$ y si $x>0$ entonces $\log x\leq x-1$ con igualdad si $x=1\,$:
  $$-\delta(\mathbf{q}\,|\,\mathbf{p})=\sum_{j\in J}p_j\log\frac{q_j}{p_j} \leq\sum_{j\in J}p_j\Big(\frac{q_j}{p_j}-1\Big) =\sum_jq_j-\sum_jp_j=1-1=0$$
con igualdad si y s\'olo si $q_j=p_j$ para todo $j\,$.
 \end{proof}
 
  Una aplicaci\'on inmediata de lo anterior es en el caso de que se desee aproximar una distribuci\'on por medio de otra:
 
  \begin{enun}
   \begin{defn}\label{discrepancia logaritmica}
     La \textit{discrepancia logar\'itmica} entre una distribuci\'on de probabilidad estrictamente positiva $\mathbf{p}\equiv(p_j\,:\,j\in J)$ sobre una partici\'on $\{E_j\,:\,j\in J\}$ y una aproximaci\'on $\mathbf{\hat{p}}\equiv(\hat{p}_j\,:\,j\in J)$ est\'a definida por:
     $$\delta(\mathbf{\hat{p}}\,|\,\mathbf{p}):=\sum_{j\in J}p_j\log\frac{p_j}{\hat{p}_j}\,.$$
   \end{defn}
 \end{enun}
 
 Es inmediato notar que la discrepancia logar\'itmica no es una m\'etrica, comenzando por que no es sim\'etrica. Se puede hacer sim\'etrica mediante:
   $$\kappa\{\mathbf{q},\mathbf{p}\}:=\delta(\mathbf{q}\,|\,\mathbf{p}) +\delta(\mathbf{p}\,|\,\mathbf{q})$$
pero a\'un as\'i no es m\'etrica ya que no cumple la desigualdad del tri\'angulo.

 \bigskip
 
 \begin{ejem}\label{aproximar binomial con poisson}
   Conocido resultado de probabilidad es que, bajo ciertas condiciones, se puede obtener una buena aproximaci\'on del modelo binomial por medio del modelo Poisson:
   \begin{equation*}
     \begin{split}
       p_j &= \binom{n}{j}\theta^j(1-\theta)^{n-j}\indic_{\{0,1,\ldots,n\}}(j)\\
 \hat{p}_j &= \exp(-n\theta)\frac{(n\theta)^j}{j!}\indic_{\{0,1,\ldots\}}(j)
     \end{split}
   \end{equation*}
de donde
   $$\delta(\mathbf{\hat{p}}\,|\,\mathbf{p})=\sum_{k=2}^n\log k +n\big[ (1-\theta)\log(1-\theta)+\theta(1-\log n)\big]-\varphi(n,\theta)$$
en donde
   \begin{eqnarray*}
     \varphi(n,\theta) &:=& \esper_{\mathbf{p}}\Big\{\log\big[(n-X)\,!\big]\Big\}\,,\quad X\sim\mathbf{p}\\
     &=& \theta^{\,n}\sum_{k=2}^n\log(k\,!)\binom{n}{k}\Big(\,\frac{1}{\theta}-1\Big)^k
   \end{eqnarray*}
Tenemos que $\delta\rightarrow 0$ conforme $n\rightarrow\infty\;$ y/o $\;\theta\rightarrow 0$, es decir, la aproximaci\'on es buena para valores grandes de $n\;$ y/o  valores de $\theta$ cercanos a cero.$\quad\diamond$
 \end{ejem}
 
 \medskip
 
 Y en general, sea $\mathbf{p}$ una distribuci\'on de probabilidad y sea $\mathcal{Q}$ una familia de distribuciones de probabilidad para aproximar $\mathbf{p}$. Bajo discrepancia logar\'itmica, la \textit{mejor aproximaci\'on} de $\mathbf{p}$ ser\'a aquella $\mathbf{q^*}\in\mathcal{Q}$ tal que:
  $$\delta(\mathbf{q^*}\,|\,\mathbf{p})=\min_{\{\,\mathbf{q}\,\in\mathcal{Q}\}} \delta(\mathbf{q}\,|\,\mathbf{p})\,.$$
  
 \bigskip
 
 En el Cap\'itulo 2 y en el presente hemos visto que la probabilidad a priori $\prob(\,\cdot\,)$ sobre un espacio de estados relevantes se \textit{actualiza} con la recolecci\'on de datos $D$ v\'ia la regla de Bayes a una medida de probabilidad $\prob(\,\cdot\,|\,D)$. Tambi\'en hemos visto que en un contexto de inferencia estad\'istica, por ejemplo, la funci\'on de puntaje logar\'itmica resulta adecuada. Esto permitir\'a calcular la utilidad esperada de recolectar los datos $D\,$:
 
 \medskip
 
 \begin{enun}
   \begin{prop}
     Si nuestras preferencias est\'an descritas en t\'erminos de una funci\'on de puntaje logar\'itmica sobre la clase de distribuciones de probabilidad definidas en una partici\'on $\{E_j\,:\,j\in J\}$, entonces el incremento esperado en utilidad proveniente de los datos $D$, cuando la distribuci\'on de probabilidad a priori $\{\prob(E_j)\,:\,j\in J\}$ es estrictamente positiva, est\'a dado por
     $$A\sum_{j\in J}\prob(E_j\,|\,D)\log\frac{\prob(E_j\,|\,D)}{\prob(E_j)}$$
en donde $A>0$ es arbitraria y $\{\prob(E_j\,|\,D)\,:\,j\in J\}$ es la probabilidad a posteriori dados los datos $D$. M\'as a\'un, este incremento esperado en utilidad es no negativo, y es cero si y s\'olo si $\prob(E_j\,|\,D)=\prob(E_j)$ para todo $j\,$.
   \end{prop}
 \end{enun}
 \begin{proof}
   Por la Definici\'on \ref{funcion de puntaje logaritmica} tenemos que la utilidad de reportar las distribuciones de probabilidad $\prob(\,\cdot\,)$ y $\prob(\,\cdot\,\,|\,D)$, bajo el supuesto de que ocurra el evento $E_j$, est\'an dadas por $A\log\prob(E_j)+B_j$ y $A\log\prob(E_j\,|\,D)+B_j$, respectivamente. De este modo, el incremento esperado en utilidad proveniente de los datos $D$ est\'a dado por
   \begin{equation*}
     \begin{split}
       &\sum_{j\in J}\Big\{ \big( A\log\prob(E_j\,|\,D)+B_j \big) - \big( A\log\prob(E_j)+B_j \big) \Big\}\prob(E_j\,|\,D) \\
       &\qquad\qquad =\;A\sum_{j\in J}\prob(E_j\,|\,D)\log\frac{\prob(E_j\,|\,D)}{\prob(E_j)}\,,
     \end{split}
   \end{equation*}
cantidad que, por la Proposici\'on \ref{perdida esperada al reportar probabilidades}, es no negativa, y cero si y s\'olo si $\prob(E_j\,|\,D)=\prob(E_j)$ para todo $j\,$.
 \end{proof}
 
 \medskip
 
 Lo anterior motiva la siguiente definici\'on:
 
 \smallskip
 
 \begin{enun}
   \begin{defn}\label{cantidad de informacion de los datos}
     La \textit{cantidad de informaci\'on de los datos} acerca de una partici\'on $\{E_j\,:\,j\in J\}$ proveniente de los datos $D$ cuando la distribuci\'on a priori sobre dicha partici\'on es $\mathbf{p}_0:=\{\prob(E_j)\,:\,j\in J\}$ se define como:
     $$I(D\,|\,\mathbf{p}_0):=\sum_{j\in J}\prob(E_j\,|\,D)\log \frac{\prob(E_j\,|\,D)}{\prob(E_j)}$$
en donde $\{\prob(E_j\,|\,D)\,:\,j\in J\}$ es la distribuci\'on de probabilidad condicional dados los datos $D\,$.
   \end{defn}
 \end{enun}
 
 Equivalentemente y de acuerdo a la Definici\'on \ref{discrepancia logaritmica} la cantidad de informaci\'on de los datos $D$ es $\delta(\mathbf{p}_0\,|\,\mathbf{p}_D)$, esto es la discrepancia logar\'itmica considerando a $\mathbf{p}_0$ como una aproximaci\'on de la distribuci\'on de probabilidad $\mathbf{p}_D:=\big(\prob(E_j\,|\,D)\,:\,j\in J\big)$.\par
 
 \medskip
 
 Lo anterior nos permite calcular la cantidad esperada de informaci\'on de un experimento $e$ antes de que los resultados del mismo sean conocidos:\par
 
 \smallskip

 \hyphenation{ocurren}
 
 \begin{enun}
   \begin{defn}\label{informacion esperada de un experimento}
     La \textit{informaci\'on esperada de un experimento} $e$ sobre una partici\'on $\{E_j\,:\,j\in J\}$ con distribuci\'on a priori $\mathbf{p}_0:=\{\prob(E_j)\,:\,j\in J\}$ est\'a dada por:
     $$I(e\,|\,\mathbf{p}_0):=\sum_i I(D_i\,|\,\mathbf{p}_0)\,\prob(D_i)$$
en donde los posibles resultados del experimento $e$, denotados por $\{D_i\}$, ocurren con probabilidades $\{\prob(D_i)\}$.
   \end{defn}
 \end{enun}
 
 \begin{ejem}\label{ejemplo e1 revisited}
   Retomando el Ejemplo \ref{e1}, tenemos como espacio de estados relevantes $\{E_0,E_1\}$ en donde $E_0$ representa el evento de que salga sol y $E_1$ el evento de que salga \'aguila. Suponiendo que la moneda fue escogida al azar (i.e. $\alpha=1$) y de acuerdo a lo obtenido en el Ejemplo \ref{e1} tenemos que
   \begin{equation*}
    \begin{split}
     &\quad \prob(E_j)= \frac{1}{2}+\frac{\alpha}{4}(-1)^{j+1}\indic_{\{0,1\}}(j) \\
     &\prob(E_j\,|\,D)=\frac{3^j\nu+2}{4(\nu+1)}\indic_{\{0,1\}}(j)=\frac{1}{4}\Big( 3^j+\frac{1}{\nu+1}(-1)^j\Big)\indic_{\{0,1\}}(j)
    \end{split}
   \end{equation*}
en donde
   $$\nu=\frac{3^{\sum_{k=1}^n x_k}}{2^n}$$
y con lo anterior obtenemos
   \begin{equation*}
     \begin{split}
  I(D\,|\,\mathbf{p}_0) &=\frac{1}{4}\bigg\{ \Big( 1+\frac{1}{\nu+1} \Big)\Big[ \log 2 - \log 3 + \log\big(1+\frac{1}{\nu+1}\big) \Big] \\
                    &\qquad + \Big( 3-\frac{1}{\nu+1} \Big)\Big[ \log 2 - \log 5 + \log\big(3-\frac{1}{\nu+1}\big) \Big] \bigg\}
     \end{split}
   \end{equation*}
En el Ejemplo \ref{e1} se vio que si el tama\~no de muestra $n\rightarrow\infty$ entonces ocurre una de dos cosas: $\nu\rightarrow\infty$ (lo cual implicar\'ia que $\theta=\frac{3}{4}$) o bien $\nu\rightarrow 0$ (lo cual implicar\'ia que $\theta=\frac{1}{2}$). En el primer caso obtenemos $\lim_{n\rightarrow\infty}I(D\,|\,\mathbf{p}_0)$ es aproximadamente igual a $0{.}03537489$ y en el segundo caso aproximadamente igual a $0{.}03226926$. Resulta muy ilustrativo analizar gr\'aficamente el comportamiento de $I(D\,|\,\mathbf{p}_0)$ para distintos valores de $n$ y compararlo con el comportamiento de $p(x=1\,|\,\mathbf{x})$, bajo ambos escenarios de comportamiento asint\'otico de $\nu\,$. (Sugerencia: simule muestras de tama\~no $n\geq 100$). Ante la pregunta de por qu\'e el supremo de $I(D\,|\,\mathbf{p}_0)$ es mayor cuando $\theta=\frac{3}{4}$ que con $\theta=\frac{1}{2}$, en este caso podemos contestar moment\'anea y parcialmente que la informaci\'on de Fisher alcanza su m\'inimo justamente en $\theta=\frac{1}{2}\,$ (ver Ejemplo \ref{ejem info fisher binomial}, Cap\'itulo 3). Por \'ultimo, para calcular la informaci\'on esperada de este experimento $I(e\,|\,\mathbf{p}_0)$ s\'olo falta calcular $\prob(D_i)$, que viene a ser en este caso la distribuci\'on predictiva a priori conjunta de $n$ observaciones:
   \begin{eqnarray*}
  p(x_1,\ldots,x_n) &=& \int_{\Theta}p(x_1,\ldots,x_n\,|\,\theta)\,p(\theta) \,d\theta \\
      &=& \frac{\nu+1}{2^{\,n+1}}
   \end{eqnarray*}
No procederemos en este caso a calcular expl\'icitamente $I(e\,|\,\mathbf{p}_0)$ porque s\'olo estamos considerando un s\'olo tipo de experimento, y tendr\'ia sentido en tanto se tuvieran distintos experimentos a realizar para la obtenci\'on de datos, se calculan sus informaciones esperadas respectivas y se selecciona aqu\'el que tenga la informaci\'on esperada mayor.$\qquad\diamond$
 \end{ejem}
 
 \bigskip

 \section{Acciones y utilidades generalizadas}
 
   Para la aplicaci\'on de los resultados de teor\'ia de la decisi\'on a la inferencia estad\'istica resulta necesario considerar conjuntos de acciones y espacios de estados infinito no numerables (como puede ser un subconjunto de $\mathbb{R}$, por ejemplo) as\'i como permitir que $\mathcal{E}$ sea un $\sigma$-\'algebra. Para ello son necesarias una serie de justificaciones formales que no analizaremos a detalle (para ello ver Bernardo y Smith (1994)).\par
 
 \medskip
 
  Consideremos un problema de decisi\'on $(\mathcal{E},\mathcal{C},\mathcal{A},\preceq)$ en donde $\mathcal{A}$ y el espacio de estados $\Theta$ son infinito no numerables, $\mathcal{E}$ un $\sigma$-\'algebra. Tenemos entonces que el conjunto de consecuencias $\mathcal{C}$ es tambi\'en infinito no numerable. Sea la funci\'on de utilidad $u:\mathcal{C}\rightarrow\mathbb{R}$. Sea $p(\theta)$ una funci\'on de densidad de probabilidades sobre $\Theta$. Tenemos entonces que la acci\'on \'optima ser\'a aqu\'ella $a^*\in\mathcal{A}$ tal que:
    $$\overline{u}(a^*)=\max_{a\in\mathcal{A}}\overline{u}(a)\,,$$
en donde
    $$\overline{u}(a):=\int_{\Theta}\! u(a,\theta)\,p(\theta)\,d\theta\,.$$
    
Cabe aclarar que $p(\theta)$ es una distribuci\'on de probabilidad condicional en la informaci\'on que se tiene en un momento dado, esto es, puede tratarse de una distribuci\'on a priori o a posteriori, por ejemplo. Retomaremos los conceptos de la secci\'on anterior pero para el caso que ahora nos ocupa.\par
  
 \smallskip
  
 \begin{enun}
   \begin{defn}
     Sea $\Theta$ un espacio de estados. Definimos como la \textit{clase de distribuciones condicionales} sobre $\Theta\,$:
   $$\mathcal{Q}:=\{\,q(\theta)\,:\,q(\theta)\geq 0,\int_{\Theta}q(\theta)\,d\theta=1\}$$
   \end{defn}
 \end{enun}
 
 \smallskip
 
 \begin{enun}
   \begin{defn}\label{funcion de puntaje continua}
     Una \textit{funci\'on de puntaje} para una familia de distribuciones de probabilidad $\mathcal{Q}$ definidas sobre un espacio de estados $\Theta$ es una funci\'on $u:\mathcal{Q}\times\Theta\rightarrow\mathbb{R}\,$. Se dice que esta funci\'on es \textit{suave} si es continuamente diferenciable como funci\'on de $\theta\,$.
   \end{defn}
 \end{enun}
 
 \smallskip
 
 \begin{enun}
   \begin{defn}\label{funcion de puntaje propia continua}
     Una funci\'on de puntaje $u$ es \textit{propia} si para cada distribuci\'on de probabilidad $p(\theta)$ definida sobre $\Theta$ se cumple:
     $$\sup_{q\in\mathcal{Q}}\int_{\Theta}u(q,\theta)\,p(\theta)\,d\theta = \int_{\Theta}u(p,\theta)\,p(\theta)\,d\theta\,,$$
en donde el supremo se alcanza si y s\'olo si $q=p$ casi seguramente, esto es, excepto posiblemente sobre conjuntos de medida cero.
   \end{defn}
 \end{enun}
 
 \smallskip
 
 \begin{enun}
   \begin{defn}\label{funcion de puntaje cuadratica continua}
     Una \textit{funci\'on de puntaje cuadr\'atica} para la familia de distribuciones de probabilidad $\mathcal{Q}$ definidas sobre $\Theta$ es cualquier funci\'on de la forma
    $$u(q,\theta)=A\Big\{ 2\,q(\theta)-\int_{\Theta}q^2(\tilde{\theta)}\,d\tilde{\theta} \Big\} +B(\theta)\,,\quad A>0\,,$$
en donde la funci\'on $B(\,\cdot\,)$ es cualquiera mientras se garantice la existencia de
    $$\overline{u}(q)=\int_{\Theta}u(q,\theta)\,p(\theta)\,d\theta\,.$$ 
   \end{defn}
 \end{enun}
 
 \smallskip
 
 \begin{enun}
   \begin{prop}
     Una funci\'on de puntaje cuadr\'atica es propia.
   \end{prop}
 \end{enun}
 \begin{proof}
   Escogemos $q\in\mathcal{Q}$ de modo que se maximice
   \begin{eqnarray*}
     \overline{u}(q) &=& \int_{\Theta}u(q,\theta)\,p(\theta)\,d\theta \\
     &=& \int_{\Theta}\Big[\,A\big\{ 2q(\theta) - \int_{\Theta}q^2(\tilde{\theta})\,d\theta \big\} + B(\theta) \,\Big] \,p(\theta)\,d\theta \\
     &=& \int_{\Theta}\Big[\, 2Ap(\theta)q(\theta)-Ap(\theta) \int_{\Theta}q^2(\tilde{\theta})\,d\tilde{\theta} + B(\theta)p(\theta) \,\Big]\,d\theta\,,
   \end{eqnarray*}
pero maximizar la expresi\'on anterior (respecto a $q\in\mathcal{Q}$) equivale a maximizar
    $$-\int_{\Theta}\big[\,p(\theta)-q(\theta)\,\big]^2\,d\theta$$
ya que, como $p$ y $B$ son fijas, entonces
   \begin{equation*}
    \begin{split}
     &-\int_{\Theta}\big[\,p(\theta)-q(\theta)\,\big]^2\,d\theta = -\int_{\Theta} \big[\,p^2(\theta)-2p(\theta)q(\theta)+q^2(\theta)\,\big]\,d\theta\\
     &\qquad\qquad =  \frac{1}{A}\int_{\Theta}\big[\,2Ap(\theta)q(\theta)-Ap^2(\theta)\,\big] \,d\theta - \int_{\Theta}p(\theta)\int_{\Theta}q^2(\tilde{\theta})\, d\tilde{\theta}\,d\theta\\
     &\qquad\qquad = \frac{1}{A}\int_{\Theta}\big[\,2Ap(\theta)q(\theta) -Ap(\theta)\int_{\Theta}q^2(\tilde{\theta})\,d\tilde{\theta} -Ap^2(\theta)\,\big]\,d\theta
    \end{split}
   \end{equation*}
y por lo tanto $\overline{u}(q)$ se maximiza sobre $\mathcal{Q}$ siempre y cuando $q=p$ casi seguramente.
 \end{proof}
 
 \smallskip
 
 \begin{enun}
   \begin{defn}\label{funcion de puntaje local continua}
     Una funci\'on de puntaje $u$ es \textit{local} si para cada $q\in\mathcal{Q}$ existen funciones $\{u_{\theta}\,:\,\theta\in\Theta\}$ tales que $u(q,\theta)=u_\theta\big(q(\theta)\big)\,$. 
   \end{defn}
 \end{enun}
 
 \medskip
 
 An\'alogamente al caso discreto, caracterizaremos las funciones de puntaje local:
 
 \smallskip
 
 \begin{enun}
   \begin{prop}
     Si $u:\mathcal{Q}\times\Theta\rightarrow\mathbb{R}$ es una funci\'on de puntaje local, suave y propia, entonces debe ser de la forma
     $$u(q,\theta)=A\log q(\theta)+B(\theta)\,,$$
en donde $A>0$ es una constante arbitraria y $B(\,\cdot\,)$ es cualquier funci\'on siempre y cuando se garantice la existencia de $\overline{u}(q)\,$.
   \end{prop}
 \end{enun}
 \begin{proof}
   Maximizamos respecto a $q\in\mathcal{Q}$ la utilidad esperada
   $$\overline{u}(q)=\int_{\Theta}u(q,\theta)\,p(\theta)\,d\theta$$
sujeto a la condici\'on $\int_{\Theta}q(\theta)\,d\theta=1\,$. Como $u$ es local, lo anterior se reduce a encontrar un punto extremo de
   $$F(q):=\int_{\Theta}u_{\theta}\big(q(\theta)\big)\,p(\theta)\,d\theta - A\Big[\int_{\Theta}q(\theta)\,d\theta-1\,\Big]\,.$$
 Para que $F$ sea estacionaria en alguna $q\in\mathcal{Q}$ es necesario que 
   $$\frac{\partial}{\partial\alpha}F\big(q(\theta)+\alpha\tau(\theta)\big) \restriction_{\alpha=0}=0$$
para cualquier funci\'on $\tau:\Theta\rightarrow\mathbb{R}$ con norma suficientemente peque\~na (ver Jeffreys y Jeffreys (1946), Cap\'itulo 10). Esta condici\'on se reduce a la ecuaci\'on diferencial
   $$D\,u_{\theta}\big(q(\theta)\big)\,p(\theta)-A=0\,,$$
en donde $D\,u_{\theta}$ denota la primera derivada de $u_{\theta}$. Como $u_{\theta}$ es propia entonces el m\'aximo de $F(q)$ debe alcanzarse cuando $q=p$ por lo que una funci\'on de puntaje local, suave y propia debe satisfacer la ecuaci\'on diferencial
   $$D\,u_{\theta}\big(p(\theta)\big)\,p(\theta)-A=0\,,$$
de donde se obtiene que $u_{\theta}\big(p(\theta)\big)=A\log p(\theta) + B(\theta)\,$.
 \end{proof}
 
 \smallskip
 
 \begin{enun}
   \begin{defn}\label{funcion de puntaje logaritmica continua}
     Una \textit{funci\'on de puntaje logar\'itmica} para las distribuciones de probabilidad $q\in\mathcal{Q}$ definidas sobre $\Theta$ es una funci\'on $u:\mathcal{Q}\times\Theta\rightarrow\mathbb{R}$ de la forma
     $$u(q,\theta)=A\log q(\theta)+B(\theta)\,,$$
En donde $A>0$ es una constante arbitraria y $B(\,\cdot\,)$ es cualquier funci\'on que garantice la existencia de $\overline{u}(q)$ para todo $q\in\mathcal{Q}\,$.
   \end{defn}
 \end{enun}
 
 \bigskip
 
 \begin{enun}
   \begin{prop}
     Si nuestras preferencias est\'an descritas por una funci\'on de puntaje logar\'itmica, la p\'erdida esperada de utilidad al usar una densidad de probabilidades $q$ en vez de $p$ est\'a dada por:
    $$\delta(q\,|\,p)=A\int_{\Theta}p(\theta)\log\frac{p(\theta)}{q(\theta)} \,d\theta\,.$$
M\'as a\'un, $\delta(q\,|\,p)\geq 0$ con igualdad si y s\'olo si $q=p$ casi seguramente.
   \end{prop}
 \end{enun}
 \begin{proof}
   An\'aloga a la del caso discreto.
 \end{proof}
 
 \smallskip
 
 \hyphenation{pro-ba-bi-li-da-des}
 
 \begin{enun}
   \begin{defn}\label{discrepancia logaritmica continua}
     La \textit{discrepancia logar\'itmica} de una densidad de probabilidades $p$ estrictamente positiva sobre $\Theta$ respecto a una aproximaci\'on $\hat{p}$ est\'a definida por:
     $$\delta(p\,|\,\hat{p}):=A\int_{\Theta}p(\theta)\log\frac{p(\theta)}{ \hat{p}(\theta)} \,d\theta\,.$$
   \end{defn}
 \end{enun}
 
 \smallskip
 
 \hyphenation{con-ti-nua}
 
 \begin{ejem}
    Utilizando discrepancia logar\'itmica, la mejor aproximaci\'on normal $N(x\,|\,\mu,\lambda)$ para cualquier variable aleatoria absolutamente continua $X$ que toma valores en todo $\mathbb{R}$ con funci\'on de densidad $f_X$ y con primeros dos momentos finitos tales que $\esper(X)=m$ y $\vari(X)=\tau^{-1}$ es aqu\'ella que utiliza $\mu=m$ y $\lambda=\tau\,$. (Recuerde que $\lambda$ es la \textit{precisi\'on}, que es el inverso de la varianza). Los detalles se dejan como ejercicio.
 \end{ejem}
 
 \bigskip
 
 \begin{enun}
   \begin{prop}
     Si nuestras preferencias est\'an descritas por una funci\'on de puntaje logar\'itmica para la clase de densidades de probabilidad $p(\theta\,|\,\mathbf{x})$ definidas sobre $\Theta$, entonces el incremento esperado en utilidad proveniente de los datos $\mathbf{x}$, cuando la densidad de probabilidades a priori es $p(\theta)$, est\'a dado por
     $$A\int_{\Theta}p(\theta\,|\,\mathbf{x})\log\frac{p(\theta\,|\,\mathbf{x})}{ p(\theta)}\,d\theta\,,$$
en donde $p(\theta\,|\,\mathbf{x})$ es la densidad a posteriori de $\theta$ dado $\mathbf{x}\,$, cantidad que resulta ser no negativa, y cero si y s\'olo si $p(\theta\,|\,\mathbf{x})=p(\theta)\,$.
   \end{prop}
 \end{enun}
 \begin{proof}
   An\'aloga a la del caso discreto.
 \end{proof}
 
 \smallskip
 
 \begin{enun}
   \begin{defn}\label{informacion de los datos continua}
     La \textit{cantidad de informaci\'on de los datos} acerca de $\Theta$ proveniente de los datos $\mathbf{x}$ cuando la distribuci\'on a priori es $p(\theta)$ se define como:
     $$I(\mathbf{x}\,|\,p(\theta)):=\int_{\Theta}p(\theta\,|\,\mathbf{x}) \log\frac{p(\theta\,|\,\mathbf{x})}{p(\theta)}\,d\theta\,,$$
en donde $p(\theta\,|\,\mathbf{x})$ es la distribuci\'on a posteriori correspondiente. Es decir,
$$I(\mathbf{x}\,|\,p(\theta)):=\delta(p(\theta)\,|\,p(\theta\,|\,\mathbf{x}))$$
   \end{defn}
 \end{enun}
 
 \smallskip
 
 \begin{enun}
   \begin{defn}\label{informacion esperada de un experimento continua}
     La \textit{informaci\'on esperada de un experimento} $e$ acerca de $\Theta$ cuando la distribuci\'on a priori es $p(\theta)$ est\'a definida por:
     $$I(e\,|\,p(\theta)):=\int_{\mathcal{X}}I(\mathbf{x}\,|\,p(\theta)) \,p(\mathbf{x}\,|\,e)\,d\mathbf{x}$$
en donde $p(\mathbf{x}\,|\,e)$ es la distribuci\'on de probabilidad sobre el conjunto $\mathcal{X}$ de los resultados posibles del experimento.
   \end{defn}
 \end{enun}

 \newpage
 
 $\S$ \textbf{EJERCICIOS}
 \bigskip
 \begin{enumerate}
   \item Una compa\~n\'ia debe decidir si acepta o rechaza un lote de art\'iculos (considere estas acciones como $a_1$ y $a_2$, respectivamente). Los lotes pueden ser de tres tipos: $E_1$ (muy bueno), $E_2$ (aceptable) y $E_3$ (malo). La funci\'on de utilidad se presenta en la siguiente tabla:
     $$\begin{array}{c|c|c|c}
        u(a_i,\theta_j) & E_1 & E_2 & E_3 \\ \hline
               a_1      &  3  &  2  &  0  \\ \hline
               a_2      &  0  &  1  &  3  \\ \hline
       \end{array}$$
La compa\~n\'ia supone que los eventos $E_1,E_2$ y $E_3$ son equiprobables.
     \begin{enumerate}
       \item Describe la estructura del problema de decisi\'on.
       \item Determina las acciones admisibles.
       \item Resuelve el problema de decisi\'on utilizando el criterio de la utilidad esperada m\'axima (Definici\'on \ref{def criterio general de decision}).
       \item Determina todas las distribuciones de probabilidad sobre el espacio de estados tales que se obtiene la misma soluci\'on del inciso anterior.
     \end{enumerate}
   \item Un alumno tiene que presentar examen final de un curso y le queda poco tiempo para estudiar. Supongamos que $\mathcal{A}:=\{a_1,a_2,a_3\}$ donde:
     \begin{equation*}
       \begin{split}
         & a_1:=\,\mbox{Estudiar con detalle la primera parte del curso y nada de la segunda}\\
         & a_2:=\,\mbox{Estudiar con detalle la segunda parte y nada de la primera} \\
         & a_3:=\,\mbox{Estudiar con poco detalle todo el curso}
       \end{split}
     \end{equation*}
Y supongamos que el espacio de estados es $\Theta:=\{E_1,E_2,E_3\}$ donde:     
     \begin{equation*}
       \begin{split}
         & E_1:=\,\mbox{El examen est\'a m\'as cargado hacia la primera parte}\\
         & E_2:=\,\mbox{El examen est\'a m\'as cargado hacia la segunda parte} \\
         & E_3:=\,\mbox{El examen est\'a equilibrado}
       \end{split}
     \end{equation*}
Aunque no se se tiene una funci\'on de probabilidad sobre $\Theta$ supongamos que resulta razonable suponer que $\prob(E_2)>\prob(E_1)$ y que $\prob(E_2)>\prob(E_3)$. Definimos ahora una funci\'on de utilidad sobre el conjunto de consecuencias $\mathcal{C}$ que de hecho ser\'a la calificaci\'on que podr\'a el estudiante obtener:
     $$\begin{array}{c|c|c|c}
         u(a_i,E_j)  &  E_1  &  E_2  &  E_3 \\ \hline
             a_1     &   9   &   2   &   5  \\ \hline
             a_2     &   2   &   9   &   5  \\ \hline
             a_3     &   6   &   6   &   7  \\ \hline
       \end{array}$$
Verifique que de acuerdo a las restricciones del problema la acci\'on $a_1$ nunca tiene la posibilidad de ser la acci\'on \'optima y determine los conjuntos de valores de $\prob(E_1),\prob(E_2)$ y $\prob(E_3)$ para los cuales $a_2\succ a_3$, $a_2\prec a_3$ y $a_2\sim a_3$. 
   \item Considere un problema de decisi\'on en donde el conjunto de acciones $\mathcal{A}$ y el espacio de estados $\Theta$ tienen un n\'umero infinito numerable de elementos. Supongamos que $\mathcal{A}:=\{a_0,a_1,a_2,\ldots\}$ y $\Theta:=\{E_1,E_2,\ldots\}$ y que la funci\'on de utilidad est\'a dada por la siguiente tabla:
     $$\begin{array}{c|c|c|c|c|c|c}
        u(a_i,E_j) & E_1 & E_2 & E_3 & E_4 & E_5 & \ldots \\ \hline
             a_0   & \frac{1}{2} & \frac{1}{2} & \frac{1}{2} & \frac{1}{2} & \frac{1}{2} & \ldots \\ \hline
             a_1   & 1 & 0 & 0 & 0 & 0 & \ldots \\ \hline
             a_2   & 1 & 1 & 0 & 0 & 0 & \ldots \\ \hline
             a_3   & 1 & 1 & 1 & 0 & 0 & \ldots \\ \hline
             a_4   & 1 & 1 & 1 & 1 & 0 & \ldots \\ \hline
             \vdots&\vdots&\vdots&\vdots&\vdots&\vdots&\ddots \\ \hline
       \end{array} $$
Demuestre que la \'unica acci\'on admisible es $a_0$ y que $a_0$ no satisface el criterio de la utilidad esperada m\'axima, sea cual sea la distribuci\'on de probabilidad sobre $\Theta$.
   %\item Una administradora de sociedades de inversi\'on decide construir un portafolios de inversi\'on sencillo que consistir\'a en invertir una cierta cantidad de dinero a un plazo de un a\~no en dos conceptos: un porcentaje en acciones de la empresa ABC y el resto en un pagar\'e bancario que paga una tasa fija anual de 6\%. Aunque no existe manera de predecir exactamente cu\'al ser\'a el rendimiento al cabo de un a\~no de las acciones de ABC, analistas financieros especializados en ABC estiman que dicho rendimiento se ubicar\'a alrededor del 15\% $\pm$ 25\%. Determine los porcentajes de inversi\'on \'optimos con una funci\'on de utilidad igual al rendimiento del portafolios.
   \item Considere el siguiente problema de decisi\'on. En un juego, se tiene un conjunto de 9 cartas que contiene: 2 Ases, 3 Reyes y 4 Sotas. Al jugador, quien paga \$150 por el derecho de jugar, se le entrega una carta al azar y, una vez con esta carta en su poder, puede optar por pedir otra carta o bien pasar. Si decide pasar, simplemente pierde su pago inicial. Si, por el contrario, pide otra carta, las recompensas se pagan de acuerdo a la siguiente tabla:
     $$\begin{array}{c|c}
         \mbox{Cartas} & \mbox{Recompensa} \\ \hline
         \mbox{2 Ases o 2 Reyes} & +\$2000 \\
         \mbox{2 Sotas o 1 As y 1 Sota} & +\$1000 \\
         \mbox{otra combinaci\'on} & -\$1000 
       \end{array}$$
     \begin{enumerate}
       \item Describa la estructura del problema y obtenga la decisi\'on \'optima para un jugador que ya pag\'o su derecho de juego...
         \begin{itemize}
           \item[$a{.}1)$] ... si resuelve decidir sin mirar la primera carta;
           \item[$a{.}2)$] ... si resuelve decidir s\'olo despu\'es de observar la primera carta;
         \end{itemize}
       \item ?`Es preferible mirar la primera carta antes de decidir si se pide una segunda carta o resulta indiferente?
     \end{enumerate}
   \item Verifique los resultados del Ejemplo \ref{aproximar binomial con poisson}.
   \item Del Ejemplo \ref{ejemplo e1 revisited} simule y grafique $I(D\,|\,\mathbf{p}_0)$ para distintos valores de $n$ y compare su comportamiento con el de $p(x=1\,|\,\mathbf{x})$ bajo los dos escenarios posibles.
   \item Demuestre que, utilizando discrepancia logar\'itmica, la mejor aproximaci\'on normal $N(x\,|\,\mu,\lambda)$ para cualquier variable aleatoria absolutamente continua $X$ que toma valores en todo $\mathbb{R}$ con funci\'on de densidad $f_X$ y con primeros dos momentos finitos tales que $\esper(X)=m$ y $\vari(X)=\tau^{-1}$ es aqu\'ella que utiliza $\mu=m$ y $\lambda=\tau\,$. (Recuerde que $\lambda$ es la \textit{precisi\'on}, que es el inverso de la varianza.)
     
 \end{enumerate}

 \chapter{Inferencia estad\'istica param\'etrica bayesiana}
  
 \hyphenation{pro-ble-mas}
 
 El objetivo de haber revisado en el cap\'itulo anterior algunos conceptos y resultados importantes de teor\'ia de la decisi\'on es justamente resolver problemas de inferencia estad\'istica como problemas de decisi\'on. En cada caso supondremos que se tiene un fen\'omeno aleatorio de inter\'es que se modela mediante un vector (o variable) aleatorio $X$ cuya distribuci\'on de probabilidad pertenece a una familia param\'etrica $\mathcal{P}:=\{p(x\,|\,\theta)\,:\,\theta\in\Theta\}\,$ y que se cuenta con una distribuci\'on de probabilidad sobre el espacio param\'etrico $\Theta$, ya sea a priori o a posteriori, denotadas $p(\theta)$ o $p(\theta\,|\,\mathbf{x})$, respectivamente. Utilizaremos $p(\theta)$ en el entendido de que puede tratarse de cualquiera de las dos anteriores, salvo especificaci\'on en contrario. De igual modo utilizaremos indistintamente la distribuciones predictiva a priori $p(x)$ y la distribuci\'on predictiva a posteriori $p(x\,|\,\mathbf{x})\,.$\par
 
 \bigskip

 \section{Estimaci\'on puntual}
 
 El problema de la estimaci\'on puntual se plantea como un problema de decisi\'on $(\mathcal{E},\mathcal{C},\mathcal{A},\preceq)$ en donde el espacio de estados es justamente el espacio param\'etrico $\Theta$ y el conjunto de acciones es tambi\'en $\mathcal{A}=\Theta$, en el sentido de que habremos de tomar la acci\'on de escoger un valor particular para $\theta$. Para evitar confusi\'on, a los elementos de $\mathcal{A}$ los denotaremos mediante $\hat{\theta}$. La funci\'on de utilidad ser\'a entonces una funci\'on $u:\Theta\times\Theta\rightarrow\mathbb{R}\,.$\par
 
 \medskip
 \hyphenation{uti-li-dad} 
 
 \begin{enun}
   \begin{defn}\label{estimacion puntual}
     La \textit{estimaci\'on puntual} de $\theta$ respecto a la funci\'on de utilidad $u(\hat{\theta},\theta)$ y a una distribuci\'on de probabilidad $p(\theta)$ sobre $\Theta$ es la acci\'on \'optima $\hat{\theta}^*\in\mathcal{A}=\Theta$ tal que
  $$\overline{u}(\hat{\theta}^*)=\max_{\hat{\theta}\in\Theta}\overline{u}(\hat{ \theta})\,,$$
en donde
   $$\overline{u}(\hat{ \theta})= \int_{\Theta}u(\hat{\theta},\theta)\,p(\theta)\,d\theta\,=:\esper_{\theta} \big[\,u(\hat{\theta},\theta)\,\big]\,.$$
   \end{defn}
 \end{enun}
 
 \smallskip
 
 \begin{ejem}\label{ejem gral estim puntual}
    Supongamos que $\Theta\subset\mathbb{R}$ y que escogemos la funci\'on de utilidad cuadr\'atica $u(\hat{\theta},\theta):=-(\hat{\theta}-\theta)^2$. Entonces
 \begin{equation*}
  \begin{split}
   &\overline{u}(\hat{\theta})=\esper_{\theta}\big[\,u(\hat{\theta}, \theta)\,\big]\\
   &\qquad = -\esper_{\theta}\big[\,(\hat{\theta}-\theta)^2\,\big] \\
   &\qquad = -\esper_{\theta}\big[\,(\hat{\theta}-\esper_{\theta}(\theta)+ \esper_{\theta}(\theta)-\theta)^2\,\big] \\
   &\qquad = -\esper_{\theta}\big[\,(\hat{\theta}-\esper_{\theta}(\theta))^2 \,\big] -2\,\esper_{\theta}\big[\,(\hat{\theta}-\esper_{\theta}(\theta)) (\esper_{\theta}(\theta)-\theta)\,\big] - \esper_{\theta}\big[\,(\theta- \esper_{\theta}(\theta))^2\,\big]
  \end{split}
 \end{equation*}
en donde el segundo t\'ermino es cero y el tercero es la varianza de $\theta$ por lo que
  $$\overline{u}(\hat{\theta})=-\Big\{\big[\,\hat{\theta}-\esper_{\theta} (\theta)\,\big]^2+\vari_{\theta}(\theta)\Big\}\,.$$
El estimador puntual de $\theta$ es $\hat{\theta}^*\in\mathcal{A}=\Theta$ tal que
  \begin{eqnarray*}
    \overline{u}(\hat{\theta}^*) &=& \max_{\hat{\theta}\,\in\,\Theta}\bigg( -\Big\{\big[\,\hat{\theta}-\esper_{\theta} (\theta)\,\big]^2+\vari_{\theta}(\theta)\Big\} \bigg) \\
    &=& \min_{\hat{\theta}\,\in\,\Theta} \Big\{\big[\,\hat{\theta}-\esper_{\theta} (\theta)\,\big]^2+\vari_{\theta}(\theta)\Big\} 
  \end{eqnarray*}
de donde se obtiene que 
   $$\hat{\theta}^*=\esper_{\theta}(\theta)=\int_{\Theta}\theta\,p(\theta) \,d\theta\,,$$
siempre y cuando dicha esperanza exista, por supuesto. $\qquad\diamond$
 \end{ejem}
 
 \medskip
 
 Generalizando el ejemplo anterior al caso en que $\Theta\subset\mathbb{R}^k$, si se tiene como funci\'on de utilidad la forma cuadr\'atica
  $$u(\mathbf{\hat{\theta}},\mathbf{\theta})=-(\mathbf{\hat{\theta}}- \mathbf{\theta})^T H (\mathbf{\hat{\theta}}-\mathbf{\theta})$$
entonces
  $$\overline{u}(\hat{\theta}) = - \int_{\Theta}(\mathbf{\hat{\theta}}- \mathbf{\theta})^T H (\mathbf{\hat{\theta}}-\mathbf{\theta}) \,p(\theta)\,d\theta\,.$$
Derivando $\overline{u}(\hat{\theta})$ respecto a $\hat{\theta}$ e igualando a cero obtenemos
   $$-2H\int_{\Theta}(\hat{\theta}-\theta)\,p(\theta)\,d\theta = 0$$
de donde $H\hat{\theta}=H\,\esper_{\theta}(\theta)\,$. Si $H^{-1}$ existe entonces
     $$\hat{\theta^*}=\esper_{\theta}(\theta)=\int_{\Theta}\theta\,p(\theta) \,d\theta\,,$$
siempre y cuando dicha esperanza exista, por supuesto. N\'otese que lo anterior, m\'as que un ejemplo, es un resultado de car\'acter general.\par

 \bigskip
 
 El resultado es an\'alogo si lo que se desea es estimar puntualmente una observaci\'on futura de $X\,,$ misma que denotaremos $\hat{x}^*.$ En este caso el espacio de estados es $Ran X$ (el conjunto de todos los valores posibles de $X$) y por lo tanto la estimaci\'on puntual de una observaci\'on futura de $X$ respecto a la funci\'on de utilidad $u(\hat{x},x)$ y a una distribuci\'on predictiva $p(x)$ es la acci\'on \'optima $\hat{x}^*\in Ran X$ tal que
  $$\overline{u}(\hat{x}^*)=\max_{\hat{x}\in\ Ran X}\overline{u}(\hat{x})\,,$$
en donde
   $$\overline{u}(\hat{x})= \int_{Ran X}u(\hat{x},x)\,p(x)\,dx\,=:\esper_{p(x)} \big[\,u(\hat{x},x)\,\big]\,.$$ 
 
 Y nuevamente, si la funci\'on de utilidad es la de el ejemplo anterior entonces:
   $$\hat{x}^*\,=\,\int_{Ran X}x\,p(x)\,dx\,.$$
 
 \bigskip
  
 \begin{ejem}\label{ejem caseta estim puntual}
    Nos remitimos al Ejercicio 3 del Cap\'itulo 3, pero aqu\'i supondremos que de acuerdo al responsable de la caseta en el mayor de los casos el n\'umero promedio de autos por minuto es 8 (y no 12). De acuerdo a un procedimiento similar al que se utiliza para resolver dicho ejercicio se obtiene como distribuci\'on a priori para $\lambda$ una distribuci\'on Gamma con hiperpar\'a-\linebreak metros $\alpha=9{.}108$ y $\beta=0{.}01012$ y por tanto la distribuci\'on a posteriori de $\lambda$ es Gamma tambi\'en:
    $$p(\lambda\,|\,\mathbf{x})=Ga(\lambda\,|\,9{.}108+\sum x_j\,,\,0{.}01012 + n)\,.$$
 \end{ejem}
   Suponiendo que como informaci\'on muestral tenemos $\mathbf{x}=(679\,,\,703)$ entonces
    $$p(\lambda\,|\,\mathbf{x})=Ga(\lambda\,|\,1391{.}108\,,\,2{.}01012)\,.$$ 
  Utilizando la funci\'on de utilidad del Ejemplo \ref{ejem gral estim puntual} obtenemos como estimaci\'on puntual de $\lambda\,:$
    $$\hat{\lambda}^*=\frac{1391{.}108}{2{.}01012}=692{.}05\,.$$
 \bigskip
 \bigskip

 \section{Contraste de hip\'otesis}
 
 Supongamos que se tienen $m$ hip\'otesis acerca del par\'ametro (o vector de par\'ametros) $\theta\,:$
   $$H_1:\theta\in\Theta_1\quad,\quad H_2:\theta\in\Theta_2\quad,\quad\ldots\quad,\quad H_m:\theta\in \Theta_m\,,$$
en donde los conjuntos $\Theta_j$ son subconjuntos del espacio param\'etrico $\Theta\,.$ Podemos suponer que los subconjuntos $\Theta_j$ son disjuntos. En caso de que no lo fueren, los redifinimos de modo que s\'i lo sean. Por ejemplo, si $\Theta_i\cap\Theta_k\neq\varnothing$ definimos un nuevo subconjunto $\Theta_j:=\Theta_i\cap\Theta_k$ y redefinimos $\Theta_i$ como $\Theta_i\setminus\Theta_j$ y a $\Theta_k$ como $\Theta_k\setminus\Theta_j\,,$ adem\'as de agregar la hip\'otesis $H_j:\theta\in\Theta_j\,.$\par
 \medskip
 En caso de que $\displaystyle{\cup_{j\,=\,1}^m\Theta_j\neq\Theta}$ definimos el subconjunto $\displaystyle{\Theta_{m+1}:=\Theta\setminus\cup_{j\,=\,1}^m\Theta_j}$ de modo que $\displaystyle{\{\Theta_j:j=1,\ldots,m+1\}}$ sea partici\'on de $\Theta\,,$ adem\'as de agregar la hip\'otesis $H_{m+1}:\theta\in\Theta_{m+1}\,.$\par
 \medskip
 Con base en lo anterior, supongamos entonces que los subconjuntos $\Theta_1\,,\ldots\,, \Theta_m$ constituyen una partici\'on del espacio param\'etrico $\Theta\,.$ El contraste de las hip\'otesis $H_1\,,\ldots\,,H_m$ se plantea como un problema de decisi\'on que consiste en escoger una de estas hip\'otesis. Definimos como espacio de estados
   $$\Phi:=\{H_1\,,\,\ldots\,,\,H_m\}\,,$$
y como medida de probabilidad $\prob$ sobre $\Phi$ utilizamos la distribuci\'on a priori o a posteriori (seg\'un sea el caso) de $\theta$ ya que
    $$\prob(H_j)=\prob(\theta\in\Theta_j)=\int_{\Theta_j}p(\theta\,|\,\mathbf{x}) \,d\theta\,.$$
    
   Consideremos el conjunto de acciones $\mathcal{A}:=\{a_1,\ldots,a_m\}$ en donde $a_j$ representa la acci\'on de actuar como si la hip\'otesis $H_j$ fuese a ocurrir. Si se define adem\'as una funci\'on de utilidad $u:\mathcal{A}\times\Phi\rightarrow\mathbb{R}$ podemos entonces resolver el problema de elegir una de las hip\'otesis como un problema de decisi\'on simple: se elige aquella $a_*\in\mathcal{A}$ tal que
     $$\overline{u}(a_*)=\max_{a_i\,\in\,\mathcal{A}}\overline{u}(a_i)$$
en donde
     $$\overline{u}(a_i)=\sum_{j\,=\,1}^m u(a_i,H_j)\prob(H_j)\,.$$
 
 \bigskip
 
  En caso de que se tengan hip\'otesis pero en relaci\'on a una observaci\'on futura de la variable o vector aleatorio $X\sim p(x\,|\,\theta)$ el procedimiento es an\'alogo:
     $$H_1:X\in\mathcal{X}_1\quad,\quad\ldots\quad,\quad H_m:X\in\mathcal{X}_m\,,$$
en donde $\mathcal{X}=Ran\,X$ y $\{\mathcal{X}_1,\ldots,\mathcal{X}_m\}$ es partici\'on de $\mathcal{X}\,.$ Como medida de probabilidad $\prob$ sobre el espacio de estados $\Phi$ utilizamos la distribuci\'on predictiva a priori o a posteriori (seg\'un sea el caso) ya que
    $$\prob(H_j)=\prob(X\in\mathcal{X}_j)=\int_{\mathcal{X}_j}p(x\,|\,\mathbf{x}) \,d\theta\,.$$
  
  \medskip
  
  Pero\ldots ?` qu\'e funci\'on de utilidad ocupar? Depender\'a de las caracter\'isticas de cada problema y de todo aquello que se desee sea tomado en consideraci\'on. Por ejemplo, si utilizamos una funci\'on de utilidad muy simple como
     $$u(a_i,H_j):=\indic_{\{i\,=\,j\}}$$
obtenemos
     $$\overline{u}(a_i)=\sum_{j\,=\,1}^m \indic_{\{i\,=\,j\}}\prob(H_j) = \prob(H_i) = \int_{\Theta_i}p(\theta\,|\,\mathbf{x})\,d\theta$$
y por lo tanto la soluci\'on \'optima ser\'a aquella $a_*\in\mathcal{A}$ tal que
     $$\overline{u}(a_*)=\max_{j\,=\,1,\ldots,m}\prob(H_j)\,,$$
es decir, bajo esta funci\'on de utilidad la decisi\'on \'optima es escoger aquella hip\'otesis que tenga la probabilidad m\'as alta de ocurrir. Dijimos que esta funci\'on de utilidad es demasiado simple porque s\'olo toma en consideraci\'on la probabilidad de cada hip\'otesis, sin tomar en cuenta aspectos de otra \'indole que pudiera ser de inter\'es tomar tambi\'en en consideraci\'on (por ejemplo, consideraciones de tipo econ\'omico) como se ilustrar\'a en el siguiente:\par
 \smallskip
 
 \begin{ejem}\label{ejem caseta hipotesis}
   Continuando con el Ejemplo \ref{ejem caseta estim puntual} supongamos, de manera simplificada, que de acuerdo a normas de la Secretar\'ia de Comunicaciones y Transportes el n\'umero de cobradores que se deben de tener en dicha caseta va de acuerdo al n\'umero de autos que llegan, y que en el caso particular de los viernes de 5 a 8 p.m. resulta como sigue:
   $$\begin{array}{c|c}\hline
      \mbox{N\'um. de autos} & \mbox{N\'um. de cobradores} \\ \hline
      \mbox{0 a 690} & 5 \\
      \mbox{691 a 750} & 10 \\
      \mbox{m\'as de 750} & 15 \\ \hline 
     \end{array}$$
  El responsable de la caseta est\'a en libertad de contratar con anticipaci\'on al n\'umero cobradores que considere pertinentes para cada viernes de acuerdo a sus expectativas de aforo vehicular. Pero si sus expectativas se ven rebasadas tendr\'a que contratar cobradores emergentes. Supongamos que un cobrador contratado con anticipaci\'on cuesta \$300 pesos pero uno contratado de \'ultima hora (emergente) cuesta \$700. De acuerdo a la informaci\'on que se tiene (Ejemplo \ref{ejem caseta estim puntual}) el responsable de la caseta desea tomar una decisi\'on \'optima en cuanto al n\'umero de cobradores a contratar con anticipaci\'on (5, 10 o 15).\par
 \smallskip
 Lo anterior se puede plantear como un contraste de hip\'otesis:
     $$H_1:X\in\{0,1,\ldots,690\}\,,\,H_2:X\in\{691,\ldots,750\}\,,\, H_3:X\in\{751,752,\ldots\}\,,$$
en donde, recordemos, $X$ representa el n\'umero de autos que llegan a la caseta. Con la informaci\'on del Ejemplo \ref{ejem caseta estim puntual} as\'i como del Ejemplo \ref{ejem fam conjugada} tenemos que la distribuci\'on predictiva a posteriori es Poisson-Gamma:
    $$p(x\,|\,\mathbf{x})=Pg(x\,|\,1391{.}108,2{.}01012,1)$$
con lo que
  $$\prob(H_1)=0{.}4849\quad,\quad \prob(H_2)=0{.}4786\quad,\quad\prob(H_3)= 0{.}0365\,.$$
  S\'olo nos falta la funci\'on de utilidad, que en este caso est\'a impl\'icita en las condiciones mismas del problema:
  $$\begin{array}{c|c|c|c|c} \hline
    \prob(H_j) & 0{.}4849 & 0{.}4786 & 0{.}0365 & { } \\ \hline
    u(a_i,H_j) &    H_1   &    H_2   &    H_3   & \overline{u}(a_i) \\ \hline
         a_1   &  -\$1500 &  -\$5000 &  -\$8500 & -\$3,430{.}60 \\
         a_2   &  -\$3000 &  -\$3000 &  -\$6500 & -\$3,127{.}75 \\
         a_3   &  -\$4500 &  -\$4500 &  -\$4500 & -\$4,500{.}00 \\ \hline
     \end{array}$$
  Por ejemplo $u(a_2,H_3)=-\$6500$ porque en este caso la acci\'on $a_2$ implica contratar 10 cobradores con anticipaci\'on y si el escenario que ocurre finalmente es $H_3$ entonces esto implica contratar 5 cobradores emergentes y por ello el desembolso total es de $10\times\$300 + 5\times\$700 = \$6500\,.$\par
  \smallskip
  De acuerdo a lo anterior tenemos que la soluci\'on \'optima es $a_2$ por tener la m\'axima utilidad esperada. N\'otese que $a_2$ implica actuar como si $H_2$ fuese a ocurrir, y $H_2$ no es precisamente la hip\'otesis que tiene la mayor probabilidad de cumplirse. Esto es porque en este caso la funci\'on de utilidad tom\'o en cuenta no s\'olo las probabilidades de los escenarios sino tambi\'en la intensidad de sus consecuencias econ\'omicas.
 \end{ejem}
 
 \bigskip

 \section{Estimaci\'on por regiones}
 
  En ocasiones, la descripci\'on de la informaci\'on sobre $\theta$ (o bien sobre una observaci\'on futura de $X$) a trav\'es de $p(\theta\,|\,\mathbf{x})$ (o bien $p(x\,|\,\mathbf{x})$) no resulta accesible para cierto tipo de usuarios de la estad\'istica, a quienes resulta preferible obtener regiones (subconjuntos) $C\subset\Theta$ (o bien $C\subset\mathcal{X}=Ran\,X$) que tengan una probabilidad dada de contener al valor correcto de $\theta$ (o de una observaci\'on futura de $X$). De la construcci\'on de estas regiones nos ocupamos en esta secci\'on.\par
 \medskip
   
 \begin{enun}
   \begin{defn}\label{region de probabilidad}
     Una regi\'on (o subconjunto) $C\subset\Theta$ tal que
      $$\int_C p(\theta\,|\,\mathbf{x})\,d\theta = \alpha$$
en donde $0\leq\alpha\leq 1$ es llamada \textit{regi\'on de probabilidad} $\alpha$ para $\theta$ con respecto a $p(\theta\,|\,\mathbf{x})\,.$
   \end{defn}
 \end{enun}
 
 N\'otese que $C$ no es necesariamente un intervalo. La soluci\'on para $C$  en la ecuaci\'on $\int_C p(\theta\,|\,\mathbf{x})\,d\theta = \alpha$ no es \'unica y por tanto podemos hablar del conjunto de soluciones
     $$\mathcal{A}:=\{C\subset\Theta:\int_C p(\theta\,|\,\mathbf{x})\,d\theta = \alpha\}\,,$$
lo cual implica la necesidad de definir un criterio adicional para elegir una regi\'on $C$ adecuada. Esto se puede resolver como un problema de decisi\'on en donde el conjunto $\mathcal{A}$ que acabamos de definir es el conjunto de acciones (es decir, cada acci\'on es una de las distintas regiones que podemos elegir), el espacio de estados es el espacio param\'etrico $\Theta$ cuya medida de probabilidad queda definida mediante $p(\theta\,|\,\mathbf{x})\,.$ S\'olo nos hace falta una funci\'on de utilidad que contenga ese criterio adicional, que puede ser, por ejemplo, el preferir regiones $C$ que tengan el menor tama\~no posible, mismo que denotaremos $\|C\|$, pero que contengan al valor correcto de $\theta\,:$
   $$u(C,\theta)=-k\|C\| + \indic_C(\theta)\,,\quad k>0\,.$$
 Mediante esta funci\'on de utilidad obtenemos la utilidad esperada para cada $C\in\mathcal{A}$ mediante
   $$\overline{u}(C)=\int_{\Theta}u(C,\theta)p(\theta\,|\,\mathbf{x})\,d\theta = -k\|C\|+\alpha\,,$$
de donde es claro entonces que la regi\'on \'optima ser\'a aqu\'ella $C^*\in\mathcal{A}$ tal que su tama\~no $\|C^*\|$ sea m\'inimo. A tal $C^*$ se le denomina \textit{regi\'on de probabilidad $\alpha$ de m\'axima densidad}.\par
 \bigskip
 
 \begin{ejem}\label{ejem estim regiones}
   Utilizaremos la informaci\'on del Ejemplo \ref{ejem caseta estim puntual}. Aunque ya dijimos que las regiones que se pueden construir no son necesariamente intervalos, supongamos en este caso que deseamos construir un intervalo de probabilidad $0{.}95$ de m\'axima densidad para $\lambda\,.$ Representemos dicho intervalo mediante $[\,\lambda_1,\lambda_2\,]\,.$ Entonces el problema consiste en encontrar los valores para $\lambda_1$ y $\lambda_2$ tal que $\prob(\lambda\in[\,\lambda_1,\lambda_2\,])=0{.}95$ y que la longitud del intervalo $[\,\lambda_1,\lambda_2\,]$ sea m\'inima. Esto es
  \begin{equation*}
    \begin{split}   
      & \mbox{minimizar:}\qquad h(\lambda_1,\lambda_2)=\lambda_2-\lambda_1 \\
      & \mbox{sujeto a:}\quad\,\int_{\lambda_1}^{\,\lambda_2}Ga(\lambda\,|\, 1391{.}108\,,\,2{.}01012)\,d\lambda = 0{.}95\,.
    \end{split}
  \end{equation*}
 \end{ejem}
 Resolviendo num\'ericamente lo anterior obtenemos como soluci\'on \'optima el intervalo $[\,655{.}88\,,\,728{.}6\,]\,.$

 %\section{Suficiencia}
 %\section{An\'alisis conjugado}
 %\section{An\'alisis asint\'otico}
 %\section{An\'alisis de referencia}
  
 \newpage
 
 $\S$ \textbf{EJERCICIOS}
 \bigskip
 \begin{enumerate}
  \item Obtenga el estimador puntual $\hat{\theta}^*\in\Theta\subset\mathbb{R}$ bajo las siguientes funciones de utilidad:
    \begin{enumerate}
      \item $u(\hat{\theta},\theta)=k|\hat{\theta}-\theta|$ con $k<0$ una constante. 
      \item $\displaystyle{u(\hat{\theta},\theta)=-\Big(\frac{\hat{\theta}-\theta}{\hat{ \theta}}\Big)^2\,.}$
    \end{enumerate}
  \item Respecto al Ejemplo \ref{e2} supongamos que de los 150 expedientes revisados 17 resultan incompletos. Obtenga estimaciones puntuales de $\theta$ bajo las siguientes funciones de utilidad:
    \begin{enumerate} 
      \item $u(\hat{\theta},\theta)=-(\hat{\theta}-\theta)^2\,.$ 
      \item $u(\hat{\theta},\theta)=-|\hat{\theta}-\theta|\,.$
      \item $\displaystyle{u(\hat{\theta},\theta)=-\Big(\frac{\hat{\theta}-\theta}{\hat{ \theta}}\Big)^2\,.}$
    \end{enumerate}
  \item Una m\'aquina produce cierto componente que debe tener una longitud especificada. Sea la variable aletoria $X$ igual al margen de error en dicha longitud y supongamos que se distribuye Normal con media cero y precisi\'on $\lambda>0$ desconocida. Suponga que no cuenta con informaci\'on a priori y que obtiene una muestra 
  $$\mathbf{x}=(0{.}033,0{.}002,-0{.}019,0{.}013,-0{.}008,-0{.}0211,0{.}009,0{.}021, -0{.}015)\,.$$
    Construya un intervalo de probabilidad $0{.}95$ de m\'axima densidad para el margen de error en la longitud de dicha componente.

 \end{enumerate}

 %\chapter{Implementac\'on num\'erica} 
 
 %\chapter{Selecci\'on de modelos}
 
 %\chapter{An\'alisis jer\'arquico}
 
 %\chapter{Estad\'istica bayesiana no param\'etrica}

 \backmatter
 
 \chapter{\bibname}
 
%Artículo -> \autor{}\anio{}\articulo{}\revista{}\volumen{}\pags{}\salta

% Libro   -> \autor{}\anio{}\libro{}\editorial{}\salta

\autor{Albert, J.}\anio{2007}\libro{Bayesian Computation with R}\editorial{Springer}\salta

\autor{Bayes, T.}\anio{1763}\articulo{An Essay Towards Solving a Problem in the Doctrine of Chances}\revista{Philos. Trans. Royal Soc. London}\volumen{61.53}\pags{370-418}\salta

\autor{Bernardo, J.M.}\anio{1979}\articulo{Reference posterior distributions for bayesian inference}\revista{Journal of the Royal Statistical Society, Serie B}\volumen{41}\pags{113-147}\salta

\autor{Bernardo, J.M. y Smith, A.F.M}\anio{1994}\libro{Bayesian Theory}\editorial{Wiley}\salta

\autor{Bernoulli, J.}\anio{1713}\libro{Ars Conjectandi}\editorial{Baseae, Impensis Thurnisiorum}\salta

\autor{Casella, G. y Berger, R.L.}\anio{2002}\libro{Statistical Inference}\editorial{Duxbury}\salta

\autor{Chen, M-H., Shao, Q-M., Ibrahim, J.G.}\anio{2000}\libro{Monte Carlo Methods in Bayesian Computation}\editorial{Springer}\salta

\autor{Congdon, P.}\anio{2005}\libro{Bayesian Models for Categorical Data}\editorial{Wiley}\salta

\autor{Datta, G.S., Mukerjee, R.}\anio{2004}\libro{Probability Matching Priors: Higher Order Asymptotics}\editorial{Springer}\salta

\autor{Gelman, A., Carlin, J.B., Stern, H.S., Rubin, D.B.}\anio{1995}\libro{Bayesian Data Analysis}\editorial{Chapman \& Hall}\salta

\autor{Ghosh, J.K., Ramamoorthi, R.V.}\anio{2003}\libro{Bayesian Nonparametrics}\editorial{Springer}\salta

\autor{Guti\'errez Pe\~na, E.A.}\anio{1995}\libro{Bayesian Topics Relating to the Exponential Family}\editorial{tesis doctoral, University of London}\salta

\autor{Guti\'errez Pe\~na, E.A.}\anio{1998}\articulo{An\'alisis bayesiano de modelos jer\'arquicos lineales}\revista{Monograf\'ias IIMAS-UNAM}\volumen{7} N\'um. 16.\salta

\autor{Jeffreys, H.}\anio{1939/1961}\libro{Theory of Probability}\editorial{Oxford University Press}\salta

\autor{Jeffreys, H. y Jeffreys, B.S}\anio{1946/1972}\libro{Methods of Mathematical Physics}\editorial{Cambridge University Press}\salta

\autor{Kahneman, D., Tversky, A.}\anio{1979}\articulo{Prospect Theory: An Analysis of Decision under Risk}\revista{Econometrica}\volumen{47}\pags{263-292}\salta

\autor{Laplace, P.S.}\anio{1774}\libro{Essai Philosophique sur les Probabilites}\editorial{Dover (a partir de la 6a edici\'on, 1951)}\salta

\autor{Le Cam, L.}\anio{1990}\articulo{Maximum Likelihood: An Introduction}\revista{International Statistical Review}\volumen{58}\pags{153-171}\salta

\autor{Lehmann, E.L. y Casella, G.}\anio{1998}\libro{Theory of Point Estimation}\editorial{Springer}\salta

\autor{Lindley, D.V.}\anio{2000}\articulo{The Philosophy of Statistics}\revista{The Statistician}\volumen{49}\pags{293-337}\salta

\autor{Lynch, S.M.}\anio{2007}\libro{Introduction to Applied Bayesian Statistics and Estimation for Social Scientists}\editorial{Springer}\salta

\autor{Marin, J.M., Robert, C.P.}\anio{2007}\libro{Bayesian Core}\editorial{Springer}\salta
 
\autor{Migon, H.S y Gammerman, D.}\anio{1999}\libro{Statistical inference: an integrated approach}\editorial{Oxford University Press}\salta

\autor{Press, S.J.}\anio{2003}\libro{Subjective and Objective Bayesian Statistics: Principles, models, and applications}\editorial{Wiley}\salta

\autor{Press, S.J., Tanur, J.M.}\anio{2001}\libro{The Subjectivity of  Scientists and the Bayesian Approach}\editorial{Wiley}\salta

\autor{Rachev, S.T., Hsu, J.S.J., Bagasheva, B.S., Fabozzi, F.J.}\anio{2008}\libro{Bayesian methods in finance}\editorial{Wiley}\salta

\autor{Robert, C.P.}\anio{2007}\libro{The Bayesian Choice, 2nd edition}\editorial{Springer}\salta

\autor{Robert, C.P., Casella, G.}\anio{2004}\libro{Monte Carlo Statistical Methods}\editorial{Springer}\salta

\autor{Rossi, P.E., Allenby, G.M., McCulloch, R.}\anio{2005}\libro{Bayesian Statisics and Marketing}\editorial{Wiley}\salta

\autor{Stigler, S.M.}\anio{1986a}\libro{The History of Statistics}\editorial{Harvard University Press (Cambridge)}\salta

\autor{Stigler, S.M.}\anio{1986b}\articulo{Laplace's 1774 memoir on inverse probability}\revista{Statist. Sci.}\volumen{1}\pags{359-378}\salta

\autor{West, M., Harrison, J.}\anio{1997}\libro{Bayesian Forecasting and Dynamic Models}\editorial{Springer}\salta

\autor{Wolpert, L.}\anio{1992}\libro{The Unnatural Nature of Science}\editorial{Harvard University Press}\salta

 %\chapter{Ap\'endice}

 \end{document}